%% file: main.tex
\documentclass[a4paper,11pt]{article}
\pdfoutput=1 

\usepackage{jcappub} 

\usepackage[T1]{fontenc} 

\usepackage[utf8]{inputenc} 
\usepackage{diagbox}

\usepackage{graphicx}
\usepackage{caption}
\usepackage{subcaption}

\usepackage[nameinlink,noabbrev]{cleveref}
\crefname{equation}{eq.}{eqs.}
\Crefname{Equation}{Eq.}{Eqs.}

\usepackage{tikz}
\usetikzlibrary{shadows,positioning}
\colorlet{colD}{red!40}
\colorlet{colIP}{cyan!40}
\colorlet{colV}{blue!40}
\colorlet{colBorder}{gray!70}
\tikzset
  {mybox/.style=
    {rectangle,rounded corners,drop shadow,minimum height=1cm,
     minimum width=2cm,align=center,fill=#1,draw=colBorder,line width=1pt
    },
   myarrow/.style=
    {draw=#1,line width=3pt,-stealth,rounded corners
    },
   mylabel/.style={text=#1}
  }

\makeatletter
\newcommand\notsotiny{\@setfontsize\notsotiny\@vipt\@viipt}
\makeatother

\usepackage{bm}

\def\aap{A\&A}%

\def\apjs{ApJS}

\def\apjl{ApJ}

\usepackage{aasmacros}

\newcommand{\Cos}[1]{c_{#1}}
\newcommand{\Sin}[1]{s_{#1}}

\newcommand{\SinS}[1]{s^2_{#1}}

\newcommand{\amp}{$\mathcal{A}$}
\newcommand{\spec}{$\mathcal{B}$}
\newcommand{\ang}{$\mathcal{C}$}

\newcommand{\myvector}[1]{#1}

\def\LB{{\it LiteBIRD}}
\def\Planck{{\it Planck}}
\def\WMAP{{\it WMAP}}
\def\COBE{{\it COBE}}
\def\fgbuster{{\tt FGBuster}}

\def\obs{{\mathrm{obs}}}

\usepackage{multirow}

\usepackage{lineno}

\title{\boldmath LiteBIRD Science Goals and Forecasts: constraining isotropic cosmic birefringence}

\input{authors_affiliations_LB_2021.27_general_jcap}

\emailAdd{alessandro.gruppuso@inaf.it}
\emailAdd{josquin@apc.in2p3.fr}

\abstract{
Cosmic birefringence (CB) is the rotation of the photons' linear polarisation plane during propagation. Such an effect is a tracer of parity-violating extensions of standard electromagnetism and would probe the existence of a new cosmological field acting as dark matter or dark energy. It has become customary to employ cosmic microwave background~(CMB) polarised data to probe such a phenomenon. Recent analyses on \Planck\ and \WMAP\ data provide a hint of detection of the isotropic CB angle with an amplitude of around $0.3^\circ$ at the level of $2.4$ to $3.6\sigma$. 
In this work, we explore the \LB\ capabilities in constraining such an effect, accounting for the impact of the more relevant systematic effects, namely foreground emission and instrumental polarisation angles.
We build five semi-independent pipelines and test these against four different simulation sets with increasing complexity in terms of non-idealities.

All the pipelines are shown to be robust and capable of returning the expected values of the CB angle within statistical fluctuations for all the cases considered.
We find that the uncertainties in the CB estimates increase with more complex simulations. However, the trend is less pronounced for pipelines that account for the instrumental polarisation angles.
For the most complex case analysed, we find that \LB\ will be able to detect a CB angle of $0.3^\circ$ 
with a statistical significance ranging from $5$ to $13 \, \sigma$, depending on the pipeline employed, where the latter uncertainty corresponds to a total error budget of the order of $0.02^\circ$. 
}

\begin{document}


\maketitle
\flushbottom

\section{Introduction}
\label{sec:intro}

Parity-violating extensions of the standard electromagnetism due to the coupling between photons and a (pseudo-)scalar field, for instance, an axion, via a Chern-Simons term~\cite{Carroll:1989vb}, can be traced with the cosmic birefringence (henceforth CB) effect, i.e.\ the in-vacuo\footnote{Here, with ``in-vacuo'' we mean from the point of view of the standard model.} rotation of the linear polarisation plane of photons during their propagation~\cite{Carroll:1991zs, Harari:1992ea, PhysRevLett.81.3067, Kostelecky:2007zz, Balaji_2003}.
Such a rotation is expected to be null in the standard Maxwellian electromagnetism, so detecting a CB angle different from zero would probe the existence of a new cosmological scalar field, possibly acting as dark matter or dark energy~\cite{Marsh:2015xka, Ferreira:2020fam}.

In particular, for ultra-light scalar fields, the Chern-Simons term predicts a CB rotation proportional to the slowly-rolling distance travelled by the photons~\cite{Harari:1992ea,Li:2008tma, Pospelov:2008gg}. 
Since cosmic microwave background (CMB) photons are linearly polarised at the last scattering surface due to Thomson scattering, and have travelled the furthest possible distance through the Universe, it has become customary to exploit polarised CMB data to probe for such an effect. 

After \COBE~\cite{1992ApJ...397..420B}, \WMAP~\cite{WMAP:2003ogi}, and \Planck~\cite{Planck:2018nkj}, \LB~\cite{2022LB_ptep} is the next CMB space-based mission aiming to measure the polarised signal at large angular scales with unprecedented precision.
This work is part of a series of papers presenting the science
achievable by \LB\ space mission and expanding on the overview
published in ref.~\cite{2022LB_ptep}.
In particular, we focus on CB and forecast \LB's\ capabilities in constraining this phenomenon, taking into account realistic systematic effects of both instrumental and astrophysical origins. 

Our present analysis is limited to the {\it isotropic} CB effect, where the rotation is constant over the whole sky and over time.
Hence, we do not consider here either the isotropic CB effect with a time-varying rotation angle due to an oscillatory motion of the axion background\footnote{Such dark-matter axions with mass scales around $10^{-21} \mathrm{eV} \lesssim m_\phi \lesssim 10^{-19} \mathrm{eV}$ have been constrained by the polarisation oscillation measurements of BICEP/Keck~\cite{BICEPKeck:2020hhe, BICEPKeck:2021sbt}, SPT~\cite{SPT-3G:2022ods}, and {\sc Polarbear}~\cite{POLARBEAR:2023ric}.} \cite{Finelli:2008jv, Fedderke:2019ajk}, or the anisotropic CB effect, where the rotation angle depends on the direction of observation~\cite{Kamionkowski:2008fp}, and which has been found to be compatible with zero in previous studies~\cite{Contreras:2017sgi, SPT:2020cxx, Namikawa:2020ffr, Gruppuso:2020kfy, Bortolami:2022whx, Zagatti:2024jxm, BICEP:2023ksh}.
In this case, the CMB angular power spectra are rotated by the isotropic CB angle $\beta$ in the following way, see 
e.g. Refs.~\cite{Lue:1998mq,Feng:2004mq,Liu:2006uh},
\begin{align}
    C_\ell^{TT,\obs} &= C_\ell^{TT},\\
    C_\ell^{TE,\obs} &= C_\ell^{TE}\cos(2\beta), \label{eq:TEobs} \\
    C_\ell^{TB,\obs} &= C_\ell^{TE}\sin(2\beta), \label{eq:TBobs}\\
    C_\ell^{EE,\obs} &= C_\ell^{EE}\cos^2(2\beta) + C_\ell^{BB}\sin^2(2\beta), \label{eq:EEobs}\\
    C_\ell^{BB,\obs} &= C_\ell^{BB}\cos^2(2\beta) + C_\ell^{EE}\sin^2(2\beta), \label{eq:BBobs}\\
    C_\ell^{EB,\obs} &= \frac{1}{2} \left(C_\ell^{EE} - C_\ell^{BB}\right)\sin(4\beta), \label{eq:EBobs}
\end{align}
where the spectra with the ``$\obs$'' label are observed and the spectra without the ``$\obs$'' tag are those that would be observed in the case of null CB. 
\Cref{eq:TBobs,eq:EBobs} 
ignore any pre-recombination $TB$ and $EB$ signals, as we consider CB the only parity-violating extension to the standard model (see e.g. ref.~\cite{Komatsu:2022nvu} for a review).  

Unfortunately, CB investigations based solely on CMB data are limited due to a degeneracy between the instrument polarisation angles ($\alpha$) and the birefringence angle ($\beta$). This degeneracy is the main obstacle in performing such analyses, since the uncertainty in $\alpha$ typically dominates the total error budget for most of the currently available CMB observations~\cite{Takahashi:2009vp, Koopman:2018vgj}. For example, the estimate of $\beta$ provided by the \Planck\ collaboration~\cite{planckXLIX2016} is compatible with the mentioned value of $0.3^\circ$ but the instrumental uncertainty is of the same order of magnitude, whereas the statistical part is at least 6 times smaller. The degree of degeneracy depends on the mass of the axion~\cite{Sherwin:2021vgb, Nakatsuka:2022epj, Galaverni:2023zhv, Greco:2024oie}. However, as typically assumed in the literature and supported by current data~\cite{Eskilt:2023nxm}, in our analysis we will consider such degeneracy to be complete. This would correspond to an axion mass small enough ($m_\phi\lesssim 10^{-32}$~eV) to make the axion oscillate only after the reionisation epoch, see e.g. Refs.~\cite{Fujita:2020ecn,Greco:2024oie}.\footnote{See Refs.~\cite{Greco:2022ufo, Greco:2022xwj} for the dependence of the anisotropic birefringence effect on the axion mass also in the cross-spectra or cross-bispectra with CMB fields.} See also Refs.~\cite{Yin:2023srb,Kochappan:2024jyf} for further theoretical interpretations of the $EB$ \emph{Planck} spectrum.

The most recent isotropic CB constraints, based on a novel technique~\cite{Minami:2019ruj, Minami:2020f3E01M, Minami:2020fin} (hereafter the Minami-Komatsu or MK technique), also exploit the information contained in the foreground emission and simultaneously recover both the $\beta$ and $\alpha$ angles. This new method was proposed to break the degeneracy between $\beta$ and $\alpha$, finding hints of a detection of the isotropic CB angle $\beta {\simeq} 0.3^\circ$~\cite{Minami:2020odp, Diego-Palazuelos:2022dsq, Eskilt:2022cff, Eskilt:2022wav} at a confidence level of $2.4$ to $3.6 \sigma$. 
The MK technique represents a step forward in the investigation of parity-violating extensions of electromagnetism. Unfortunately, a deeper knowledge of the foregrounds themselves is necessary to provide a robust conclusion, and current results may yet be plagued by unknown systematics~\cite{Clark:2021kze, Diego-Palazuelos:2022dsq, Cukierman:2022odh, Diego-Palazuelos:2023}.
Nevertheless, in ref.~\cite{Cosmoglobe:2023pgf}, using only reprocessed \Planck-LFI and \WMAP\ data, the authors still found $\beta$ compatible with the previously reported value of approximately $0.3^\circ$, primarily derived from \Planck-HFI maps. It is interesting that this compatibility was found despite the very different astrophysical and instrumental systematics present in these data sets. Recently~\cite{2025arXiv250314452L} found that their global polarization angle shows a $2.5~\sigma$ departure from zero.
See also ref.~\cite{Gruppuso:2025ywx} for a recent review of the latest constraints on the isotropic and anisotropic CB effects.

The paper is organised as follows.
In \cref{simulations}, we describe the simulations used, which contain realistic systematic effects coming both from astrophysical signals and from the instrument.
We employ five different pipelines to estimate $\beta$, described in \cref{pipelines}.
In particular, we divide them into two categories: in \cref{ssec:pipelineA_methods}, we present the techniques that do not explicitly consider instrumental $\alpha$, and, in \cref{ssec:pipelineBCD_methods}, we describe those that simultaneously estimate $\beta$ and $\alpha$.
\Cref{analyses} presents the results from employing the different pipelines on the simulations and in \cref{discussionone} we discuss the impact of foregrounds on our results.
In \cref{discussion} we compare and discuss the results obtained, and finish with the main conclusions from this work.

\section{Simulations}
\label{simulations}

We consider four different sets of simulations, each with increasing complexity in terms of non-idealities. 

We call these sets ``Phases'' and label them from 1 to 4. For each phase, we have produced $2200$ polarised maps, i.e., $100$ for each of the 22 \LB\ frequency channels. 
These are all provided at HEALPix\footnote{\url{http://healpix.sourceforge.net}}~\cite{Gorski:2004by} resolution $N_{\rm side}=512$ (which means that the number of pixels needed to cover a CMB map is $N_{\rm pix}=12N_{\rm side}^2$). We convolved each frequency map with the Gaussian of its nominal beam width. See Table 3 of ref.~\cite{2022LB_ptep} for specifics of all the \LB\ frequency bands. Moreover, we assume that each frequency map is obtained with a Dirac delta function bandpass. 

The simulated frequency maps $d_f$, in CMB units, are expressed as 
\begin{equation}
    \centering
        \mathbf{d}_{f} = \mathbf{R}^{\rm inst}(\alpha_f)\mathbf{B}(\theta_{\rm beam}^f)\left[\mathbf{R}^{\rm biref}(\beta)\mathbf{A_{\rm CMB}}, \mathbf{A_{\rm dust}}^f, \mathbf{A_{\rm sync}}^f \right]\left[\begin{array}{c}\mathbf{s}_{\rm CMB}\\\mathbf{s}_{\rm dust}\\\mathbf{s}_{\rm sync}\end{array}\right] + \mathbf{n}_{f},
    \label{eq:data_modeling}
\end{equation}
where boldface quantities are spanning the $I$, $Q$, and $U$ Stokes parameters, and $f$ labels the frequency band (in the rest of the text, we use the quantity $\nu$ for the frequency); $\mathbf{R}^{\rm inst}(\alpha_f)$ is a rotation matrix modeling the imperfect $\alpha_f$ calibration, and $\mathbf{R}^{\rm biref}(\beta)$ models the effect of CB. Note that $\mathbf{R}^{\rm biref}$ is only applied to the CMB component $s_{\rm CMB}$. For a given sky component, $\mathbf{R^x}$ is defined as: 
\begin{eqnarray}
    \centering
        \mathbf{R^x}(\gamma) 
        \equiv 
        \left(\begin{array}{ccc} 
        1 & 0 & 0 \\ 
        0 & \cos(2\gamma) & -\sin(2\gamma) \\ 
        0 & \sin(2\gamma) & \phantom{-}\cos(2\gamma) \end{array}
        \right) \, 
        \mbox{  with $\mathbf{x}\,\in\, \left\{{\rm inst, biref} \right\}$}.
    \label{eq:rotation_IQU_sims}
\end{eqnarray}
$\mathbf{B}$ is a $3 \times 3$ diagonal matrix beam operator that takes as input the beam size $\theta_{\rm{beam}}^f$ and is applied to each frequency 
band and each Stokes parameter independently; $\mathbf{A_{\rm CMB}}$, $\mathbf{A_{\rm dust}}$ and $\mathbf{A_{\rm sync}}$ contain the dust and synchrotron spectral energy densities (SEDs), parameterised by sets of non-linear parameters $\beta_{\rm d}$, $T_{\rm d}$ (parameterizing a modified blackbody) and $\beta_{\rm s}$ (parameterizing the synchrotron's power law).
The analytical parameterization of the SEDs in CMB units are given by
\begin{align}
\begin{split}
    \centering
        \mathbf{A}^X_{\rm CMB}(\nu,\nu_0) &= 1
        ,\\
        \mathbf{A}^X_{\rm dust}(\nu,\nu_0, \beta_{\rm d},T_{\rm d}) &= \frac{1}{u(\nu)} \left(\frac{\nu}{\nu_0}\right)^{\beta_{\rm d}-2} \frac{B_\nu(T_{\rm d})}{B_{\nu_0}(T_\mathrm{d})},
        \\
        \mathbf{A}^X_{\rm sync}(\nu,\nu_0,\beta_{\rm s}) &= \frac{1}{u(\nu)} \left(\frac{\nu}{\nu_0}\right)^{\beta_{\rm s}}, 
        \label{eq: sync SED}
\end{split}
\end{align}
where $X\,\in\,\{I,Q,U\}$, $B_\nu$ is the Planck blackbody function, and $u(\nu) = x^2e^{x}/(e^{x}-1)$, with $x=h\nu/(k_\mathrm{B}T_{\rm CMB})$, is the unit conversion factor transforming from antenna temperature to thermodynamic units.
For a given frequency band $f$ and a given Stokes parameter, $\mathbf{A_{\rm CMB}}$, $\mathbf{A_{\rm dust}}$, and $\mathbf{A_{\rm sync}}$ are simply three scalars. 
The quantity $\mathbf{s}_{\rm x}$ from \cref{eq:data_modeling} are the component maps for the CMB, dust, and synchrotron, defined at the reference frequency $\nu_0=150$ GHz, while $\mathbf{n}$ is the additional Gaussian white noise. 

\begin{table}
    \centering
    \begin{tabular}{|c|c|c|c|} \hline
             Phase  & $\alpha_f$ [deg] & $\beta$ [deg] & $\beta_{\rm d}$, $T_{\rm d}$, $\beta_{\rm s}$ \\
                \hline
        1 & 0 & $0.0$ &  Spatially constant (\texttt{pysm} \texttt{d0s0}) \\
        2 & 0 & $0.0$ & Spatially varying (\texttt{pysm} \texttt{d1s1})\\
        3 & Random for each band & $0.0$ & Spatially varying (\texttt{pysm} \texttt{d1s1})\\
        4 & Random for each band & $0.3$ & Spatially varying (\texttt{pysm} \texttt{d1s1}) \\ \hline
    \end{tabular}
    \caption{Properties of the various simulations phases, in terms of miscalibration ($\alpha_f$) and birefringence ($\beta$) angles,  and foreground spectral parameters ($\beta_{\rm d}$, $T_{\rm d}$, and $\beta_{\rm s}$). In the case of Phase 3 and 4, the random $\alpha_f$ are drawn from uniform distributions $\alpha_f\in[-\sigma_{\alpha_f}, \sigma_{\alpha_f}]$ which widths are specified in \cref{tab:alpha_f} and derived in Ref-~\cite{Vielva2022:jcna}.}
    \label{tab:phase_description}
\end{table}

In the following, we describe these simulations in detail, with properties summarized in \cref{tab:phase_description}.

\begin{figure}
    \centering
    \includegraphics[width=0.7\textwidth]{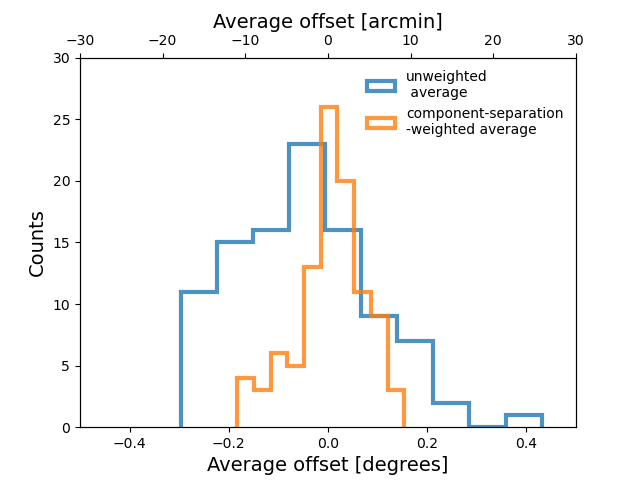}
    \caption{Average offset across frequency bands computed from the 100 simulations, before ($(1/22)\times\sum_f \alpha_f$) and after ($\sum_f \mathbf{W}_f \alpha_f$, with the $\mathbf{W}_f$ weights defined in \cref{ssec:pipelineA_methods}) component separation.}
    \label{fig:histogram_offset_phase4}
\end{figure}

\begin{itemize}
    \item \textbf{Phase 1:} These are made of CMB maps generated from the \Planck\-2018 best fit~\cite{Planck2020cosmo}, combined with white noise based on the \LB\ latest instrument model~\cite{2022LB_ptep}, as well as foreground emission based on the \texttt{pysm}~\footnote{\url{https://github.com/galsci/pysm}} \texttt{d0s0} model~\cite{PYSM3}. In this case, dust and synchrotron are considered ``simple'' because their SEDs are spatially constant.
    \item \textbf{Phase 2:} 
    These are similar to Phase 1 but this time the foreground emission is based on the \texttt{d1s1} model from \texttt{pysm}: the dust and synchrotron SEDs are spatially varying, as estimated from the analysis of \Planck's total intensity data with \texttt{Commander}~\cite{diffuse_foregrounds_Planck2018}. 
    \item \textbf{Phase 3:} These are built on top of Phase 2, with the inclusion of randomly distributed instrumental miscalibrations following ref.~\cite{Vielva2022:jcna}. Each of the 100 simulations corresponds to independent realisations of the $\alpha_f$ parameter that creates the rotation matrix $\mathbf{R}^{\rm inst}$ in \cref{eq:data_modeling}. Similarly to ref.~\cite{Krachmalnicoff2022:jakv}, the $\alpha_f$ values are drawn from independent uniform distributions centered at zero and with a width $\alpha_f\in[-\sigma_{\alpha_f}, \sigma_{\alpha_f}]$, where the $\sigma_{\alpha_f}$ for each frequency band is taken from ``Case 2.3'' summarized in Table 3 of~\cite{Vielva2022:jcna} that we reproduce in Table.~\ref{tab:alpha_f}. For each simulation, the average miscalibration across frequencies, $\langle \alpha_f \rangle_f$, gives a non-zero global offset to the whole instrument by a statistical fluctuation roughly within the limits considered in Table 2 of ref.~\cite{Vielva2022:jcna}. \Cref{fig:histogram_offset_phase4} shows the distribution of average offsets before ($\langle \alpha_f \rangle_f \equiv (1/22)\times\sum_f\alpha_f$) and after component separation ($\langle \alpha_f \rangle_f \equiv\sum_f \mathbf{W}_f \alpha_f$, where the weights $\mathbf{W}_f$ will be defined in \cref{ssec:pipelineA_methods}) across the 100 simulations. The smaller spread of the weighted $\langle \alpha_f \rangle_f$ shows how the combination of frequency bands during component separation reduces the impact of relative miscalibrations.
    \item \textbf{Phase 4:} These are built on top of Phase 3 with the inclusion of an isotropic $\beta = 0.3^\circ$, an amplitude that is representative of what was found in Refs.~\cite{Minami:2020odp, Diego-Palazuelos:2022dsq, Eskilt:2022cff, Eskilt:2022wav}.
\end{itemize}

\begin{table}[]
    {\footnotesize
    \centering
    \begin{tabular}{c|c|c|c|c|c|c|c|c|c|c|c|c|c|c|c|c|c|c|c|c|c|c}
     Instrument  & 
LFT & LFT & LFT & LFT & LFT & LFT & LFT & LFT & LFT & LFT & LFT  \\
Frequency [GHz] & 40 & 50 & 60 & 68a & 68b & 78a & 78b & 89a & 89b & 100 & 119 \\  
$\sigma_{\alpha_f}$ [arcmin] & 147.8 & 118.2 & 47.9 & 41.8 & 106.5 & 25.5 & 38.6 & 15.9 & 87.4 & 11.3 & 6.2 \\
\\
Instrument & LFT &MFT & MFT & MFT & MFT & MFT & HFT & HFT & HFT & HFT & HFT \\
Frequency [GHz] & 140&100 & 119 & 140 & 166 & 195 & 195 & 235 & 280 & 337 & 402 \\
$\sigma_{\alpha_f}$ [arcmin]& 5.6 &7.6 & 3.4 & 4.3 & 3.3 & 5.3 & 11.6 & 12.3 & 20.1 & 50.9 & 237.6

    \end{tabular}}
    \caption{$\alpha_f$ requirements for all frequency bands in the Low-, Mid-, and High-Frequency Telescopes (LFT, MFT, HFT) of \LB. These requirements on \LB's absolute angle calibration per band ensure an unbiased measurement of the tensor-to-scalar ratio when grouping detectors per frequency and assuming that all frequencies are uncorrelated (Case 2.3 from Table 3 of ref.~\cite{Vielva2022:jcna}).}
    \label{tab:alpha_f}
\end{table}

We verified that, at the angular scales relevant for our analysis, \texttt{pysm}'s \texttt{d0} and \texttt{d1} models contain a small non-null $EB$ signal that is roughly compatible with \textit{Planck}'s dust measurements within the statistical uncertainty of \textit{Planck} polarisation data. Although enough to introduce a noticeable bias in our analysis, such $EB$ correlation is not purposely built into \texttt{d0} and \texttt{d1} to reproduce the physical properties of dust emission. Hence, our simulations likely underestimate the true foreground complexity that can arise from the coupling of magnetic misalignments~\cite{Huffenberger2020, Clark:2021kze, Cukierman:2022odh} and the spatial variation of the emission properties~\cite{Pelgrims2021, Ritacco2023, Vacher2023b, Vacher2023}. We leave the study of more complex dust models (e.g., \cite{HerviasCaimapo2024, Vansyngel_dust_model, MKD_dust_model, Vacher2024}) for future work.

\section{Techniques employed}
\label{pipelines}

We employ five different pipelines to estimate $\beta$ in the presence of foreground contamination and systematic effects, as described in \cref{simulations}. These pipelines are categorized into two groups. The first category, detailed in \cref{ssec:pipelineA_methods}, includes two pipelines that neither characterise nor control the effect of the instrumental $\alpha_f$. The second category, discussed in \cref{ssec:pipelineBCD_methods}, comprises three pipelines: one that implements the MK technique and two that fit $\alpha_f$ while performing component separation.

\begin{figure}
\includegraphics[width=\textwidth]{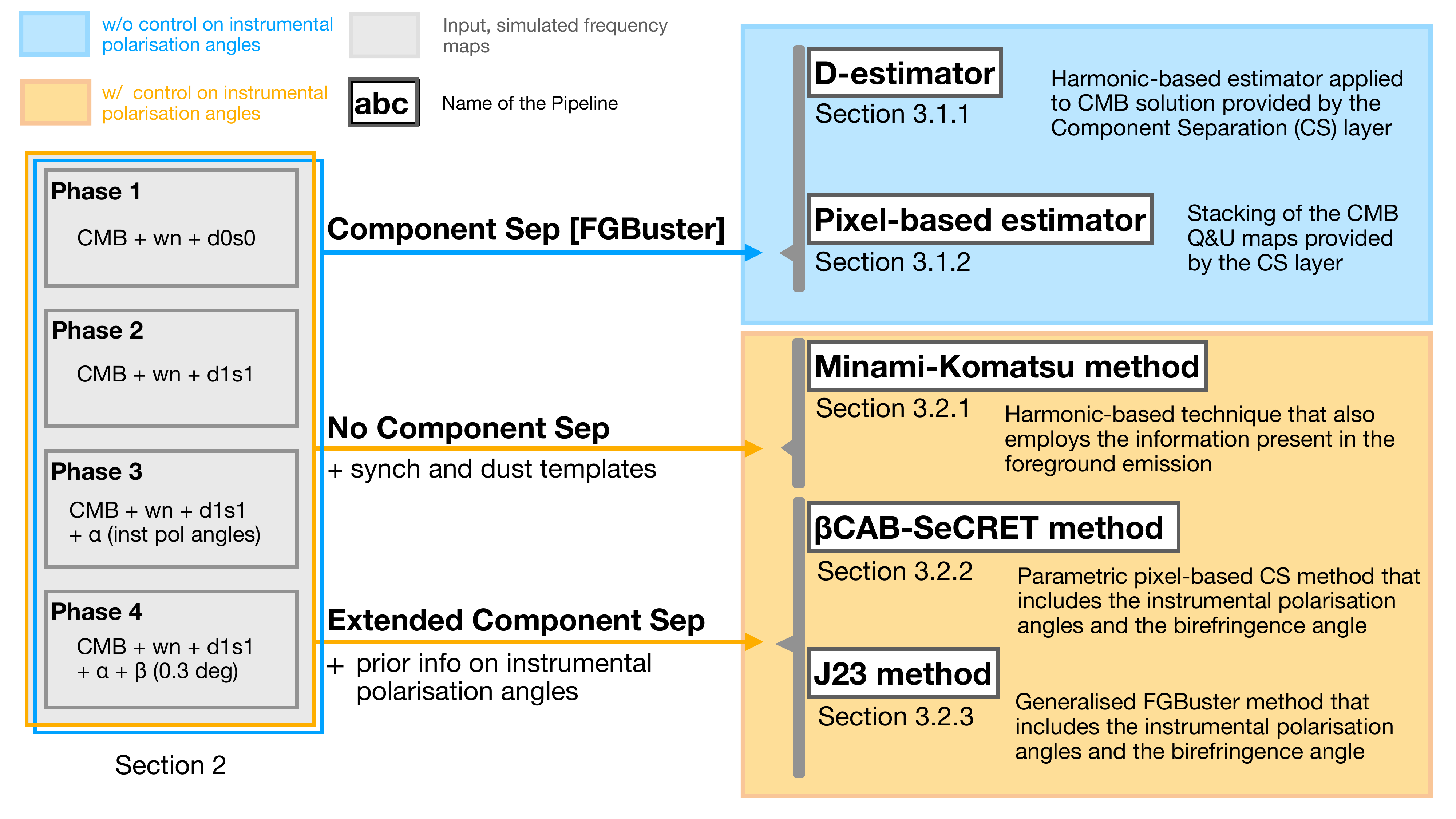}
    \caption{Overview of the pipelines employed to estimate CB. On the left, ``wn'' stands for white noise, \texttt{d0s0} and \texttt{d1s1} represent the foreground models, and $\alpha$ indicates the instrumental polarisation angles (see also \cref{simulations}). On the right, the pipelines' names are displayed in white boxes accompanied by a brief description. We indicate below each pipeline the section where a full description can be found. Pipelines in the blue block do not account for $\alpha_f$, while those in the orange block do.}
    \label{fig:overview}
\end{figure}

In order to improve readability, before delving into the details of each technique, we provide a concise description of the pipelines used to deliver the forecasts on CB. This is done through \cref{fig:overview}, which highlights the main features of each pipeline. In the left part of this figure, the inputs considered (shown in light grey) are divided into four different sets of simulations (i.e., Phases). Each of these Phases has been analysed by five pipelines, whose names are displayed in white boxes in \cref{fig:overview}. The pipelines in the blue area do not account for $\alpha_f$, while those in the orange area do. The former are applied to the CMB solutions provided by \fgbuster, whereas the latter either extend their component separation to handle $\alpha_f$ or have no component separation at all since the method is applied directly to the frequency maps.
In \cref{fig:overview}, below each pipeline name, the section where the pipeline is described in detail is indicated, while a brief description of the pipeline is provided next to it.

\subsection{Pipelines without control of instrumental polarisation angles}
\label{ssec:pipelineA_methods}

\begin{figure}
\includegraphics[width=\textwidth]{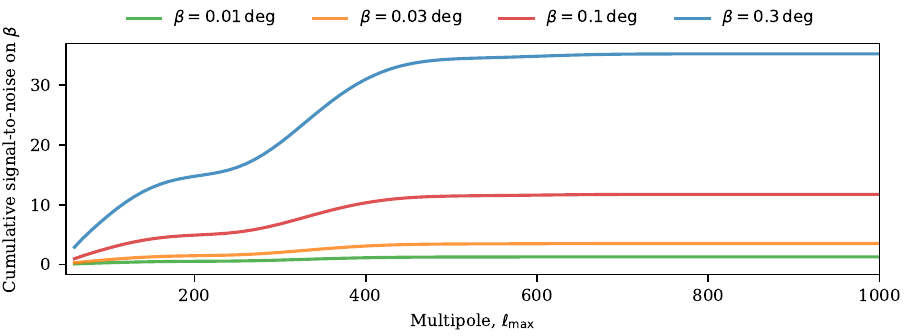}
    \caption{Forecasted cumulative signal-to-noise on $\beta$ from $EB$ correlation (see \cref{eq:EBobs}) taking into account cosmic variance, noise, and foreground residuals from Phase 1 as estimated from \fgbuster. The sky fraction is assumed to be $f_\mathrm{sky}=60\%$, and $\ell_{\rm min} = 50$. }
\label{fig:validation_D_estimator1}
\end{figure}

\begin{figure}
\includegraphics[width=\textwidth]{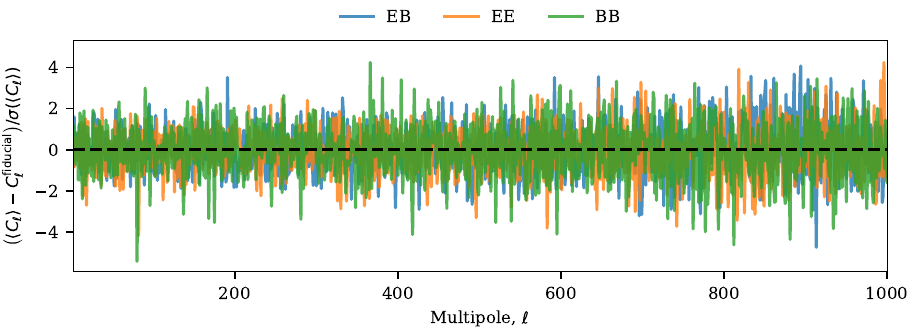}
    \caption{Significance of the averaged bias on the reconstructed angular power spectra. The low significance of fluctuations confirms that the choices made (such as neglecting the synchrotron contribution) are valid.}
\label{fig:validation_D_estimator2}
\end{figure}

These pipelines are both based on the CMB solution provided by the considered component-separation layer, i.e.\ \fgbuster\footnote{\url{https://github.com/fgbuster/fgbuster}}, a maximum-likelihood parametric map-based tool~\cite{Stompor:2008sf}. 
In order to increase the signal-to-noise ratio of the CB effect, we need to access high CMB multipoles. This is possible by neglecting the frequency channels that do not possess sufficiently high angular resolution. However, this choice has to be balanced against our ability to remove the foreground emission. 
The low-frequency channels have the lowest angular resolution, but trace the synchrotron emission the best.
Hence, as a trade-off, we consider only medium and high-frequency channels with a full width at half maximum (FWHM) smaller than 30 arcmin. In \cref{fig:validation_D_estimator1}, we display the signal-to-noise ratio we expect from this choice. In \cref{fig:validation_D_estimator2}, we show that the bias induced by the synchrotron emission is subdominant in our results.

We proceed in two steps.
\begin{enumerate}
    \item As in ref.~\cite{2022LB_ptep}, we degrade the resolution of the input, frequency-dependent maps, \cref{eq:data_modeling}, to a common $80$ arcmin resolution. This procedure allows an unbiased determination of the spectral indices, which would otherwise suffer from frequency-dependent resolution. We then optimise the so-called spectral likelihood to estimate the non-linear parameters of the synchrotron and dust SEDs, $\beta_\mathrm{s}$, $\beta_\mathrm{d}$ and $T_\mathrm{d}$ in \cref{eq:data_modeling}.
    \item We build a weighting operator $\mathbf{W}\equiv\left(\mathbf{A}^{\mathrm{T}}\mathbf{N}^{-1}\mathbf{A}\right)\mathbf{A}^{\mathrm{T}}\mathbf{N}^{-1}$,  estimated using the noise covariance and the SEDs estimated in step (1), which we use to estimate the foreground-cleaned $Q$ and $U$ CMB maps $\tilde{s}_{\rm CMB}$ as well as foreground-cleaned CMB spectra. Using the notation from \cref{eq:data_modeling}, the estimated sky signal can be written as 
    \begin{align}
        \tilde{s}_{kc}^{\rm fgs\text{-}clean} = \sum_f\mathbf{W}_{cf}d_{kf},
        \label{eq:Wd_def}
    \end{align}
    where $d_{kf}$ corresponds to the input, noisy, and foreground-contaminated frequency maps (not convolved to a common resolution), $k$ denotes the Stokes parameter and $c$ is the index of the sky components (CMB, dust, synchrotron). Similarly, the component-separated spectra can be estimated as
    \begin{align}
        \tilde{C}^{cc', {\rm fgs\text{-}clean}}_\ell = \sum_{f,\,f'}\mathbf{W}_{cf}^TC_\ell^{ff'}\mathbf{W}_{c'f'},
        \label{eq:WCW_def}
    \end{align}
    where $C_\ell^{ff'}$ are the frequency cross-power spectra estimated from the input maps. Note that we apply the $f_{\rm sky}=60~\%$ Galactic mask derived by the \Planck\ collaboration\footnote{\texttt{HFI\_Mask\_GalPlane-apo0\_2048\_R2.00.fits} from \url{https://pla.esac.esa.int/##maps}.} to the foreground-clean maps, \cref{eq:Wd_def}, prior to the estimation of $C_\ell^{ff'}$. Finally, we only consider the CMB solutions to \cref{eq:Wd_def,eq:WCW_def}, i.e.\ $c=c'={\rm CMB}$.
\end{enumerate}
The weights constructed in this approach are also applied to foreground-only, noise-only, and foregrounds+noise maps to estimate the noise and statistical-foreground-residuals bias~\cite{Stompor:2016} present in the reconstructed, foreground-cleaned CMB map. In the case of real data sets, we would rely on similar foreground + noise simulations or use data splits. As noted in ref.~\cite{Errard:2011}, statistical foreground-residuals do not depend on the actual amplitude of foregrounds, but rather on the number of degrees of freedom considered in the component separation, making such an approach robust in the case of complex foregrounds.

The CMB solution $\tilde{s}^{\rm fgs\text{-}clean}$ coming out from \fgbuster\ is considered as the input of two estimators, one is harmonic-based, using the so-called ``$D$-estimator'' defined on the observed polarised spectra~\cite{QUaD:2008ado, Gruppuso:2016nhj} (\cref{sssec:d_estimator}), and the other is pixel-based and consists of the stacking of $Q$ and $U$ maps at the location of the $T$ and $E$ peaks~\citep{komatsu2010,planckXLIX2016,Sullivan:2025btc} (\cref{sssec:stack-estimator}). 
Because of their different domains, in principle, these two independent estimators respond to residual systematic effects differently. Hence, having these two compatible estimates is a benefit from the point of view of the robustness of the analysis.
However, in these pipelines, the estimate of $\beta$ is degenerate with $\alpha_f$. This means that the total error budget has to be complemented with additional (and independent) information from the instrument calibration. Reversing the argument, forecasts based on these pipelines provide a reference to assess requirements on the knowledge of $\alpha_f$. This can be achieved by requiring that the uncertainty on $\alpha_f$ be subdominant with respect to the statistical uncertainty of $\beta$.

\subsubsection{Harmonic-based, \texorpdfstring{$D$}{D}-estimator}
\label{sssec:d_estimator}

To build the $D$-estimator, we need a model for the observed, foreground-cleaned CMB (see \cref{eq:WCW_def}).
From \cref{eq:TEobs,eq:TBobs} and from \cref{eq:EEobs,eq:BBobs,eq:EBobs} we can build harmonic-based estimators based on the observed $TB$ and $EB$ spectra, respectively. Since the former has a signal-to-noise ratio smaller than the latter~\cite{Gruppuso:2016nhj, planckXLIX2016}, throughout this work we employ only $D_{\ell}^{EB, \obs}$, which is defined as
\begin{equation}\label{DEB}
  		D_{\ell}^{EB, \obs}(\hat \beta) \equiv C^{EB, \obs}_{\ell} \cos(4 \hat \beta) - \frac{1}{2} (C^{EE, \obs}_{\ell} - C^{BB, \obs}_{\ell}) \sin(4 \hat \beta) \, , 
\end{equation}
where $\hat \beta $ is the estimate for the birefringence angle.
The main property of $D_{\ell}^{EB, \obs}(\hat \beta)$ is that its expectation value goes to zero when $ \hat \beta = \beta$, i.e.,
\begin{equation}
    \langle D_{\ell}^{EB, \obs}(\hat \beta = \beta) \rangle = 0 \, ,
\end{equation}
see e.g. ref.~\cite{Gruppuso:2016nhj} for further details.
Therefore, this suggests that it is enough to look for the zeros of $D_{\ell}^{EB, \obs}(\hat \beta)$ to find $\beta$
\cite{QUaD:2008ado,Gruppuso:2016nhj, planckXLIX2016}. This is typically done through standard $\chi^2$-minimisation techniques.

In the following, we will consider the observed spectra to be the noise-debiased, estimated, foreground-cleaned CMB spectrum:
\begin{equation}
    \centering
        C_{\ell}^{XY,\, \obs} \equiv \tilde{C}_\ell^{XY,{\rm fgs\text{-}clean}} - \tilde{N}^{XY}_\ell,
\end{equation}
where $\tilde{N}_\ell$ is the estimated noise bias combined with statistical foreground residuals,
and $\tilde{C}^{XY,{\rm fgs\text{-}clean}}_\ell$ the CMB$\times$CMB component-separated spectra defined in \cref{eq:WCW_def}, with $X,Y\,\in\left\{ E,B\right\}$.
In addition, we perform a complementary analysis without any noise and foreground residual debiasing, i.e., assuming $C_{\ell}^{XY,\, \obs} = \tilde{C}_\ell^{XY,{\rm fgs\text{-}clean}}$.

\subsubsection{Pixel-based estimator}
\label{sssec:stack-estimator}
Our second estimator, which also does not attempt to remove any miscalibration angle, is a pixel-based approach. This method consists of the stacking of $Q$ and $U$ maps at the peak location of the $E$-mode and temperature maps~\citep{komatsu2010,jow2019,contreras2017,planckXLIX2016,Sullivan:2025btc}.
At each of the $T$ and $E$ extrema, we calculate the transformed Stokes parameters,
\begin{align}
\begin{split}
    Q_r(\theta)&=-Q(\theta)\cos(2\phi)-U(\theta)\sin(2\phi) \, ,
    \\
    U_r(\theta)&= \phantom{-}Q(\theta)\sin(2\phi)-U(\theta)\cos(2\phi) \, , \label{eq:urmap}
\end{split}
\end{align}
where $\phi$ is the angle from local east, with north pointing towards the Galactic north pole (anti-clockwise), and $\theta$ is a distance radially outward from the center of the extrema, introduced in ref.~\cite{Kamionkowski1997}. These can be interpreted as a local $E$ and $B$ transformation, where the $Q_r(\theta)$ measures the tangential and radial components of the polarisation at the radial distances $\theta$ from the peaks, and $U_r(\theta)$ the polarisation at $45^\circ$ to the tangential and radial directions.
The expected profiles of the transformed Stokes parameters are given for $T$ extrema by
\begin{align}
\begin{split}
    \langle Q_r^T \rangle(\theta)&=-\int \frac{\ell d\ell}{2\pi}W^T_\ell W^P_\ell \left(\bar{b}_{\nu}+\bar{b}_{\zeta}\ell^2\right)C_\ell^{TE,\obs}J_2(\ell\theta) \, , \\
    \langle U_r^T \rangle(\theta)&=-\int \frac{\ell d\ell}{2\pi}W^T_\ell W^P_\ell \left(\bar{b}_{\nu}+\bar{b}_{\zeta}\ell^2\right)C_\ell^{TB,\obs}J_2(\ell\theta) \, ,
\end{split}
\end{align}
and for $E$ extrema by 
\begin{align}
\begin{split}
    \langle Q_r^E \rangle(\theta)&=-\int \frac{\ell d\ell}{2\pi}W^E_\ell W^P_\ell \left(\bar{b}_{\nu}+\bar{b}_{\zeta}\ell^2\right)C_\ell^{EE,\obs}J_2(\ell\theta) \, , \\
    \langle U_r^E \rangle(\theta)&=-\int \frac{\ell d\ell}{2\pi}W^E_\ell W^P_\ell \left(\bar{b}_{\nu}+\bar{b}_{\zeta}\ell^2\right)C_\ell^{EB,\obs}J_2(\ell\theta)\, , 
\end{split}
\end{align}
where $W_\ell^{X}$ with $X=P,E$ and $T$ are the beam and window pixel functions for the $P$ polarisation (observed $Q$ and $U$ maps), $E$, and $T$ 
maps respectively, $J_2(\ell\theta)$ are the second-order Bessel functions of the first kind, and $\bar{b}_{\nu}$ and $\bar{b}_{\zeta}$ are the scale-dependent bias parameters as described in ref.~\cite{komatsu2010}. The $E$-mode map is produced from the observed $Q$ and $U$ maps at the same resolution as the $T$, $Q$, and $U$ maps. While this does not have to be the case, we found no improvements with additional smoothing of the $E$-mode map, so we maintain the input resolution. In the case of no parity-violating terms, we expect $\langle U_r^X \rangle(\theta)$ to be zero (as there is no $C_\ell^{EB}$ or $C_\ell^{TB}$ predicted by $\Lambda\mathrm{CDM}$). However, with CB modifying the power spectra as given in \cref{eq:TBobs,eq:EBobs}, we instead have,
\begin{align}
\begin{split}
    \langle U_r^T \rangle(\theta)&=-\sin(2\beta)\int \frac{\ell d\ell}{2\pi}W^T_\ell W^P_\ell \left(\bar{b}_{\nu}+\bar{b}_{\zeta}\ell^2\right)C_\ell^{TE}J_2(\ell\theta) \, , \\
    \langle U_r^E \rangle(\theta)&=-\frac{1}{2}\sin(4\beta)\int \frac{\ell d\ell}{2\pi}W^E_\ell W^P_\ell \left(\bar{b}_{\nu}+\bar{b}_{\zeta}\ell^2\right)(C_\ell^{EE}-C_\ell^{BB})J_2(\ell\theta) \, ,
    \label{eq:Urfinal}
\end{split}
\end{align}
which is different from zero for non-null $\beta$. 
It is then simple to perform a least-squares analysis on the resulting stacked profiles compared with the theory to obtain the final $\beta$. See Refs.~\cite{planckXLIX2016,contreras2017,jow2019} for similar analyses and details. 

\subsection{Pipelines with control of instrumental polarisation angles}
\label{ssec:pipelineBCD_methods}

\subsubsection{Template-based extension of the Minami-Komatsu estimator}
\label{pipelineB method}
\input{PipelineB/pipelineB_methodology.tex}

\subsubsection{CAB-SeCRET and \texorpdfstring{$\beta$}{b}CAB-SeCRET}

\label{pipelineC_method}
\input{PipelineC/pipelineC_methodology}

\subsubsection{Generalised pixel-based parametric component separation, \texttt{J23}}
\label{pipelineD method}
\input{PipelineD/pipelineD_methodology}

\section{Analyses}
\label{analyses}

\subsection{Pipelines without control of instrumental polarisation angles}
\label{pipelineA analysis}

\subsubsection{D-estimator analysis}
\label{subsection:harmonic}

As mentioned in \cref{sssec:d_estimator}, CB is estimated looking for the $\hat \beta$ angle that minimises the following $\chi^2(\hat \beta)$:
\begin{equation}
    \chi^2(\hat \beta) = \sum_{\ell,\ell'}D_\ell^{EB,\obs}  (\hat \beta ) M_{\ell\ell'}^{-1} D_{\ell'}^{EB,\obs}(\hat \beta ),
\end{equation}
where $M_{\ell\ell'} = \langle D_{\ell}^{EB,\obs}(\beta)D_{\ell'}^{EB,\obs}(\beta)\rangle$ is the covariance matrix of $D_\ell^{EB,\obs}(\hat \beta = \beta)$, which is defined in \cref{DEB}. It is possible to show that the matrix $M_{\ell\ell'}^{EB}$ reduces to the $EB$ covariance matrix, i.e., without the CB effect included~\cite{Gruppuso:2016nhj} when the noise of the $E$ and $B$ CMB fields are equivalent. 
Since this condition is always satisfied in all the cases we consider, we take $M_{\ell\ell'}$ to be the $EB$ covariance. This covariance is built through simulations:
we do that by performing an aggressive harmonic binning because of the low number of simulations available (details below). 
Moreover, for each of the simulations employed to sample the $\chi^2(\hat \beta)$, the covariance $M_{\ell\ell'}$ is built from the remaining simulations (i.e., excluding the specific simulation considered in $D_{\ell}^{EB,\obs}$).
In the following, we discuss in detail the results obtained for each Phase.

\begin{itemize}
    \item \textbf{Phase 1:} 
    In \cref{fig:spectra_phase1}, we show the spectra and the combination of the spectra needed to build $D_{\ell}^{EB,\obs}$. Blue and orange curves are for the simulations in which CMB + noise + foregrounds and only noise + foregrounds are considered, respectively. Green is for the simulations in which the spectra have been debiased with noise + foregrounds. It is interesting to observe that the expected averages for these spectra are recovered in both the debiased and unbiased cases. Moreover, the foreground residuals show up at low multipoles, while the noise (coupled to the beam) presents itself at high multipoles. In between these regions, around $20\leq\ell\leq 500$, the $EB$ correlation is compatible with zero, which is expected in the absence of CB. 
\begin{figure}
\includegraphics[width=1\textwidth]{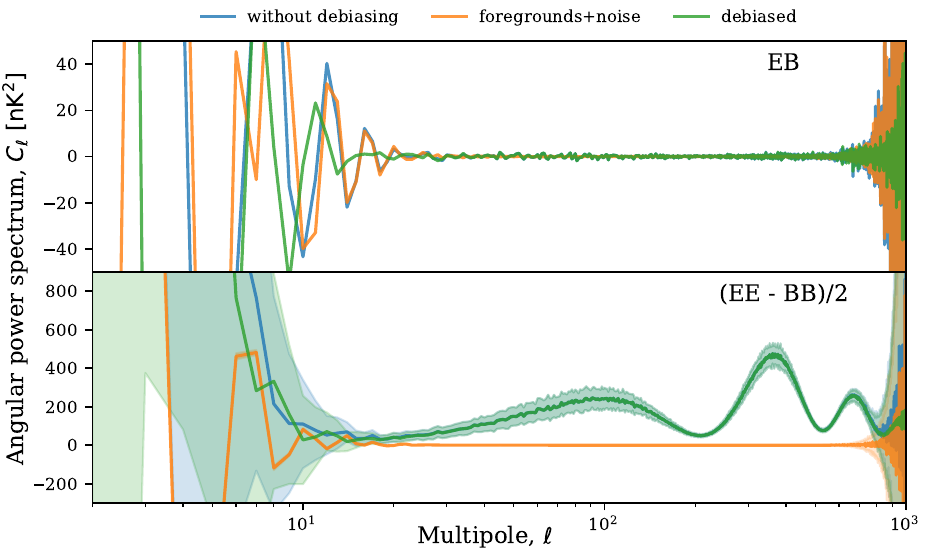}
    \caption{Upper panel: $EB$ spectra. Lower panel: $(EE-BB)/2$ spectra after component separation using \fgbuster. In all the panels, Phase 1 simulations of CMB + noise + foregrounds are displayed in blue and simulations of only noise + foregrounds are shown in orange. The spectra debiased with foreground and noise are given in green. The shaded regions represent the standard deviation of the spectra across the 100 simulations.}
    \label{fig:spectra_phase1}
\end{figure}
For each simulation, we estimate $\beta$ and construct the histograms given in \cref{fig:histogram_phase1}. These are obtained by building the covariance through simulations, setting a harmonic binning $\Delta \ell = 100$ in the range $[100,700]$, displayed in orange, or $[50,750]$, displayed in blue. In the same \cref{fig:histogram_phase1}, we show histograms of $\beta$ obtained through spectra with and without noise + foregrounds debiasing. In particular, the latter are given with filled histograms, and the former are provided with unfilled histograms.
In all the cases, the average of $\beta$ is found to be compatible with zero with an uncertainty of $0.011^\circ$ at $68~\%$ C.L.. Moreover, approximating the covariance $M_{\ell\ell'}$ to be diagonal, we find the same statistical efficiency for all cases.

    \item \textbf{Phase 2:}
Considering the same setting adopted for Phase 1, we estimate $\beta$ for each simulation and construct the histograms given in \cref{fig:histogram_phase1}. We find an uncertainty on $\beta$ of $0.011^\circ$ at $68~\%$ C.L. for all the cases considered. Again similarly to Phase 1, when we approximate the covariance $M_{\ell\ell'}$ to be diagonal, we find the same statistical efficiency for all the cases considered.
\begin{figure}
    \centering
    \includegraphics[width=0.85\textwidth]{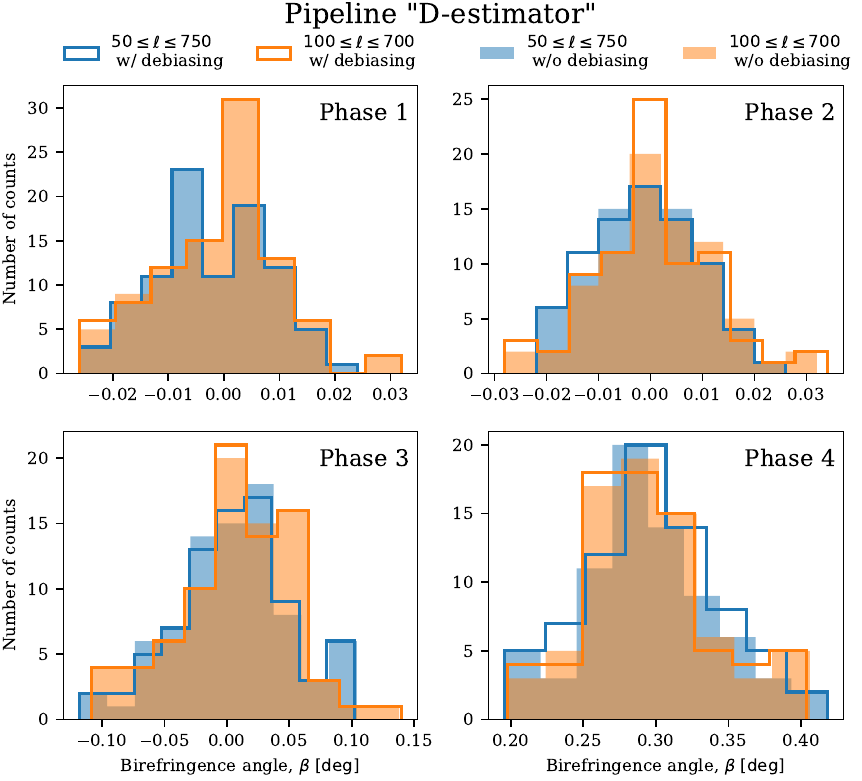}
    \caption{Histograms of $\beta$ with and without debiasing, as obtained for all simulation phases. For each phase, we considered two harmonic ranges: $[100,700]$, shown in orange, and $[50,750]$, shown in blue. Shaded histograms represent the case without debiasing, while empty histograms indicate the case with debiasing.} 
\label{fig:histogram_phase1}
\end{figure}
    \item \textbf{Phase 3:}
Following the same approach considered above, we construct the histograms given in \cref{fig:histogram_phase1}.
We find an uncertainty on $\beta$ of $0.046^\circ$ at $68~\%$ C.L. for all the cases considered. 
Moreover, approximating the covariance $M_{\ell\ell'}$ to be diagonal, we find a general increase in the uncertainty on $\beta$ of a factor of $1.6$ to $1.7$. 
This seems to indicate that the inclusion of $\alpha_f$ different from zero (albeit compatible with requirements) provides an increase of the uncertainty of $\beta$ compared to Phases 1 and 2, and also non-negligible off-diagonal correlations, which, if not taken into account, make the uncertainty even larger.
    \item \textbf{Phase 4:}   
In \cref{fig:spectraPhase4}, we show the spectra needed to build $D_{\ell}^{EB,\obs}$. 
As in the previous cases, the averages of the recovered spectra are compatible with the fiducial spectra (which, in this case, are rotated with respect to the $\Lambda$CDM case by $\beta=0.3^\circ$) for both the noise + foregrounds debiased and the non-debiased cases.
\begin{figure}
\includegraphics[width=\textwidth]{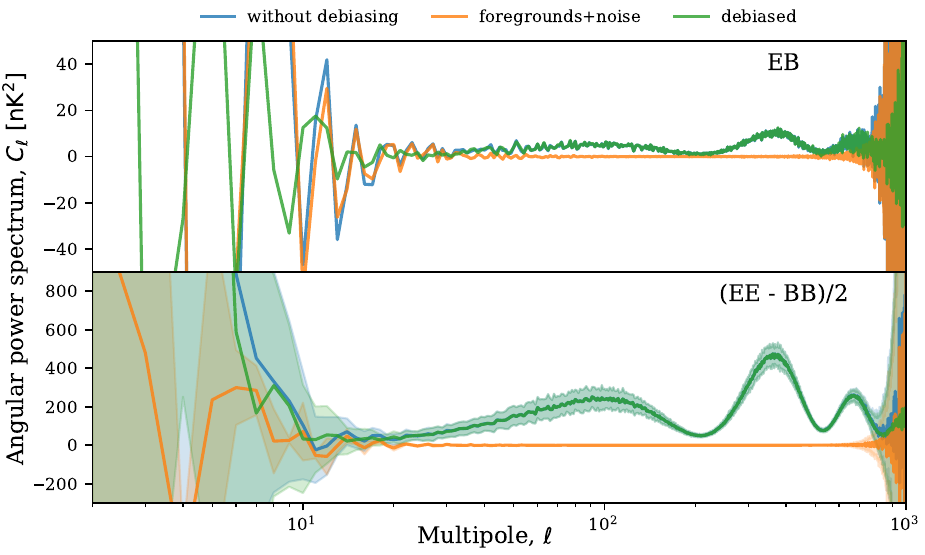}
    \caption{Same as \cref{fig:spectra_phase1} but for the Phase 4 simulations.}
    \label{fig:spectraPhase4}
\end{figure}
Employing the same harmonic setting as above, we build the histograms shown in \cref{fig:histogram_phase1}.
When we consider the harmonic range $\ell\in[100,700]$ we find $\beta=0.296^\circ \pm 0.045^\circ$ and $\beta=0.295^\circ \pm 0.045^\circ$ for the non-debiased and debiased cases, respectively. We hence retrieve the fiducial value of $\beta^{\rm fid} = 0.3^\circ$. Similarly, for the harmonic range $\ell\in[50,750]$, we obtain $\beta=0.296^\circ \pm 0.047^\circ$ and $\beta=0.297^\circ \pm 0.047^\circ$ for the non-debiased and noise + foregrounds debiased cases, respectively.
Moreover, as for the Phase 3 case, we note that considering a diagonal matrix, we obtain an increase of the uncertainty on $\beta$ of a factor of $1.6$ to $1.7$.
\end{itemize}

\subsubsection{Pixel-based analysis}
\label{subsection:pixel}
To determine $\beta$ using the pixel-based approach we begin by finding all the hot and cold extrema using the nearest-neighbour method on both $E$ maps (produced from noisy $Q$ and $U$), and the fiducial $T$ maps used to generate the \LB\ simulations (clean with no noise). We chose a threshold of zero such that we only retain positive hot spots and negative cold spots.
Including all peaks comes at the cost of more noise peaks adding no additional information to the analysis (increasing required computation time and slightly increasing the error bar), whereas including fewer peaks (with a more constraining threshold) comes at the risk of ignoring true peaks. Thresholds of zero have been used for previous analyses and we find that more constraining thresholds do not improve our results. 

From \cref{eq:Urfinal}, we take $\theta$ out to $5^\circ$, where the profile converges towards zero. Higher angular sizes tested did not change the results. Peaks within $5^\circ$ of the mask are discarded since they often lack enough pixels to both be identified as a peak and to correctly determine the value for $\beta$ (the sample of peaks near a masked region had poorer return of the fiducial $\beta$ and with much higher variance than other peak samples). 

\begin{figure}[tbp]
\centering
    \includegraphics[width=\linewidth]{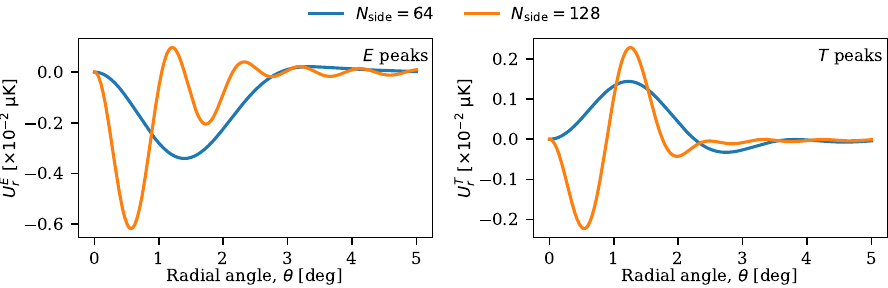}
\caption{Theory profile for $E$ and $T$ peaks for different $N_{\rm side}$, used in the pixel-based analysis. Phase 1 and 2 have $N_{\rm side}=64$ and Phase 3 and 4 have $N_{\rm side}=128$. Note that the theory profiles are shown here assuming $\beta=0.3^\circ$, otherwise theory profiles for $\langle U_r^X \rangle(\theta)$ would be consistent with zero.
}
\label{fig:pix_theory}
\end{figure}

For each peak, $m$, we obtain an estimate of the angle $\beta_m$, and a goodness of fit $\sigma_{\beta_m}^2$ using a standard least-squares analysis fit. First, we calculate the theoretical value $\tilde{U}_r^{X}(\theta_p)$, $X=\{T,E\}$, for all the pixels at $\theta<5^\circ$ from the central peak, using \cref{eq:Urfinal}. The values for the bias parameters from the final maps used can be found in \cref{tab:pix_biasparams}, and the final theory profiles are shown in \cref{fig:pix_theory}.

\begin{table}[tbp]
    \centering
    \begin{tabular}{c c c | c c c c}\hline
         Phase & $N_{\rm side}$ &Beam FWHM & $b_\nu^E$ & $b_\zeta^E$&  $b_\nu^T$ & $b_\zeta^T$ \\ & & [arcmin] & & & &\\\hline
         1 & 64 & 80 & 0.226 &2.85$\times10^{-4}$ & 0.0022& 1.47$\times10^{-6}$ \\
         2 & 64 & 80 & 0.226 &2.86$\times10^{-4}$ &0.0022 &1.47$\times10^{-6}$ \\
         3 & 128 & 40 &0.157 &2.67$\times10^{-5}$ &0.0018 &4.06$\times10^{-7}$ \\
         4 & 128 & 40 &0.157 &2.66$\times10^{-5}$ &0.0018 &4.06$\times10^{-7}$ \\\hline
    \end{tabular}
    \caption{Stacking theory bias parameters for various data sets. Note that the bias parameters are very dependent on the map features (such as the beam, window function, and $N_{\rm side}$), and the underlying power spectrum in the map (including the noise). }
    \label{tab:pix_biasparams}
\end{table}

Next, we compute $\hat{U}_r(\theta_\mathrm{pix})$ at the location of the $T$ or $E$ peak on the $Q$ and $U$ maps, using \cref{eq:urmap}. The peak values are calculated using
\begin{align}
\begin{split}
    \beta_m&=\frac{1}{2}\frac{\sum_p \hat{U}_r(\theta_p)\tilde{U}^X_r(\theta_p)}{\sum_p\tilde{U}^X_r(\theta_p)\tilde{U}^X_r(\theta_p)}\, , 
    \\
    \sigma_{\beta_m}^2&=\frac{1}{N-1}\frac{\sum_p (\hat{U}_r(\theta_p) -2\beta_m\tilde{U}_r^{X}(\theta_p))^2}{\sum_p\tilde{U}^X_r(\theta_p)\tilde{U}^X_r(\theta_p)}\, , \label{eq:sigbetals}
\end{split}
\end{align}
where $N$ is the total number of pixels in the sum, and $\sigma_{\beta_m}^2$ is the sum of the squares of the residuals and gives a goodness of fit to each peak calculation. 
Note that \cref{eq:sigbetals} implicitly assumes that the noise covariance matrix around each peak is diagonal and independent of $\theta_m$. This approximation is enough to describe the statistics of the peaks' surroundings in early simulation Phases but becomes insufficient for later Phases, and especially higher map resolutions, where component separation introduces more complicated residuals and pixel correlations. The construction of a more optimal estimator is left for future work.

We weight the $\beta_m$ using the residuals to obtain the final map result:
\begin{align}
\begin{split}
    \beta=\frac{\sum_m \beta_m/\sigma_{\beta_m}^2}{\sum_m 1/\sigma_{\beta_m}^2}.
    \label{eq:pixaverage}
\end{split}
\end{align}
These results are calculated on maps provided using the \fgbuster\ method at $N_{\rm side}=64$ with an FWHM beam of 80 arcminutes for all Phases, at $N_{\rm side}=128$ with an FWHM beam of 40 arcminutes for Phase 3 and 4, and at $N_{\rm side}=512$ with an FWHM beam of around 26~arcminutes (this is an approximation because the true beam for the highest resolution data is not a perfect Gaussian) also for Phases 3 and 4, as well as on the \texttt{CAB-SeCRET} maps at $N_{\rm side}=64$ with a beam of FWHM of 132 arcminutes for all Phases. The Galactic plane mask with $f_{\rm sky}=60~\%$ was applied to the data (as in pipeline \texttt{CAB-SeCRET} and the harmonic approach). 

\begin{figure}[tbp]
    \centering
    \includegraphics{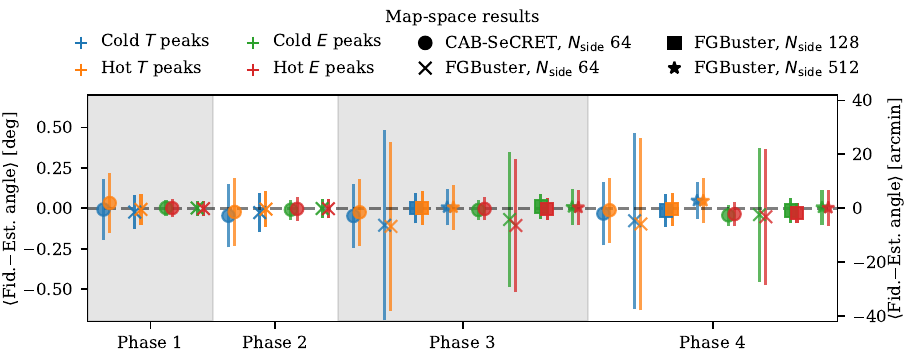}
    \caption{Comparison of the pixel-based results using different \fgbuster\ runs at different resolutions, and results using the \texttt{CAB-SeCRET} maps. \texttt{CAB-SeCRET} maps are used at $N_{\rm side}=64$ with a beam FWHM of 132 arcminutes. The fiducial angle (`Fid.') is zero for Phases 1, 2 and 3, and 0.3$^\circ$ for Phase 4. `Est.' is the central value of the estimated $\hat{\beta}$ from the 100 simulations, the error bars are given by the standard deviation from the 100 simulations. The best results were achieved using \fgbuster\ $N_{\rm side}=64$ for Phases 1 and 2, and \fgbuster\ $N_{\rm side}=128$ for Phases 3 and 4. The $N_\mathrm{side}$ 128 resolution \fgbuster\ maps outperformed the highest resolution maps due to the additional noise complexity in the highest resolution maps.
    }
    \label{fig:pix_compresults}
\end{figure}

\begin{figure}[tbp]
    \centering
    \includegraphics[width=0.9\linewidth]{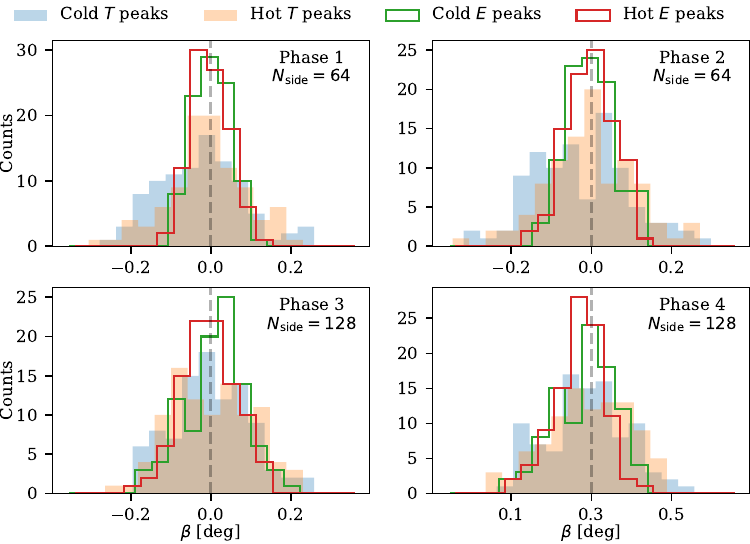}
\caption{Results for 100 \LB\ simulation maps from the pixel-based analysis, generated using \fgbuster. Details regarding each phase may be found in the text and numerical averages in \cref{tab:mapspace_results}.}
\label{fig:pix_hist}
\end{figure}

Results for each of these maps are summarised in \cref{fig:pix_compresults}. The maps producing the best results came from the \fgbuster\ method, for Phases 3 and 4 the $N_{\rm side}=128$ results outperformed the $N_{\rm side}=512$ results. Simple tests show that in maps with similar noise properties, increasing the resolution leads to better constraints on the results, however, in this case the $N_{\rm side}=512$ maps have more complex noise properties. To combine the various frequency maps (with their distinct resolutions and beam functions) the method for producing the $N_{\rm side}=128$ and the $N_{\rm side}=512$ maps using \fgbuster\ are very different\footnote{While the $N_{\rm side}=128$ component maps were obtained from a regular run of \fgbuster\, in pixel space, the $N_{\rm side}=512$ maps were estimated by applying the $\mathbf{W}$ operator (\cref{eq:W_definition}) in harmonic space in order to have an optimal weighting of each frequency map by its associated beam.}. As a result, the $N_{\rm side}=512$ maps perform less well for the stacking approach in particular. Other component separation techniques may have improvements at the $N_{\rm side}=512$ resolution in future work. The large beam of the \texttt{CAB-SeCRET} maps reduces the ability of the peak-stacking method to estimate the final result with high accuracy. Histograms showing the results for the \fgbuster\ Phase 1 and 2 at $N_{\rm side}=64$, and Phase 3 and 4 at $N_{\rm side}=128$ are presented in \cref{fig:pix_hist}.

As shown in \cref{tab:mapspace_results}, most of the constraining power from the stacking method comes from the extrema in the $E$ maps, with a small boost from the extrema in the $T$ maps. This is expected because the $E$ peaks correspond to the $EB$ signal from the power spectrum, whereas the $T$ peaks correspond to the $TB$ signal from the power spectrum. The justification for the higher signal-to-noise of $EB$ with respect to $TB$ can be derived from the Fisher information matrix \cite{planckXLIX2016,Abghari:2022bet}. The consistency of $\beta$ estimates can be assessed by stacking only hot or only cold peaks and the comparison of the various data splits can help ascertain to what degree foregrounds may be affecting results, since foregrounds might be expected to change cold or hot peaks differently. Furthermore, the local nature of the estimator can be used to test the isotropy of the signal by targeting peaks in different regions of the sky.

\begin{center}
\small\addtolength{\tabcolsep}{-3pt}
\begin{table}[tbp]
    \centering
    \begin{tabular}{c|c c c c c}\hline
         & $T$ cold peaks & $T$ hot peaks & $E$ cold peaks & $E$ hot peaks & All  \\
        Phase &$\beta$ [deg]&$\beta$ [deg]&$\beta$ [deg]&$\beta$ [deg]&$\beta$ [deg]\\ \hline
        1 & $ -0.024 \pm 0.104 $ & $ -0.008 \pm 0.096 $ & $ \phantom{-}0.001 \pm 0.046 $ & $ -0.003 \pm 0.048 $ & $ -0.003 \pm 0.030 $\\
        2 &$ -0.027 \pm 0.121 $ & $ -0.005 \pm 0.110 $ & $ -0.002 \pm 0.057 $ & $ -0.003 \pm 0.060 $ & $ -0.005 \pm 0.037 $\\
        3 &$\phantom{-}0.000 \pm 0.094 $ & $ \phantom{-}0.001 \pm 0.103 $ & $ \phantom{-}0.011 \pm 0.076 $ & $ -0.005 \pm 0.066 $ & $\phantom{-} 0.002 \pm 0.040 $\\
        4 &$ \phantom{-}0.285 \pm 0.101 $ & $ \phantom{-}0.294 \pm 0.101 $ & $ \phantom{-}0.285 \pm 0.076 $ & $ \phantom{-}0.272 \pm 0.063 $ & $ \phantom{-}0.281 \pm 0.040 $\\\hline
    \end{tabular}
    \caption{Comparison of the pixel-based results for the best data set from each phase (i.e., \fgbuster, $N_{\rm side}$ 64 for Phases 1 and 2, and \fgbuster\ $N_{\rm side}$ 128 for Phases 3 and 4), showing the average results from hot and cold spots for the $E$ and $T$ peaks (see \cref{eq:pixaverage} for calculation). Results between each of the data splits should be consistent with one another. Inconsistencies would be an indication of effects due to uncontrolled systematics, such as foreground contamination. }
    \label{tab:mapspace_results}
\end{table}
\end{center}

\subsection{Pipelines with control of instrumental polarisation angles}
\label{Pipelineswithcontrolofinstrumentalpolarisationangles}

\subsubsection{Template-based Minami-Komatsu estimator}
\label{pipelineB analysis}
\input{PipelineB/pipelineB_analysis.tex}

\subsubsection{CAB-SeCRET}
\label{sec:pipelineC_analysis}
\input{PipelineC/pipelineC_analysis}

\subsubsection{Pipeline \texttt{J23}}
\label{sec:pipelineD_analysis}
\input{PipelineD/pipelineD_analysis}

\section{Discussion about the impact of foreground contamination}
\label{discussionone}

Our current understanding of Galactic emission is inherently limited by the data presently available, and a comprehensive grasp of the sky's true complexity will remain elusive until new data emerges. See ref.~\cite{Pan-ExperimentGalacticScienceGroup:2025vcd} and references therein, for example refs.~\cite{Meisner2014,Vacher2023,Liu2025}, where new models of polarised Galactic dust and synchrotron emission at CMB frequencies have been developed drawing on the latest observational constraints. We acknowledge the limitations of our simulations, particularly concerning the foreground models included since \texttt{pysm}'s \texttt{d0} and \texttt{d1} models do not account for the line-of-sight variations in the dust SED, which can induce frequency-dependent changes in the polarisation angle as has been observed in \textit{Planck} data~\cite{Pelgrims2021, Ritacco2023}. This frequency decorrelation may impact component separation methods and the characterization of primordial signals, including the $EB$ spectrum~\cite{Vacher2023b,Vacher2024}. Regarding the synchrotron models \texttt{s0} and \texttt{s1}, they show less spatial variability than what has been observed in ground-based measurements~\cite{SPASS2018, QUIJOTE2023:viii}. Nevertheless, current constraints on the synchrotron \textit{EB} signal are consistent with zero, suggesting a smaller impact compared to thermal dust~\cite{synchrotronEB2022,QUIJOTE:2023iv}. However, synchrotron emission remains relevant for accurately characterising polarisation angles. While our analysis does not explicitly model these complexities, our conclusions remain robust within the limitations of the foreground models considered. Future work should explore more sophisticated sky models (e.g., those of refs.~\cite{HerviasCaimapo2024, Vansyngel_dust_model, MKD_dust_model, Vacher2023, Vacher2024}) as well as other instrumental effects due to, e.g., a non-ideal half-wave plate~\cite{Monelli2023}, to refine our understanding of these astrophysical and instrumental effects.

However, we expect that these additional foreground complexities will not substantially alter the estimated isotropic birefringence angle presented in this work. The main argument supporting this statement is the following: independently of their complexity, foreground models are expected to primarily impact the lowest multipoles. 
We have, however, shown for two out of the five pipelines presented here that our results are weakly dependent on the lowest $\ell$s. Regarding the $D$-estimator, we excluded the first 50 or 100 multipoles from the analysis in \cref{subsection:harmonic} (see \cref{fig:histogram_phase1}). 
Of course, this statement assumes that the component separation layer effectively handles foregrounds at multipoles larger than $100$ and smaller than $750$, thereby minimising their impact on our results within this range, which we believe to be reasonable. 
Similarly, although the MK estimator (see \cref{pipelineB method}), benefits from accessing the foreground signal at large angular scales to break the degeneracy between CB and polarisation angles, it can still tolerate the exclusion of the lowest multipoles if the foreground model is not reliable. 
We tested that an $\ell_\mathrm{min}\approx 100$ cut leads to only $<0.7\,\sigma$ shifts in $\beta$ compared to the results presented in \cref{pipelineB analysis}. Furthermore, we have seen that an incorrect modeling of the dust $EB$ correlation could bias the CB angle measured with the MK estimator. However, ref.~\cite{HerviasCaimapo2024} showed that the difference between $C_\ell^{EB,\mathrm{dust}}$ models can be partially absorbed by the extra degrees of freedom the $\mathcal{A}$ amplitudes provide, resulting in $<1\,\sigma$ shifts in $\beta$ as long as a reasonable template for $EB$ is provided. Alternative ways to estimate the dust $EB$ contribution from measurements of the magnetic misalignment angle also exist~\cite{Clark:2021kze, Cukierman:2022odh}.

In general, the component separation layer employed to provide the input for the methods in \cref{ssec:pipelineA_methods} and all the techniques presented in \cref{ssec:pipelineBCD_methods} can be extended to parametrize the foreground complexity. Of course, the increase in the number of parameters employed comes along with an increase in their statistical uncertainties which needs to be quantified. Ultimately, a robust assessment will require end-to-end simulations incorporating realistic noise properties, scanning strategy, and a representative level of instrumental systematics. Such simulations will be essential to draw firm conclusions regarding the impact of foreground complexity on the measurement of the birefringence angle which can be the subject for future analyses.

\section{Conclusion}
\label{discussion}

\input{alternative_conclusions}

\acknowledgments
This work is supported in Japan by ISAS/JAXA for Pre-Phase A2 studies, by the acceleration program of JAXA research and development directorate, by the World Premier International Research Center Initiative (WPI) of MEXT, by the JSPS Core-to-Core Program of A. Advanced Research Networks, and by JSPS KAKENHI Grant Numbers JP15H05891, JP17H01115, and JP17H01125. The Canadian contribution is supported by the Canadian Space Agency. The French \textit{LiteBIRD} phase A contribution is supported by the Centre National d’Etudes Spatiale (CNES), by the Centre National de la Recherche Scientifique (CNRS), and by the Commissariat à l’Energie Atomique (CEA). The German participation in \textit{LiteBIRD} is supported in part by the Excellence Cluster ORIGINS, which is funded by the Deutsche Forschungsgemeinschaft (DFG, German Research Foundation) under Germany’s Excellence Strategy (Grant No. EXC-2094 - 390783311). The Italian \textit{LiteBIRD} phase A contribution is supported by the Italian Space Agency (ASI Grants No.~2020-9-HH.0 and 2016-24-H.1-2018), the National Institute for Nuclear Physics (INFN) and the National Institute for Astrophysics (INAF). Norwegian participation in \textit{LiteBIRD} is supported by the Research Council of Norway (Grant No.~263011) and has received funding from the European Research Council (ERC) under the Horizon 2020 Research and Innovation Programme (Grant agreement No.~772253 and 819478). The Spanish \textit{LiteBIRD} phase A contribution is supported by MCIN/AEI/10.13039/501100011033, project refs. PID2019-110610RB-C21, PID2020-120514GB-I00, PID2022-139223OB-C21 (funded also by European Union NextGenerationEU/PRTR), and by MCIN/CDTI ICTP20210008 (funded also by EU FEDER funds). Funds that support contributions from Sweden come from the Swedish National Space Agency (SNSA/Rymdstyrelsen) and the Swedish Research Council (Reg.~no.~2019-03959). The UK  \textit{LiteBIRD} contribution is supported by the UK Space Agency under grant reference ST/Y006003/1 - LiteBIRD UK: A major UK contribution to the LiteBIRD mission - Phase1 (March 25). The US contribution is supported by NASA grant no.~80NSSC18K0132. We acknowledge funding from the SCIPOL project\footnote{\url{scipol.in2p3.fr}} funded by the European Research Council (ERC) under the European Union’s Horizon 2020 research and innovation program (PI: Josquin Errard, Grant agreement No.~101044073).
NB, AG, and PN acknowledge support by the MUR PRIN2022 Project BROWSEPOL: Beyond standaRd mOdel With coSmic microwavE background polarisation-2022EJNZ53 financed by the European Union-Next Generation EU. RS acknowledges the support of the Natural Sciences and Engineering Research Council of Canada (NSERC).
EdlH acknowledges support from IN2P3.
This research used resources from the National Energy Research Scientific Computing Center (NERSC), a U.S. Department of Energy Office of Science User Facility. We acknowledge the use of the \texttt{healpy}~\cite{healpy}, \texttt{pysm}~\cite{PYSM3}, \fgbuster~\cite{Stompor:2008sf}, \texttt{NaMaster}~\cite{namaster}, \texttt{emcee}~\cite{emcee}, \texttt{scipy}~\cite{2020SciPy-NMeth}, \texttt{numpy}~\cite{numpy}, and \texttt{matplotlib}~\cite{matplotlib} software packages.

\bibliographystyle{JHEP}
\bibliography{bibliography}

\appendix
\section{Definition of the template-based Minami-Komatsu estimator}
\label{pipelineB appendix}
\input{PipelineB/pipelineB_appendix.tex}

\end{document}

%% file: authors_affiliations_LB_2021.27_general_jcap.tex
\author[1]{E.\,de\,la\,Hoz,}
\author[2]{P.\,Diego-Palazuelos,}
\author[3]{J.\,Errard,}
\author[4,5]{A.\,Gruppuso,}
\author[6]{B.\,Jost,}
\author[7,8]{R.\,M.\,Sullivan,}
\author[9,10]{M.\,Bortolami,}
\author[11,6]{Y.\,Chinone,}
\author[7,12]{L.\,T.\,Hergt,}
\author[2,6]{E.\,Komatsu,}
% ...
\affiliation[1]{CNRS-UCB International Research Laboratory, Centre Pierre Binétruy, UMI2007, Berkeley, CA 94720, USA}
\affiliation[2]{Max Planck Institute for Astrophysics, Karl-Schwarzschild-Str. 1, D-85748 Garching, Germany}
\affiliation[3]{Université Paris Cité, CNRS, Astroparticule et Cosmologie, F-75013 Paris, France}
\affiliation[4]{INAF - OAS Bologna, via Piero Gobetti, 93/3, 40129 Bologna, Italy}
\affiliation[5]{INFN Sezione di Bologna, Viale C. Berti Pichat, 6/2 – 40127 Bologna, Italy}
\affiliation[6]{Kavli Institute for the Physics and Mathematics of the Universe (Kavli IPMU, WPI), UTIAS, The University of Tokyo, Kashiwa, Chiba 277-8583, Japan}
\affiliation[7]{Department of Physics and Astronomy, University of British Columbia, 6224 Agricultural Road, Vancouver, BC V6T1Z1, Canada}
\affiliation[8]{Institute of Theoretical Astrophysics, University of Oslo, Blindern, Oslo, Norway}
\affiliation[9]{Dipartimento di Fisica e Scienze della Terra, Università di Ferrara, Via Saragat 1, 44122 Ferrara, Italy}
\affiliation[10]{INFN Sezione di Ferrara, Via Saragat 1, 44122 Ferrara, Italy}
\affiliation[11]{International Center for Quantum-field Measurement Systems for Studies of the Universe and Particles (QUP), High Energy Accelerator Research Organization (KEK), Tsukuba, Ibaraki 305-0801, Japan}
\affiliation[12]{Université Paris-Saclay, CNRS/IN2P3, IJCLab, 91405 Orsay, France}
% ...

%% file: PipelineB/pipelineB_methodology.tex
This pipeline is based on the MK estimator introduced in refs.~\cite{Minami:2019ruj, Minami:2020f3E01M, Minami:2020fin} and applied to data in refs.~\cite{Minami:2020odp, Diego-Palazuelos:2022dsq, Eskilt:2022wav, Eskilt:2022cff}. As we mentioned in the introduction, thanks to the use of the information contained in the foreground emission it is possible to estimate at the same time $\beta$ and $\alpha_f$ corresponding to each frequency channel $\nu_f$ (with $f=1,...,22$ in the case of \LB). This technique is robust against several instrumental systematics~\cite{Diego-Palazuelos:2023}, including the miscalibration of $\alpha_f$ and other systematic effects that produce spurious $EB$ correlations (e.g., intensity-to-polarisation leakage, beam leakage, or cross-polarisation effects~\cite{Yadav:2010psy, Miller:2009syn, Hu:2003dsf}). However, it is sensitive to the $EB$ correlation of Galactic foreground emission and its modelling~\cite{Diego-Palazuelos:2022dsq, Diego-Palazuelos:2023, Clark:2021kze, Cukierman:2022odh, HerviasCaimapo2024}. Hence, compared to the pipelines from \cref{ssec:pipelineA_methods}, this approach has the advantage of being independent of additional knowledge of the instrument calibration, at the price of needing a precise model for the $EB$ correlation of the polarised foreground emission. 

The MK pipeline is applied directly on frequency maps rather than the CMB solution provided by the component-separation layer to exploit the information contained in foreground emission. Taking foregrounds into account, the $EB$ spectrum obtained from the cross-correlation of the $i$ and $j$ frequency channels is, within the small-angle approximation ($|\alpha_f|\lesssim5^\circ$ and $|\beta|\lesssim 5^\circ$),
%
\begin{align}
\begin{split}
    C_\ell^{E_iB_j,\obs} \approx & \phantom{+} 2\alpha_j C_\ell^{E_iE_j,\obs} - 2\alpha_i C_\ell^{B_iB_j,\obs}+ 2\beta \left( C_\ell^{E_iE_j,\mathrm{CMB}} - C_\ell^{B_iB_j,\mathrm{CMB}}\right)\\
    & +{\cal A^\mathrm{sync}} C_\ell^{E_iB_j,\mathrm{sync}}  + {\cal A^\mathrm{dust}}  C_\ell^{E_iB_j,\mathrm{dust}} \\
    & +{\cal A^\mathrm{sync\times dust}} \left( C_\ell^{E_i^\mathrm{sync} B_j^\mathrm{dust}} + C_\ell^{E_i^\mathrm{dust} B_j^\mathrm{sync}} \right)  ,
    \label{eq:pipelineBrotatedEB}
\end{split}
\end{align}
%
where the superscripts indicate whether the spectra come from the correlation of the $E$ and $B$ modes of the observed frequency maps, the synchrotron and dust models, or the CMB spectra in the case of null CB, multiplied by the beam and pixel window function of each channel. The $\mathcal{A}^{\rm x}$ are multiplicative factors added \textit{ad hoc} to control the amplitude of the synchrotron ($\mathcal{A}^\mathrm{sync}$) and dust ($\mathcal{A}^\mathrm{dust}$) contributions as well as their cross-correlation ($\mathcal{A}^\mathrm{sync\times dust}$). This constitutes an extension of the methodology compared to previous works~\cite{Diego-Palazuelos:2022dsq, Diego-Palazuelos:2023} where neither the synchrotron emission nor its cross-correlation with dust were considered. Although \cref{eq:pipelineBrotatedEB} is a general expression considering a potential $EB$ contribution from both synchrotron and dust, no significant synchrotron $EB$ correlation has been found yet~\cite{Martire:2022lsh, QUIJOTE:2023iv} and we only have moderate evidence of non-null dust $EB$ correlation induced by the misalignment between dust filaments and the Galactic magnetic field~\cite{Huffenberger2020, Clark:2021kze, Cukierman:2022odh}.

Following refs.~\cite{Diego-Palazuelos:2022dsq,Diego-Palazuelos:2023}, we leave $\mathcal{A}^{\rm x}$ as a free amplitude parameter and obtain the synchrotron and dust $EB$ spectrum from the templates produced in component-separation analyses that fit parametric models to CMB data (e.g., \texttt{Commander}~\cite{Eriksen:2008dfa}, \texttt{FGBuster}~\cite{Stompor:2008sf}, or \texttt{B-SeCRET}~\cite{delaHoz:2020ggq}). In particular, we use the \texttt{d0s0} and \texttt{d1s1} foreground models provided by \texttt{pysm}~\cite{Thorne_2017,PYSM3}. We take $\mathcal{A}^{\rm x}$ as a single overall amplitude and use the SEDs provided by the parametric models to scale the foreground template to the target frequencies. See refs.~\cite{Minami:2019ruj, Minami:2020f3E01M, Diego-Palazuelos:2022dsq, Eskilt:2022cff, Cukierman:2022odh, HerviasCaimapo2024} for alternative ways to model the foreground $EB$ correlation.

From \cref{eq:pipelineBrotatedEB}, we build a Gaussian likelihood to simultaneously fit $(\mathcal{A}^{\rm x}, \beta, \alpha_f)$ and find the maximum-likelihood solution for all parameters following the semi-analytical iterative algorithm presented in refs.~\cite{delaHoz:032D, Diego-Palazuelos:2023}. To emulate real analyses~\cite{Minami:2020odp, Diego-Palazuelos:2022dsq, Eskilt:2022cff, Eskilt:2022wav}, we consider a minimal mask covering extragalactic point sources and the regions of brightest CO emission that leaves a sky fraction of $f_\mathrm{sky}=93.2~\%$ free for the analysis after applying a $0.5^\circ$ smoothing apodisation. For brevity, we limit our study to only one mask, since a good correction of the dust-$EB$ bias is possible independently of the choice of mask when the foreground model that generated the simulations is also used as the dust template in the MK estimator~\cite{Diego-Palazuelos:2023}. In \cref{pipelineB analysis}, we also explore how our imperfect knowledge of foreground emission impacts the efficiency of the MK estimator by using templates of a simpler sky model to describe a more complex sky. We use \texttt{NaMaster}\footnote{\url{https://github.com/LSSTDESC/NaMaster}}~\cite{namaster} to calculate the pseudo-$C_\ell$ of the masked sky and bin the power spectra into $N_{\rm b}=21$ uniform bins of $\Delta\ell=20$ in the range $\ell\in[21,440]$. In the presence of an unexpectedly high $1/f$ correlated noise contribution or an unreliable dust $EB$ model, the minimum multipole can be raised to $\ell_\mathrm{min}\approx 100$ values with minimal impact on the results presented in \cref{pipelineB analysis}. Extending the analysis to higher multipoles does not report an advantage as the MK-estimator signal-to-noise on $\beta$ already saturates around $\ell_\mathrm{max}\approx400$ for \LB's instrumental configuration. See \cref{pipelineB appendix} for more details on the definition of the estimator.

%% file: PipelineC/pipelineC_methodology.tex
\texttt{CAB-SeCRET} is shorthand for Calibration Angles \texttt{B-SeCRET}, representing an extension of the standard \texttt{B-SeCRET} method by incorporating $\alpha_f$ into the component-separation process. The component-separation analysis is performed using \texttt{B-SeCRET}  (Bayesian-Separation of Components and Residuals Estimate Tool) \cite{delaHoz:2020ggq}, an alternative parametric component-separation approach to \texttt{FGBuster}. \texttt{B-SeCRET} utilizes an affine-invariant ensemble sampler for Markov chain Monte Carlo (MCMC) to sample the posterior distribution, from which we can recover the marginalized parameters' posterior distribution. By integrating $\alpha_f$ into the component-separation process, they are marginalized over, and their uncertainties are propagated into the CMB maps, and hence into the estimation of $\beta$. Instead of sampling the entire posterior distribution simultaneously, we partition the parameter space into three subspaces: amplitudes (\amp); spectral parameters (\spec); and rotation angles (\ang). We then sample their conditional distributions. This iterative process continues until convergence is reached.

\texttt{B-SeCRET} is a Bayesian method hence prior information is applied. Specifically, multivariate Gaussian priors are employed for $\alpha_f$ ($\mathcal{N}(\myvector{\alpha_f}, \mathbf{C}_{\alpha_f})$). These priors are derived from multi-frequency data by nulling the $EB$ power spectra, as described in Refs.~\cite{delaHoz:032D, Krachmalnicoff2022:jakv}.  However, this method assumes $\beta=0$. Consequently, if $\beta$ is non-zero, the recovered $\alpha_f$ become $\myvector{\alpha}_f + \beta$. In such cases, $\alpha_f$ estimates from the template-based MK estimator, as detailed in \cref{pipelineB method}, are utilized as prior information. 
The former set of priors is denoted as $\Pi(\myvector{\alpha_f}, \beta=0)$, while the latter is denoted as $\Pi(\myvector{\alpha_f},\beta)$. Alternatively, other prior information, such as instrument calibration before launch, external calibration with Tau~A, or an artificial calibration source, could potentially be utilized.

Prior to conducting the component-separation analysis, the computational load for fitting the maps at $N_{\rm side} = 512$ is reduced by downgrading them to $N_{\rm side} = 64$. Additionally, all maps are subjected to convolution with the same beam as follows.

\begin{itemize}
\item[i)] The original maps at $N_{\rm side} = 512$ are transformed into spherical harmonics coefficients.
\item[ii)] Pixel window function and beam deconvolution is applied to each frequency channel based on the beam's FWHM, as specified in Table 3 of ref.~\cite{2022LB_ptep}.
\item[iii)] The spherical harmonic coefficients of each channel are convolved with a Gaussian beam with $\mathrm{FWHM}=2.2^{\circ}$ and the corresponding pixel window function, and then transformed into $N_{\rm side} = 64$ maps. The FWHM corresponds to $2.4$ times the effective pixel size at $N_{\rm side} = 64$ to account for the pixel window function effects.
\end{itemize}
Additionally, we apply the same \Planck\ $f_{\rm sky}=60~\%$ Galactic-plane mask discussed in \cref{ssec:pipelineA_methods} to mitigate the influence of large residuals from foregrounds on other parameters. 

We consider two cases of this pipeline, \texttt{CAB-SeCRET}, where $\alpha_f$ are included in the parametric model and $\beta$ is estimated from the cleaned CMB map, and \texttt{$\beta$CAB-SeCRET}, where $\beta$ is included as another parameter in \ang. 

\paragraph{CAB-SeCRET.} 
\texttt{CAB-SeCRET} includes only the $\alpha_f$ as model parameters (\ang $= \left\{\alpha_{f}, \forall f\in N_{\nu}\right\}$) as in ref.~\cite{delaHoz:032D}. Hence, the signal is modelled as follows for a given pixel $p$:
\begin{equation}
    \begin{pmatrix}
		Q^{\rm{obs}}(f, \alpha_f, \gamma) \\
	    U^{\rm{obs}}(f, \alpha_f, \gamma) 
	\end{pmatrix}_{p}
	=
	\begin{pmatrix}
		\cos(2\alpha_f) & -\sin(2\alpha_f)\\
	    \sin(2\alpha_f) & \phantom{-}\cos(2\alpha_f)\\
	\end{pmatrix}
	\begin{pmatrix}
		Q(f,\gamma) \\
	    U(f,\gamma)
	\end{pmatrix}_{p} 
    =
    \mathbf{R}(\alpha_f)
    \begin{pmatrix}
				{Q} (f,\gamma)\\
			    {U} (f,\gamma)
	\end{pmatrix}_{p} \, ,
 \label{eq:pipeline_c1_model}
\end{equation}
where $\mathbf{R}(\alpha_f)$ is the rotation matrix that accounts for the $Q$ and $U$ mixing due to non-perfect calibration,
$\gamma = \mathcal{A} \cup \mathcal{B}$, and
\begin{equation}
    \begin{pmatrix}
		{Q}(f, \gamma) \\
	    {U}(f, \gamma)
	\end{pmatrix}_{p}
	=
	\begin{pmatrix}
		a_{\rm CMB}^{\scriptscriptstyle{Q}} \\
	    a_{\rm CMB}^{\scriptscriptstyle{U}}
	\end{pmatrix}_{p}
	+ 
	\begin{pmatrix}
		a_{\rm s}^{\scriptscriptstyle{Q}} \\
	    a_{\rm s}^{\scriptscriptstyle{U}}
	\end{pmatrix}_{p}
	\dfrac{1}{u(f)}\left(\dfrac{f}{f_{\rm s}}\right)^{\beta_{\rm s}} 
	+ 
	\begin{pmatrix}
		a_{\rm d}^{\scriptscriptstyle{Q}} \\
	    a_{\rm d}^{\scriptscriptstyle{U}}
	\end{pmatrix}_{p}
	\dfrac{1}{u(f)}
	\left(\dfrac{f}{f_{\rm d}}\right)^{\beta_{\rm d}-2}
    \dfrac{B_f\left(T_{\rm d}\right)}{B_{f_{\rm d}}\left(T_{\rm d}\right)}
    \, ,
 \label{eq:pipeline_c1_astrophysical_model}
\end{equation}
is the parametric model used to describe the astrophysical components considered in the simulations, taking $f_{\rm s} = 30$\,GHz and $f_{\rm d} = 353$\,GHz as pivot frequencies for the synchrotron and dust emission.

The amplitude parameters
\begin{equation}
    \mathcal{A} = \bigcup\limits_{p \in N_{\rm pix}}\left\{a_{\rm CMB}^{\scriptscriptstyle{Q}}, a_{\rm CMB}^{\scriptscriptstyle{U}}, a_{\rm s}^{\scriptscriptstyle{Q}},  a_{\rm s}^{\scriptscriptstyle{U}}, a_{\rm d}^{\scriptscriptstyle{Q}}, a_{\rm d}^{\scriptscriptstyle{U}}\right\}_p
    \label{eq:pipeline_c_amp}
\end{equation} 
 vary from pixel to pixel, while the spectral parameters
 \begin{equation}
     \mathcal{B} = \bigcup\limits_{c \in N_{c}}\left\{\beta_{\rm s},\beta_{\rm d},T_{\rm d}\right\}_c
 \end{equation} 
 are considered uniform within specific sky regions (i.e., clusters of pixels). Additionally, individual Gaussian priors are applied on \spec: $\beta_{\rm s}\sim \mathcal{N}(-3.1,0.3)$, $\beta_{\rm d}\sim \mathcal{N}(1.56,0.1)$, and $T_{\rm d}\sim \mathcal{N}(21,3)$ \cite{delaHoz:2020ggq,delaHoz:032D}. Reducing the number of parameters by clustering the \spec\ yields significantly smaller statistical uncertainties for all parameters, including CB, although it introduces biases in $\beta$ when the assumption of constant \spec\ is invalid. Different clustering strategies are investigated in \cref{sec:pipelineC_analysis} using Phase 2 simulations where spatial variations of the spectral parameters are present.

From the cleaned CMB maps, we determine CB as the residual rotation angle in the CMB, following the approach outlined in ref.~\cite{Krachmalnicoff2022:jakv}.
This involves finding the $\beta$ value that minimises the harmonic estimator derived from \cref{DEB}. Initially, we calculate the angular power spectrum and their covariance matrix using \texttt{NaMaster}. The power spectra are then binned uniformly with a bin width of $\Delta\ell=10$ within the range $\ell\in[12,191]$, employing the same weighting scheme as ref.~\cite{Krachmalnicoff2022:jakv}. We choose to use all the available bins at $N_{\rm side}=64$, except the first bin due to the high uncertainties recovered with a pseudo-$C_{\ell}$ method. Additionally, we apply apodisation to the mask using a C2 kernel with an apodisation scale of $22^\circ$, as suggested in ref.~\cite{eclipse}. 

The CB angle is obtained by minimising the following log-Gaussian distribution:
\begin{equation}
   -2 \log \mathcal{L} = D^{EB, \mathrm{obs}}(\beta)\mathbf{M}^{-1}D^{EB, \mathrm{obs}}(\beta) + \log \left|\mathbf{M}\right|  + N_{\rm b} \log\left(2\pi\right) ,
    \label{eq:pipeline_c1_loglikelihood}
\end{equation}
where $D^{\rm{EB}, \rm{obs}}(\beta)$ is the vector of harmonic-based estimators in \cref{DEB} evaluated for each multipole bin $b$, $N_{\rm b}$ is the number of multipole bins, $\left|...\right|$ is the determinant operator, and
\begin{align}
\begin{split}
    \mathbf{M} = &\cos^2(4\beta) \mathbf{\Sigma}^{\scriptscriptstyle EB, EB} + \cos(4\beta)\sin(4\beta) \left(\mathbf{\Sigma}^{\scriptscriptstyle EB, BB} - \mathbf{\Sigma}^{\scriptscriptstyle EB, EE}\right) \\ 
    &+ \dfrac{1}{4} \sin^2(4\beta) \left(\mathbf{\Sigma}^{\scriptscriptstyle EE, EE} + \mathbf{\Sigma}^{\scriptscriptstyle BB, BB} - 2\mathbf{\Sigma}^{\scriptscriptstyle EE, BB}\right) \, ,
    \label{eq:pipeline_c1_covmat}
    \end{split}
\end{align}
where $\mathbf{\Sigma}$ represents the covariance matrix of the cross-power spectra, i.e.,
\begin{equation}
\Sigma_{bb'}^{\scriptscriptstyle{ XY, X'Y'}} \equiv \left<\Delta C_{b}^{XY} \left.\Delta C_{b'}^{X'Y'}\right.\right>.
\end{equation}
We employ \texttt{emcee}\footnote{\url{https://github.com/dfm/emcee}.}~\cite{emcee} to sample the likelihood in \cref{eq:pipeline_c1_loglikelihood}. The median and standard deviation of the resulting chains provide estimates for $\beta$ and $\sigma_{\beta}$, respectively.

\paragraph{$\beta$CAB-SeCRET.} 

In this pipeline, CB is incorporated into the model as a parameter, i.e., \ang $= \left\{\beta\right\} \cup \left\{\alpha_{f}, \forall f\in N_{\nu}\right\}$. Thus, \cref{eq:pipeline_c1_model,eq:pipeline_c1_astrophysical_model} lead to
\begin{equation}
    \begin{pmatrix}
		Q^{\rm{obs}}(f, \beta, {\alpha_f}, \gamma) \\
	    U^{\rm{obs}}(f, \beta, {\alpha_f}, \gamma) 
	\end{pmatrix}_{p}
	=
    \mathbf{R}(\beta + {\alpha_f})
    \begin{pmatrix}
		a_{\rm CMB}^{\scriptscriptstyle{Q}} \\
	    a_{\rm CMB}^{\scriptscriptstyle{U}}
    \end{pmatrix}_{p}
    +
    \mathbf{R}(\alpha_f)\left[
    \begin{pmatrix}
				{Q} (f,\gamma)\\
			    {U} (f,\gamma)
    \end{pmatrix}_{p} 
    -
    \begin{pmatrix}
		a_{\rm CMB}^{\scriptscriptstyle{Q}} \\
	    a_{\rm CMB}^{\scriptscriptstyle{U}}
    \end{pmatrix}_{p}\right] \, .
 \label{eq:pipeline_c2_model}
\end{equation}

In this scenario, the cleaned CMB map undergoes no additional rotation, since any rotation has been accounted for and removed during the component separation process. Consequently, $\beta$ is directly obtained from the component-separation step.

For \texttt{$\beta$CAB-SeCRET}, we utilise the results from the template-based MK estimator as prior information, denoted as $\Pi(\myvector{\alpha_f},\beta)$. It is important to note that this approach differs from the template-based MK pipeline in that we simultaneously allow for variations in the foreground model and CB. This simultaneous variation helps mitigate biases arising from potentially incorrect modelling of foreground templates, as demonstrated in \cref{subsubsec:comp_pipeline_c1_c2}.

%% file: PipelineD/pipelineD_methodology.tex
Similarly to the previous section, the pipeline presented here includes the effect of $\alpha_f$ in its data model.
This allows us to fit them in the component-separation step, and therefore correct their 
effects on the CMB map. 
This pipeline is a generalisation of \texttt{FGBuster}~\cite{Stompor:2008sf}, which is the method used to obtain the cleaned CMB
maps employed in \cref{sssec:d_estimator,sssec:stack-estimator}.
The method we use here is therefore pixel-based and parametric. The details of its
generalisation for the inclusion of $\alpha_f$ and CB are described in ref.~\cite{Jost:2023}. For conciseness, we will call this pipeline \texttt{J23}. 
The data model used to describe the frequency maps $\mathbf{d}_{hp}$ 
is as follows:
\begin{equation}
    \label{eq:pipeD_DataModel}
    \mathbf{d}_{hp} = \mathbf{X}_{hh'}(\{\alpha_f\}) \mathbf{A}_{h'o}(\{\beta_{\rm d}, T_{\rm d}, \beta_{\rm s}\}) \mathbf{s}_{op} + \mathbf{n}_{hp}.
\end{equation}
Here $p$ describes pixels. Index $h$ spans both $\{Q,U\}$ Stokes parameters and frequency bands (the number of $h$ indices is then $n_{h} = n_{\rm{Stokes}}\,n_{\rm{frequency}}$). Similarly, $o$ covers $\{Q,U\}$ and sky components ($n_{o} = n_{\rm{Stokes}}\,n_{\rm{component}}$).
$\mathbf{s}_{op}$ is the vector storing the $\{Q,U\}$ maps for each considered sky component: $\{\rm{CMB}, \rm{dust}, \rm{synchrotron}\}$. 
$\mathbf{X}_{hh'}$ is an $n_{\rm{Stokes}}\,n_{\rm{frequency}} \times n_{\rm{Stokes}}\,n_{\rm{frequency}}$ 
block diagonal matrix that encodes the effect of $\alpha_f$, each block being a 
rotation matrix $\mathbf{R}^{\rm inst} (\alpha_f)$ 
acting on the $\{Q,U\}$ maps at frequency $f$. 
$\mathbf{A}_{ho}$ is the mixing matrix, encoding the frequency dependence of the different components, which is an $n_{\rm{Stokes}}\,n_{\rm{frequency}} \times n_{\rm{Stokes}}\,n_{\rm{component}}$ matrix. 
We assume: (i) the frequency scaling to be the same between the $Q$ and $U$ parameters of a given component;
(ii) no intrinsic $Q$-$U$ mixing, making every other element of a column equal to 0;
(iii) that the CMB scaling is known and that we are working in CMB 
units, thus the non-zero elements
of the first two columns of $\mathbf{A}$ are  all equal to 1; and
(iv) that the dust frequency scaling is a standard modified blackbody law parameterised
by the dust spectral index $\beta_\mathrm{d}$ and the dust temperature $T_\mathrm{d}$, 
and that the synchrotron is a power law with index $\beta_\mathrm{s}$ 
(see \cref{eq: sync SED}
for a description of those laws).
Finally, $\mathbf{n}_{hp}$ is the white Gaussian noise present in the frequency maps, 
same as in \cref{eq:data_modeling}.
For conciseness, we define $\mathbf{\Lambda}_{ho} (\beta_{\rm d}, T_{\rm d}, \beta_{\rm s}, \{\alpha_f\}) \equiv \mathbf{X}_{hh'}(\{\alpha_f\}) \mathbf{A}_{h'o}(\beta_{\rm d}, T_{\rm d}, \beta_{\rm s})$.

Once we have the data model we can proceed to the analysis. The pipeline goes through 5+1 steps that are detailed below.
\paragraph{Step 0: Pre-processing.} 
Contrary to \cref{eq:data_modeling}, the 
data model in \cref{eq:pipeD_DataModel} does not explicitly include the effect of the beam.
To deal with this,
the frequency maps are first deconvolved according to their respective beam
(Table 3 of ref.~\cite{2022LB_ptep})
and convolved to the same common 
Gaussian beam with FWHM of $80$ arcmin used in ref.~\cite{2022LB_ptep}, as well as in the standard \texttt{FGBuster} approach in section~\ref{ssec:pipelineA_methods}. 
This is done so that we do not suffer from frequency dependence of the resolution. Furthermore, convolving to a smaller FWHM would boost the noise of channels with the poorest resolution. Extending the data model to include the beams at different frequencies is explored in ref.~\cite{rizzieri2024mainbeamtreatmentparametric}, and is a possible extension of this method, but beyond the scope of the work presented here.
The maps are also downgraded to $N_{\rm side}=64$, deconvolved by the pixel window function 
corresponding to the old $N_{\rm side}=512$, and convolved with the new one. 
The data vector after pre-processing is denoted $\bar{\mathbf{d}}$. This operation
will notably affect the white noise, which 
is then 
denoted $\bar{\mathbf{n}}$.
We apply \Planck's Galactic masks with either $f_\mathrm{sky} = 60\%$ or $f_\mathrm{sky} = 40\%$ depending on the case, as will be described in \cref{sec:pipelineD_analysis}.

\paragraph{Step 1: Component separation.} \label{paragraph:pipeD_compsep}
Unlike \texttt{B-SeCRET}, \texttt{FGBuster}
and its generalisation use a ridge-likelihood, called \textit{spectral likelihood}, $\mathcal{L}^{\rm spec}$,
where the only free parameters are $\{\beta_{\rm d}, T_{\rm d}, \beta_{\rm s}\,, \{\alpha_f\}\}$ encoding
$\mathbf{\Lambda}$.
Here, we assume those parameters to be constant across the sky.

While it was already shown in ref.~\cite{Jost:2023} that the \texttt{J23} pipeline correctly retrieves the relative angles between 
frequencies, the degeneracy between the global 
$\alpha$ offset and $\beta$ remains.
To lift it, additional information must be added to the likelihood. Here, it takes the form of 
Gaussian priors on the $\{\alpha_f\}$ parameters added to the likelihood. 
The generalized spectral likelihood that we use is then
\begin{align}
\begin{split}
    \label{eq:generalised_spectral_likelihood}
    &-2 \log (\mathcal{L}^\mathrm{spec}(\{\beta_{\mathrm{fg}}, \alpha_f\})) = \\ 
    &+ \rm{constant} - \sum_{p} \mathrm{tr} \left(\mathbf{\overline{N}}^{-1} \mathbf{\Lambda} 
     \left( \mathbf{\Lambda}^{\sf T} \mathbf{\overline{N}}^{-1} \mathbf{\Lambda}  \right)^{-1} 
      \mathbf{\Lambda}^{\sf T} \mathbf{\overline{N}}^{-1} \bar{\mathbf{d}}_{p} \bar{\mathbf{d}}_{p}^{\sf T}\right)  
    + \sum_{f} \frac{\left(\alpha_f - \tilde{\alpha}_f \right)^2}{\sigma_{\alpha_f}^2} \, ,
\end{split}
\end{align}
with constant being the constant ensuring the positivity of the right-hand side. We have omitted the $o$ and $h$ indices for conciseness, and $f$ spans the frequency bands.
Although the pre-processing correlates the initial white noise, here we still assume a white and scale-invariant noise in our analysis. $\mathbf{\overline{N}}$ is therefore a pixel-independent and diagonal $n_{\mathrm{Stokes}}\,n_{\mathrm{frequency}} \times n_{\mathrm{Stokes}}\,n_{\mathrm{frequency}}$ matrix. The diagonal elements of $\mathbf{\overline{N}}$ are computed from $\left< \bar{\mathbf{n}} \bar{\mathbf{n}}^{\sf T}\right>$, where we average over noise simulations and pixels. This approximate modelling of the noise properties does not seem to introduce a significant bias in our results, which, as will be discussed in \cref{sec:pipelineD_analysis}, are dominated by the approximations in foreground modelling.

The second term in \cref{eq:generalised_spectral_likelihood} corresponds to the Gaussian priors.
We assume the priors on each $\alpha_f$ to be independent. The prior on the angle $\alpha_f$
is centred on $\tilde{\alpha}_f$ and has a variance of $\sigma_{\alpha_f}^2$. 
Our approach here is to assume priors coming from an independent ground or in-flight calibration.
As such, we assume these priors match the current requirements of the \LB~mission and take
$\sigma_{\alpha_f}$ from \cref{tab:alpha_f}.
The central values of the priors $\tilde{\alpha}_f$ 
are \textit{not} chosen to be the true values of the input $\alpha_f$ but drawn from a normal
distribution centred at zero with variance $\sigma_{\alpha_f}^2$.\footnote{In ref.~\cite{Jost:2023}, 
the calibration measurements were averaged over, leading 
to an additional factor of $2$ in the variance and no bias in the prior.}

To estimate $\{\beta_{\rm d}, T_{\rm d}, \beta_{\rm s}\,, \{\alpha_f\}\}$ we must minimise 
\cref{eq:generalised_spectral_likelihood}. In practice, we initialise the
minimisation by performing an MCMC sampling with \texttt{emcee}~\cite{emcee}. Once the MCMC converges,
we use the mean of the chains to initialise
the minimisation of \cref{eq:generalised_spectral_likelihood},
which is performed using the \texttt{L-BFGS-B} method from \texttt{scipy}~\cite{2020SciPy-NMeth}.
We denote $\hat{\beta}_{\rm d}, \hat{T}_{\rm d}, \hat{\beta}_{\rm s}, \{\hat{\alpha}_f\}$
the best estimates of the generalized spectral likelihood that are obtained at the end of this process.

\paragraph{Step 2: CMB map estimation.}
With the best estimate of spectral parameters and $\alpha_f$ at hand, we can estimate
the generalized mixing matrix $\mathbf{\hat{\Lambda}}_{ho} \equiv \mathbf{\Lambda}_{ho}(\hat{\beta}_{\rm d}, \hat{T}_{\rm d}, \hat{\beta}_{\rm s}, \{\hat{\alpha}_f\})$.
From there, we can build a weighting operator $\mathbf{\hat{W}}_{oh}$ and apply it to the data
similarly as in \cref{eq:Wd_def}:
\begin{equation}
    \mathbf{\hat{W}}_{oh}=\left( \mathbf{\hat{\Lambda}}^{\sf T}_{oh'} \mathbf{\overline{N}}^{-1}_{h'h''} \mathbf{\hat{\Lambda}}_{h''o'}  \right)^{-1}  \mathbf{\hat{\Lambda}}^{\sf T}_{o'h'''} \mathbf{\overline{N}}^{-1}_{h'''h}.
    \label{eq:W_definition}
\end{equation}
This operator can be used to retrieve all three components' maps, but we focus solely on its 
CMB part, $\mathbf{\hat{W}}^{\rm{CMB}}_{kh}$, where the $k$ index represents the $\{Q,U\}$ Stokes parameters ($\mathbf{\hat{W}}^{\rm{CMB}}_{kh}$ shape is then $n_{\rm{Stokes}} \times n_{\rm{Stokes}}\,n_{\rm{frequencies}}$).
The $\{Q,U\}$ CMB maps after generalised component separation are then
\begin{equation}
    \hat{\mathbf{s}}^{\rm{fgs-clean}}_{{\rm CMB,}\ kp} = \mathbf{\hat{W}}^{\rm{CMB}}_{kh} \bar{\mathbf{d}}_{hp}.
\end{equation}
\paragraph{Step 3: Spectra estimation.}
Once we have the cleaned CMB maps, we can estimate their angular power spectra using \texttt{NaMaster}.
To do so, we use an apodised version of the mask used for component separation, based on a ``smooth'' apodization with a $5^\circ$ size.
The spectra are binned uniformly with $\Delta \ell = 2$ in the range $\ell \in [30, 125]$.
In this step, the common beam of $80$ arcmin is also corrected for. We denote the estimated 
CMB spectra $\mathbf{D}_b$, with $b$ representing
the multipole at the centre of a given bin. $\mathbf{D}_b$ is a $2 \times 2$ matrix with diagonal 
elements storing the $EE$ and $BB$ auto-spectra and the off-diagonal $EB$, assuming the $EB$ cross-spectrum
is equal to the $BE$ one.
\paragraph{Step 4: Noise post-component separation.}
To estimate the noise left in the angular power spectrum after all those steps,
one needs to take into account the weighting operator, as well 
as the different beam convolution and deconvolution that have been applied to the white noise. After cancelling the effect of the common beam in the spectra estimation step, only the 
deconvolution is left. The noise power spectra in the CMB maps are then given by
\begin{equation}
    \mathbf{C}_b^{\rm{noise}} = \left[ \mathbf{\hat{W}} \mathbf{N}_b \mathbf{\hat{W}}^{\sf T} \right]_{\rm{CMB}},
\end{equation}
where $\mathbf{N}_b$ is a $2 \,n_{\rm{frequency}} \times 2 \,n_{\rm{frequency}}$ matrix describing the noise power spectra for each frequency band. It is diagonal because we assume no $E-B$ correlation in the noise. Furthermore, we assume the noise has the same characteristics in $E$ and $B$. We can therefore describe the elements of $\mathbf{N}_b$ with respect to its bin and frequency $f$ as
\begin{equation}
    \mathbf{N}_{b , f} = (w_f)^{-2} e^{ 
        b (b + 1) {\rm FWHM}_f^2/8 \log{2}
    }.
\end{equation}
where $(w_f)^{-1}$ is the sensitivity at frequency $f$ in 
$\mathrm{\mu K}$-rad
and 
${\rm FWHM}_f$ the full width at half maximum of the beam of the same channel, in radians.

\paragraph{Step 5: Cosmological likelihood.} \label{paragraph:pipeD_cosmolikelihood}
With the noise covariance post-component separation at hand, we can build
a model covariance for the data: 
\begin{equation}
    \mathbf{C}_b^{\rm theory} (\beta) = \mathbf{R}(\beta)
    \begin{pmatrix}
        C_b^{EE,\, \rm prim} &    0 \\
        0         &     C_b^{BB,\, \rm lens} 
    \end{pmatrix}
    \mathbf{R}^{-1}(\beta) + \mathbf{C}_b^{\rm noise}.
\end{equation}
The first term describes the CMB power spectra. We used the \Planck-2018 best-fit power spectra for the primordial CMB $EE$ power spectrum, $C_b^{EE, \rm prim}$, and for the lensed $BB$ power spectrum, $C_b^{BB, \rm lens}$.
Contrary to ref.~\cite{Jost:2023}, the tensor-to-scalar ratio $r$ is not 
allowed to vary and is fixed to $r=0$. The $BB$ power spectrum is hence limited
to the lensing, and we fix $A_{\rm lens}=1$. We assume no intrinsic $EB$ in the CMB,
making the diagonal terms null. An $EB$ signal can however appear via the effect
of CB through the $\mathbf{R}(\beta)$ operators. 

We can then compare our data $\mathbf{D}_b$ to this model covariance using
the following Gaussian likelihood~\cite{Tegmark_1997}:
\begin{equation}
    \label{eq:pipelined:likelihood_cosmo}
    -2 \log{\mathcal{L}^{\rm cosmo}(\beta) } = \sum_b \Delta\ell 
    f_\mathrm{sky} ( 2 b + 1 )  
    \left[ 
        \mathbf{C}_b^{\rm theory\, -1}(\beta) \mathbf{D}_b + \log(|\mathbf{C}_b^{\rm theory}(\beta)|)
    \right].
\end{equation}
We approximate the effective number of modes accessible after masking 
and binning with the use of the $\Delta \ell f_\mathrm{sky} (2 b + 1)$ factor, therefore neglecting any mode coupling coming from the analysis of a partial sky. We can then proceed similarly as for the generalised spectral likelihood: first
exploring the likelihood distribution with MCMC sampling, and then minimising 
\cref{eq:pipelined:likelihood_cosmo} starting from the mean of the converged MCMC chain. From there we obtain the final best estimate
of $\beta$ for each simulation.

%% file: PipelineB/pipelineB_analysis.tex
As mentioned in \cref{pipelineB method}, the accuracy of the template-based MK estimator depends on our ability to model the $EB$ correlation of Galactic foreground emission. Therefore, we focus our discussion on studying the impact that foreground modeling has on our measurement of CB. Here, we limit our analysis to \texttt{s0}, \texttt{s1}, \texttt{d0}, and \texttt{d1} models and leave the study of further synchrotron and dust models to future work.

\begin{figure}[tbp]
    \centering
    \includegraphics[width=1.0\textwidth]{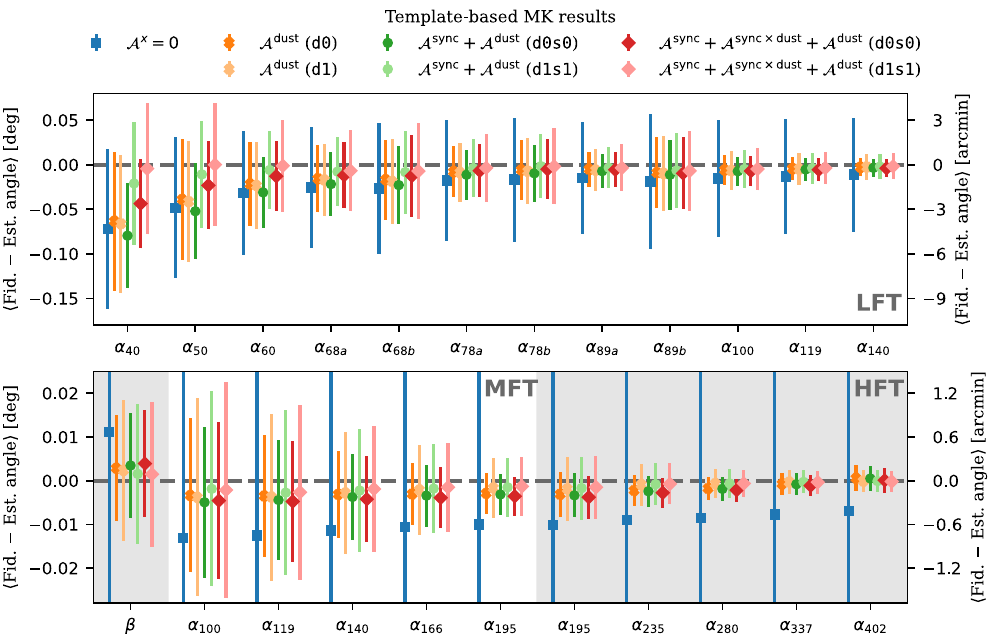}
    \caption{Average difference between the input and estimated angles from Phase 2 simulations when the foreground $EB$ correlation is ignored or modelled with different templates. Uncertainties show the $1\sigma$ dispersion over simulations. Providing a dust template suffices to obtain an unbiased estimate of $\beta$, and MFT's and HFT's $\alpha$ at the $0.12\sigma$ to $0.26\sigma$ level (depending on the frequency band). Extra information from a synchrotron template is needed to correctly recover $\alpha$ at the lower LFT frequencies. The $EB$ correlation in \texttt{d0} is a good enough approximation of that in \texttt{d1} to effectively correct the dust $EB$ bias. In contrast, \texttt{s0} proves an insufficient description of \texttt{s1}, limited in its improvement on the calibration of $\alpha$.}
    \label{fig:pipeline B Phase 2 all angs all templates}
\end{figure}

We start our discussion with Phase 2 simulations (where $\beta,\alpha_f=0$) because this is the case that allows us to isolate the contribution of the foreground $EB$. \Cref{tab: pipeline B mean values P2} shows the mean CB and template amplitudes recovered with the MK estimator when the foreground $EB$ correlation is ignored or modelled through different templates. The corresponding $\alpha_f$ are shown in \cref{fig:pipeline B Phase 2 all angs all templates}. Error bars correspond to the $68~\%$ C.L. uncertainties calculated from the values' dispersion. As described in ref.~\cite{Diego-Palazuelos:2022dsq}, we expect foreground emission to bias our estimates to $\hat{\beta}\approx\beta^{\rm fid}-\gamma_\ell$ and $\hat{\alpha}_f\approx\alpha_f^{\rm fid}+\gamma_\ell$, where $\gamma_\ell\approx C_\ell^{EB,\mathrm{fgs}}/(C_\ell^{EE,\mathrm{fgs}} - C_\ell^{BB,\mathrm{fgs}})$. Note that $\hat{\alpha}_f+\hat{\beta}$ remains unaffected, since we are trying to use foregrounds to break the degeneracy between both rotations. Hence, the small $C_\ell^{EB,\mathrm{fgs}}>0$ found in the \texttt{s0}, \texttt{s1}, \texttt{d0}, and \texttt{d1} models produces the $\hat{\beta}<\beta^{\rm fid}$ and $\hat{\alpha}>\alpha^{\rm fid}$ biases seen in \cref{fig:pipeline B Phase 2 all angs all templates}. Similar biases were also found in refs.~\cite{Diego-Palazuelos:2022dsq, Diego-Palazuelos:2023}.

When neglected ($\mathcal{A}^\mathrm{dust},\mathcal{A}^\mathrm{sync},\mathcal{A}^\mathrm{sync\times dust}=0$ in \cref{eq:pipelineBrotatedEB} and ``$\mathcal{A}_{\rm x}=0$'' entries in \cref{tab: pipeline B mean values P2} and \cref{fig:pipeline B Phase 2 all angs all templates}), the $EB$ correlation in \texttt{d1} produces a bias of $\delta_\beta=\langle\beta^\mathrm{fid}-\hat{\beta}\rangle=0.011^\circ$ ($0.17\sigma$ bias). 
This bias is lower than the $\delta_\beta\approx0.05^\circ$ to $0.15^\circ$ biases reported in almost-full-sky analyses of \Planck\ data\footnote{For almost-full-sky analyses ($f_\mathrm{sky}\gtrsim 90~\%$), we expect the local misalignments between different dust filaments and the Galactic magnetic field to average out along the line of sight and give a small global $EB$ correlation~\cite{Clark:2021kze}. Depending on the dust model, a $\delta_\beta\approx0.05^\circ$ to $0.15^\circ$ bias is expected for this $f_\mathrm{sky}$~\cite{Diego-Palazuelos:2022dsq, Eskilt:2022cff}. Far from the Galactic plane, a larger spatial and angular coherence is seen in the misalignment of dust filaments and magnetic fields~\cite{Cukierman:2022odh}, leading to $\delta_\beta\approx 0.55^\circ$ biases for $f_\mathrm{sky}\approx60~\%$.}~\cite{Diego-Palazuelos:2022dsq}, partially because of the simplified setup of our simulation framework, but also thanks to the configuration of \LB\ itself. The numerous cross-correlations between dust- and synchrotron-dominated bands possible within \LB's 22 frequency channels help suppress the impact of dust $EB$~\cite{Eskilt:2022cff, Eskilt:2022wav}. Neglecting dust $EB$ also biases $\alpha_f$ estimates~\cite{delaHoz:032D, Diego-Palazuelos:2023}. Due to their higher noise levels and lower resolutions~\cite{2022LB_ptep}, $\alpha_f$ estimation on the LFT's lower frequencies relies mostly on foreground-contaminated scales, leading to the larger biases and uncertainties seen in those bands~\cite{delaHoz:032D}.

\begin{table}[tbp]
\setlength{\tabcolsep}{3.5pt}
\centering
\begin{tabular}{|cccccc|}
\hline
{\small Foreground template} & {\small$\beta$ [$\times 10^{-2}$deg]} & {\small$\mathcal{A}^{\rm sync}$} & {\small$\mathcal{A}^{\rm sync\times dust}$} & {\small$\mathcal{A}^{\rm dust}$} & {\small $\Delta$AIC}\\
\hline
{\small $\mathcal{A}^{\rm x}=0$} & {\small $-1.12 \pm 6.42$}  & ... & ... & ... & {\small $618$}\\
{\small $\mathcal{A}^{\rm dust}$ (\texttt{d0})} & {\small $-0.29 \pm 1.22$} & ... & ... & {\small $0.9997 \pm 0.0006$} & {\small $343$}\\
{\small $\mathcal{A}^{\rm dust}$ (\texttt{d1})} & {\small $-0.24 \pm 1.61$} & ... & ... & {\small $1.0002 \pm 0.0005$} &  {\small $252$} \\
{\small $\mathcal{A}^{\rm sync} + \mathcal{A}^{\rm dust}$ (\texttt{d0s0})} & {\small $-0.35 \pm 1.21$} & {\small $0.93 \pm 0.05$} & ... & {\small $0.9998 \pm 0.0006$} & {\small 315} \\
{\small $\mathcal{A}^{\rm sync} + \mathcal{A}^{\rm dust}$ (\texttt{d1s1})} & {\small $-0.16 \pm 1.59$} & {\small $0.97 \pm 0.06$} & ... & {\small $1.0001 \pm 0.0005$} &  {\small 214}  \\
{\small $\mathcal{A}^{\rm sync} + \mathcal{A}^{\rm sync\times dust} + \mathcal{A}^{\rm dust}$ (\texttt{d0s0})} & {\small $-0.39 \pm 1.22$} & {\small $0.94 \pm 0.06$} & {\small$1.01 \pm 0.02$} & {\small$0.9998 \pm 0.0006$} & {\small $211$} \\
{\small $\mathcal{A}^{\rm sync} + \mathcal{A}^{\rm sync\times dust} + \mathcal{A}^{\rm dust}$ (\texttt{d1s1})} & {\small $-0.15 \pm 1.66$} & {\small $0.99\pm 0.08$} & {\small$1.00\pm 0.02$} & {\small$1.0001 \pm 0.0005$} &  {\small $0$}  \\
\hline
\end{tabular}
\caption{Mean CB angle and template amplitudes recovered from Phase 2 simulations. Uncertainties are calculated as the standard deviation over 100 simulations ($68~\%$ C.L.). At every row, we only fit for the amplitudes indicated in the first column. Note the sign change compared to \cref{fig:pipeline B Phase 2 all angs all templates}, which shows $\delta_\beta=\langle 0 - \hat{\beta} \rangle = -\langle\hat{\beta}\rangle$. We use the Akaike information criterion (AIC) to compare the goodness of fit of the different models, evaluating $\Delta\mathrm{AIC}=\mathrm{AIC}_i-\mathrm{AIC}_\mathrm{ref}$ with the $\mathcal{A}^\mathrm{sync} + \mathcal{A}^\mathrm{sync \times dust} + \mathcal{A}^\mathrm{dust}$ (\texttt{d1s1}) case as reference.}
\label{tab: pipeline B mean values P2}
\end{table}

Providing a dust template ($\mathcal{A}^\mathrm{dust}\neq0$ while $\mathcal{A}^\mathrm{sync},\mathcal{A}^\mathrm{sync\times dust}=0$ in \cref{eq:pipelineBrotatedEB} and “$\mathcal{A}^\mathrm{dust}$ (\texttt{d1})”  entries in \cref{tab: pipeline B mean values P2} and \cref{fig:pipeline B Phase 2 all angs all templates}) is enough to reduce the dust-$EB$ bias to $\delta_\beta=0.0024^\circ$ and bring both the $\beta$ and $\alpha_f$ angles of dust-dominated frequency bands in to agreement with zero at the $0.12$ to $0.26\sigma$ level. As discussed in ref.~\cite{Diego-Palazuelos:2023}, the mode-by-mode subtraction of the foreground signal produced by the template also leads to an overall reduction of uncertainties. For the lower frequencies of the LFT, $\alpha_f$ estimation remains dominated by the instrumental configuration, since the dust template does not provide additional constraining power at the frequencies where dust is subdominant. 

We expect synchrotron radiation to have a negligible impact on the MK estimator~\cite{Eskilt:2022cff, Eskilt:2022wav} since no significant synchrotron $EB$ has been found to date~\cite{Martire:2022lsh, QUIJOTE:2023iv}. This is confirmed by the small impact including a synchrotron template ($\mathcal{A}^\mathrm{sync},\mathcal{A}^\mathrm{dust}\neq0$ while $\mathcal{A}^\mathrm{sync\times dust}=0$ in \cref{eq:pipelineBrotatedEB} and “$\mathcal{A}^\mathrm{sync} + \mathcal{A}^\mathrm{dust}$ (\texttt{d1s1})” entries in \cref{tab: pipeline B mean values P2} and \cref{fig:pipeline B Phase 2 all angs all templates}) has on the CB measurement once we include a dust template. Nevertheless, the additional information provided by the synchrotron template helps to characterise the large angular scales of the observed signal, allowing a better estimation of $\alpha_\mathrm{40}$, $\alpha_\mathrm{50}$, $\alpha_\mathrm{60}$, $\alpha_\mathrm{68a}$, and $\alpha_\mathrm{68b}$ despite the LFT's limited angular resolution. Likewise, providing additional information on the correlation between dust and synchrotron emission ($\mathcal{A}^\mathrm{sync},\mathcal{A}^\mathrm{dust},\mathcal{A}^\mathrm{sync\times dust}\neq0$ in \cref{eq:pipelineBrotatedEB} and “$\mathcal{A}^\mathrm{sync} + \mathcal{A}^\mathrm{sync\times dust} + \mathcal{A}^\mathrm{dust}$ (\texttt{d1s1})” entries in \cref{tab: pipeline B mean values P2} and \cref{fig:pipeline B Phase 2 all angs all templates}) further improves the estimate of $\alpha_f$ in the LFT's low-frequency bands. With the observed signal and $\alpha_f$ better constrained, more information about $\beta$ can be gathered from low-frequency bands. Thus, providing a synchrotron template improves the overall performance of the estimator and reduces the foreground-induced bias to $\delta_\beta=0.0015^\circ$.

An accurate estimation of $\beta\approx0.3^\circ$ is still possible under a small mismodelling of foreground emission, like assuming \texttt{d0s0} templates to describe an \texttt{d1s1} sky. Models \texttt{d0} and \texttt{d1} are based on the same dust template derived from \Planck\ 2015 data, but scaled with different SEDs: a fixed $\beta_\mathrm{d}=1.54$ and $T_\mathrm{d}=20$K for \texttt{d0}; and the spatially varying $T_\mathrm{d}$ and $\beta_\mathrm{d}$ obtained by \texttt{Commander} from the analysis of \Planck\ 2015 data~\cite{planckX2015} for \texttt{d1}. Likewise, \texttt{s0} and \texttt{s1} share the synchrotron template derived from Haslam and \WMAP\ data~\cite{Remazeilles2015, Bennett2013:pvna}, but are scaled with different SEDs: a fixed $\beta_\mathrm{s}=-3.0$ for \texttt{s0}; and the spatially varying $\beta_\mathrm{s}$ obtained by~\cite{Miville-Deschenes2008} from the analysis of Haslam and \WMAP\ data for \texttt{s1}. See refs.~\cite{Thorne_2017, PYSM3} for further detail. At the angular scales used in our analysis, only $5~\%$ to $15~\%$ differences are found between the $EB$ spectra of \texttt{d0} and \texttt{d1}, and \texttt{s0} and \texttt{s1}.

Qualitatively, \texttt{d0} offers a good enough description of \texttt{d1} to satisfactorily correct its $EB$ correlation as the minor increase in the $\delta_\beta$ bias and the good agreement of $\mathcal{A}^\mathrm{dust}$ with unity show in \cref{tab: pipeline B mean values P2} (“$\mathcal{A}^\mathrm{dust}$ (\texttt{d0})” entries). On the contrary, in our joint analysis of dust and synchrotron templates, \texttt{s0} proves insufficient to describe \texttt{s1} (“$\mathcal{A}^\mathrm{sync} + \mathcal{A}^\mathrm{dust}$ (\texttt{d0s0})” and “$\mathcal{A}^\mathrm{sync} + \mathcal{A}^\mathrm{sync\times dust} + \mathcal{A}^\mathrm{dust}$ (\texttt{d0s0})” entries in \cref{tab: pipeline B mean values P2} and \cref{fig:pipeline B Phase 2 all angs all templates}): modeling \texttt{s1} with \texttt{s0} significantly increases the $\delta_\beta$ bias, gives $\mathcal{A}^\mathrm{sync}$ values incompatible with unity within $1\sigma$, and no longer improves the estimation of $\alpha_f$ at synchrotron-dominated frequencies. In addition to increasing the bias on $\beta$ and $\alpha_f$, assuming simplified synchrotron and dust models in the calculation of the covariance matrix also leads to roughly a $25~\%$ underestimation of uncertainties. 

Since it is based on a maximum-likelihood algorithm, our MK estimator allows the assessment of the goodness of fit and the comparison of different foreground models through statistics like the Akaike information criterion (AIC). We complement our discussion by showing the $\Delta$AIC values obtained with different foreground templates in \cref{tab: pipeline B mean values P2}. The recovered $\Delta$AIC reflect the inadequacy of neglecting foreground $EB$ and confirm $\mathcal{A}^\mathrm{sync}+\mathcal{A}^\mathrm{sync\times dust}+\mathcal{A}^\mathrm{dust}$ as the best description of the foreground emission present in the simulations.

\begin{table}[tbp]
\setlength{\tabcolsep}{4pt}
\centering
\begin{tabular}{|ccccc|}
\hline
{\small Foreground template} & {\small$\beta$ [$\times 10^{-2}$ deg]} &{\small $\mathcal{A}^\mathrm{sync}$} & {\small$\mathcal{A}^\mathrm{sync\times dust}$} & {\small$\mathcal{A}^\mathrm{dust}$}\\
\hline
{\small $\mathcal{A}^x=0$} & {\small $18.36 \pm 6.62$}  & ... & ... & ... \\
{\small $\mathcal{A}^\mathrm{dust}$ (\texttt{d0})} & {\small $29.20 \pm 1.81$} & ... & ... & {\small $1.002 \pm 0.003$} \\
{\small $\mathcal{A}^\mathrm{dust}$ (\texttt{d1})} & {\small $29.25 \pm 1.98$} & ... & ... & {\small $0.998 \pm 0.002$} \\
{\small $\mathcal{A}^\mathrm{sync} + \mathcal{A}^\mathrm{dust}$ (\texttt{d0s0})} & {\small $29.10 \pm 1.88$} & {\small $0.92 \pm 0.06$} & ... & {\small $0.999 \pm 0.004$} \\
{\small $\mathcal{A}^\mathrm{sync} + \mathcal{A}^\mathrm{dust}$ (\texttt{d1s1})} & {\small $29.28 \pm 2.11$} & {\small $0.96\pm 0.09$} & ... & {\small $0.999 \pm 0.001$}  \\
{\small $\mathcal{A}^\mathrm{sync} + \mathcal{A}^\mathrm{sync \times dust} + \mathcal{A}^\mathrm{dust}$ (\texttt{d0s0})} & {\small $29.23 \pm 1.46$} & {\small $0.92 \pm 0.07$} & {\small$0.99 \pm 0.03$} & {\small$0.999 \pm 0.004$}  \\
{\small $\mathcal{A}^\mathrm{sync} + \mathcal{A}^\mathrm{sync \times dust} + \mathcal{A}^\mathrm{dust}$ (\texttt{d1s1})} & {\small $29.27 \pm 2.24$} & {\small $0.96\pm0.13$} & {\small$0.99\pm 0.04$} & {\small$0.999 \pm 0.002$}  \\
\hline
\end{tabular}
\caption{Same as \cref{tab: pipeline B mean values P2} but for Phase 4 simulations. Although sufficient to provide an unbiased CB estimate, the current MK estimator is not suitable for model comparison through statistics like the Akaike information criterion in the more complex scenario of Phase 4.}
\label{tab: pipeline B mean values P4}
\end{table}

\begin{figure}[tbp]
    \centering
    \includegraphics[width=1.0\textwidth]{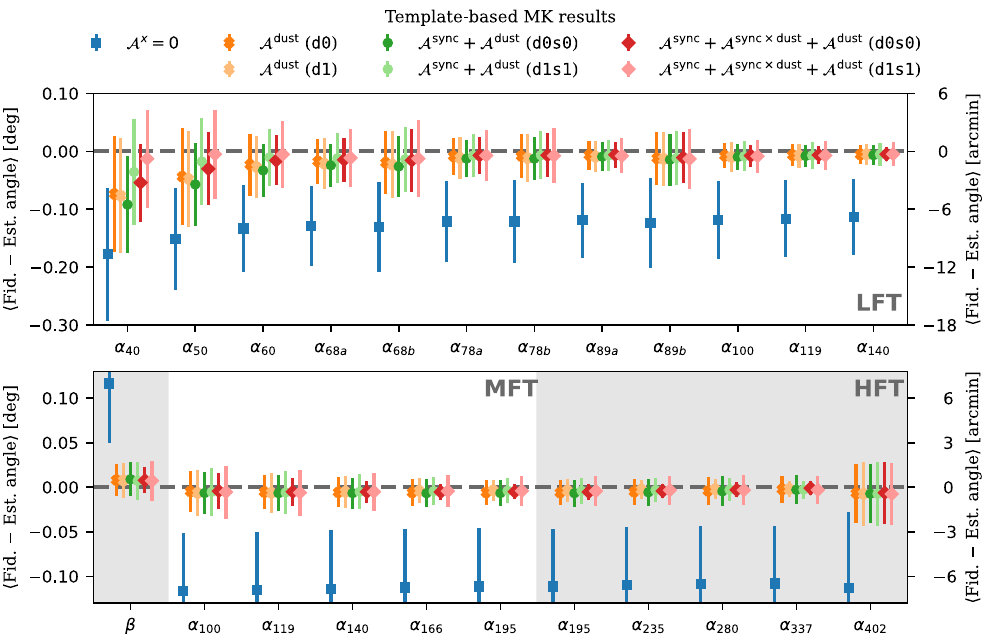}
    \caption{Same as \cref{fig:pipeline B Phase 2 all angs all templates}, but for Phase 4 simulations. For Phase 4, providing a dust template allows an unbiased estimate of $\beta$, and MFT's and HFT's $\alpha$ at the $0.19\sigma$ to $0.29\sigma$ level. Similar conclusions on the need for a synchrotron template to accurately calibrate LFT's $\alpha$ and the (in)adequacy of (\texttt{s0}) \texttt{d0} models are obtained for Phase 4.}
    \label{fig:pipeline B Phase 4 all angs all templates}
\end{figure}

As \cref{fig:pipeline B Phase 4 all angs all templates} and \cref{tab: pipeline B mean values P4} show, these conclusions qualitatively hold when moving to the more general scenario of Phase 4 simulations, where $\beta$ and $\alpha_f$ take non-zero values. In this case, the observed signal is more complex than in Phase 2 simulations, since the $E$ and $B$ modes of Galactic foregrounds are rotated into each other. The complexity of the covariance matrix describing such a signal also increases, gaining numerous terms that depend on the cross-correlations between observations and foreground templates (see \cref{pipelineB appendix}). In this more realistic scenario, providing an accurate foreground template is crucial. Without one, the estimated $\beta$ would suffer a bias of $\delta_\beta=0.1164^\circ$ ($1.8\sigma$ bias). 

With a foreground template, the MK estimator accurately measures $\beta\approx0.3^\circ$ with a small bias of $\delta_\beta \approx 0.008^\circ$ ($40~\%$ of the statistical uncertainty). Additionally, we recover a stable $\delta_\beta$ value for all the sky models considered, since now the inaccuracy of the approximations made in the likelihood dominates over the effects of foreground mismodelling. Those approximations consist of assuming that the statistics of angular power spectra follow a Gaussian distribution and using an approximate covariance matrix derived from the observed signal on a realisation-by-realisation basis. We further simplify the covariance matrix by neglecting $\ell$-to-$\ell'$ correlations; assuming that the spherical harmonic coefficients of CMB, noise, and foreground signals are Gaussian and isotropic; and discarding some of the more noisy and unstable $EB$ terms of the observed signal. While these approximations proved sufficient for the simpler framework of Phase 2, they fall short for Phase 4 simulations. For instance, the $\mathbf{d}^{\sf{T}}\mathbf{C}^{-1}\mathbf{d}$ product of $C_\ell$ data vector and covariance matrix that followed the expected $\chi^2$ distribution for Phase 2 simulations, no longer does so for Phase 4. In practice, this means that, although sufficient to provide an unbiased estimate of $\beta$, the implementation of the MK estimator presented here is not suitable for model comparison through statistical estimators like AIC in the high signal-to-noise regime offered by \LB. Therefore, we must explore extensions of the estimator beyond the Gaussian likelihood (see, e.g., ref.~\cite{Verde2003:oamf}) and a more robust calculation of its covariance matrix if we want to reliably asses the goodness of fit of different foreground models once \LB\ data arrive. We leave that study to future work.

Finally, these results demonstrate that \LB\ is capable of self-calibrating instrumental $\alpha_f$ and measuring isotropic CB at the same time, as long as a reasonable foreground model is provided. With the help of synchrotron and dust templates, we can recover unbiased estimates of $\alpha_f$ with uncertainties of around $2$~arcmin for all frequency bands, fulfilling the requirements needed to achieve an unbiased measurement of the tensor-to-scalar ratio with \LB~\cite{Vielva2022:jcna}. This study extends previous results presented in refs.~\cite{Minami:2019ruj, Minami:2020fin, delaHoz:032D, Krachmalnicoff2022:jakv}.

%% file: PipelineC/pipelineC_analysis.tex
In this section, we discuss the results obtained with \texttt{CAB-SeCRET} and $\beta$\texttt{CAB-SeCRET}, focusing on the subtleties intrinsic to this pipeline. First, we present the results obtained for each phase using \texttt{CAB-SeCRET} and discuss the impact of mismodelled spectral parameters and biased priors on $\alpha_f$. Then, we compare the Phase 4 results recovered with \texttt{CAB-SeCRET} and $\beta$\texttt{CAB-SeCRET}.\\

\begin{figure}
    \centering
    \includegraphics[width=1.0\textwidth]{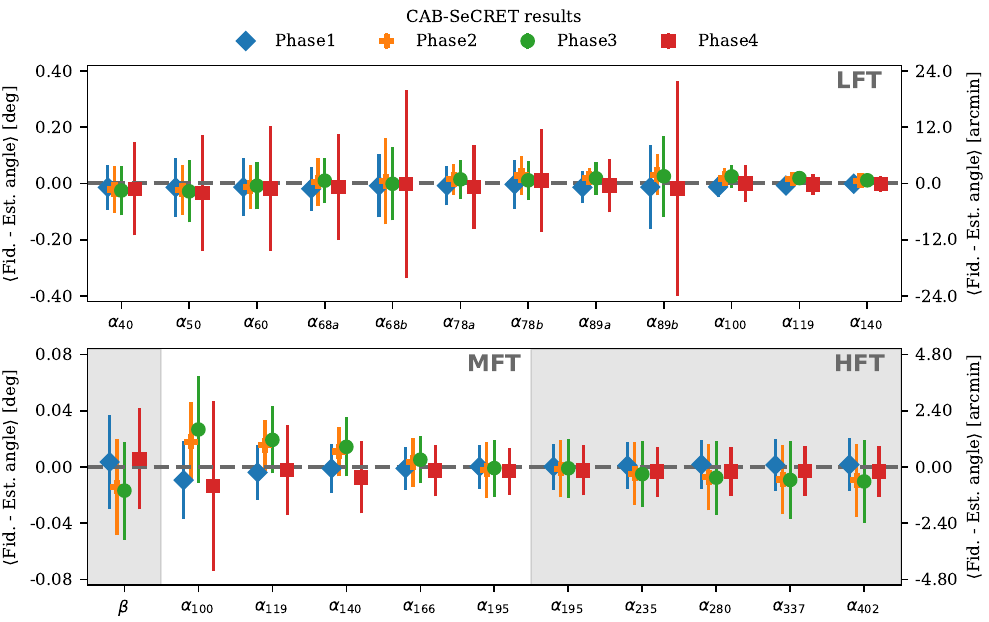}
    \caption{Average difference between the input and estimated angles obtained with \texttt{CAB-SeCRET}. Note that $\alpha_f$ are calculated at the component-separation step, while $\beta$ is calculated from the cleaned CMB map. Blue diamonds, orange pluses, and green circles correspond, respectively, to Phase 1, Phase 2, and Phase 3 results obtained using the $\Pi(\myvector{\alpha_f}, \beta=0)$ priors. Red squares show the Phase 4 results when we apply the $\Pi(\myvector{\alpha_f}, \beta)$ priors. Uncertainties correspond to the standard deviation of the simulations ($68~\%$ C.L.). Note the difference in the $y$-axis scale between the top and bottom panels.}
    \label{fig:pipeline_c_comparison_angles_c1_std_MCMC}
\end{figure}

\paragraph{CAB-SeCRET results.}
This pipeline takes into account the effect of nonzero $\alpha_f$ in the component-separation analysis, as inaccuracies in $\alpha_f$ calibration can introduce substantial biases in CB detection. \Cref{fig:pipeline_c_comparison_angles_c1_std_MCMC} shows the average difference between the input and recovered $\beta$ and $\alpha_f$ obtained with \texttt{CAB-SeCRET} for each phase. We apply the $\Pi(\myvector{\alpha_f},\beta=0)$ priors for Phases 1 to 3 and switch to $\Pi(\myvector{\alpha_f},\beta)$ for Phase 4. Had we employed $\Pi(\myvector{\alpha_f},\beta)$ for Phases 1 to 3 as well, we would have observed a slight increase in $\sigma_{\beta}$. However, this increase is deemed insignificant upon comparing the uncertainty observed in Phase 4 to that of the preceding Phases, as illustrated in \cref{tab:pipeline_c_bias_unc}. 

Subsequently, we examine the outcomes for each phase, in order of ascending complexity.

\begin{itemize}
    
    \item \textbf{Phase 1:} For Phase 1, we assume that the \spec\ parameters are uniform across the entire observable sky after applying the $f_{\rm sky} = 60\%$ \textit{Planck} Galactic mask. This modelling approach reproduces that of the simulated foregrounds, ensuring that we do not obtain biased results due to incorrect modeling of the sky. Our findings indicate that the $\alpha_f$ overall align with the values injected in the simulation, i.e., $\myvector{\alpha}_f = 0$, as shown in \cref{fig:pipeline_c_comparison_angles_c1_std_MCMC}. Furthermore, we demonstrate that the $\beta$ left in the clean CMB map is consistent with zero, with an uncertainty of $\sigma_{\beta} \simeq 0.034^\circ$ and a bias of $\delta_{\beta} \approx -0.1\sigma_{\beta}$.

    \item \textbf{Phase 2:} For this simulation set, the assumption of uniform \spec\, parameters no longer holds. Relaxing this condition implies the addition of more free parameters, increasing the uncertainty of all model parameters, including $\beta$. Thus, we need to find a trade-off between a tolerable $\delta_\beta$ and the $\sigma_\beta$ necessary to achieve a detection if $\beta$ is of the order of $0.3^\circ$. We consider three different cases. 
    \begin{itemize}
        \item[i.] As for Phase 1, we assume that \spec\, parameters are uniform, meaning the resolution of \spec\ is $N_{\rm side}^{\mathcal{B}}=0$.
        \item[ii.] We allow \spec\ to vary from pixel to pixel ($N_{\rm side}^{\mathcal{B}}=64$). This case aligns the most with the fiducial foreground model, hence, we expect negligible biases.
        \item[iii.] An intermediate case, where \spec\ are assumed to be uniform in superpixels of $N_{\rm side}^{\mathcal{B}} =8$. Here, we incorporate spatial variations of \spec\ but with fewer parameters, approximately a factor of $3 \times 64$ fewer than in the previous one.
    \end{itemize}
    As anticipated, the value of $\sigma^{\rm MC}_{\beta}$ increases with the number of \spec\ parameters ($N_{\rm side}^{\mathcal{B}}$). Explaining the value of $\sigma_{\beta}$ is more challenging, since it accounts for both the dispersion due to a larger number of \spec\ parameters and the bias resulting from imperfect modelling of the foreground emission. Regarding the bias, if $N_{\rm side}^{\mathcal{B}}=0$ like in the previous Phase, we recover a biased value of $\delta_{\beta}=-2.379\sigma_{\beta}$ due to imperfect modelling of contaminants. When including all possible parameters at the current resolution, i.e., $N_{\rm side}^{\mathcal{B}}=64$, allowing the model to capture the complexity of the spatial variations of spectral indices, the uncertainty increases to $\sigma_{\beta}= 0.084^\circ$, corresponding to a $\beta$ detection of $3.6\sigma$, similar to current constraints. With $N_{\rm side}^{\mathcal{B}} =8$, we find a compromise between bias, $\delta_{\beta}=0.420\sigma_{\beta}$, and competitive uncertainty, $\sigma_{\beta}= 0.0341^\circ$. We use $N_{\rm side}^{\mathcal{B}}=8$ in the following phases.

    \begin{table}
    \centering
    \begin{tabular}{|c|c|c|ccccc|}
    \hline
         Phase & Method & Prior & $N_{\rm side}^{\mathcal{B}}$ & $\delta_{\beta} / \sigma_{\beta}$  & $\sigma_{\beta}$ [deg] & $\delta^{\rm MC}_{\beta}/ \sigma^{\rm MC}_{\beta}$ & $\sigma^{\rm MC}_{\beta}$  [deg]\\
    \hline
         1 & \texttt{CAB-SeCRET} & $\Pi(\myvector{\alpha_f},\beta=0)$ & 0 &  -0.104 &  0.0336 & -0.124 & 0.0282\\ 
         2 & \texttt{CAB-SeCRET} & $\Pi(\myvector{\alpha_f},\beta=0)$ & 0 &   -2.378 & 0.0357 & -2.891 & 0.0294 \\
         2 & \texttt{CAB-SeCRET} & $\Pi(\myvector{\alpha_f},\beta=0)$ & 8 &  0.420 & 0.0341 & 0.407 & 0.0352 \\
         2 & \texttt{CAB-SeCRET} & $\Pi(\myvector{\alpha_f},\beta=0)$ & 64 & -0.218  & 0.0840 & -0.279 &  0.0656\\
         3 & \texttt{CAB-SeCRET} & $\Pi(\myvector{\alpha_f},\beta=0)$ & 8 & 0.483 & 0.0349 & 0.479 & 0.0352 \\
         4 & \texttt{CAB-SeCRET} & $\Pi(\myvector{\alpha_f},\beta)$& 8 & 8.414 & 0.0373 & 10.056 & 0.0312\\
         4 & $\beta$\texttt{CAB-SeCRET} & $\Pi(\myvector{\alpha_f},\beta)$ & 8 & -0.160 & 0.0359 & -0.167 & 0.0345\\   
    \hline
    \end{tabular}
    \caption{Biases and uncertainties for each phase obtained using either \texttt{CAB-SeCRET} or $\beta$\texttt{CAB-SeCRET}. The biases are expressed in terms of the uncertainty on $\beta$. $\sigma_{\beta}$ and $\sigma^{\rm MC}_{\beta}$ represent, respectively, the standard deviation calculated from simulations and from the MCMC chains obtained by minimizing \cref{eq:pipeline_c1_loglikelihood}.}
    \label{tab:pipeline_c_bias_unc}
\end{table}

    \item \textbf{Phase 3:}  \Cref{fig:pipeline_c_comparison_angles_c1_std_MCMC} demonstrates that the estimates for both $\alpha_f$ and $\beta$ match the input values when non-zero $\alpha_f$ are introduced in the simulations. Furthermore, we observe that $\delta_{\beta}$ and $\sigma_{\beta}$ from Phase 3 closely resemble the values from Phase 2 ($N_{\rm side}^{\mathcal{B}}=8$), confirming a robust marginalisation over $\alpha_f$.

\begin{figure}
    \begin{subfigure}{.49\textwidth}
        \centering
    \includegraphics[width=\textwidth]{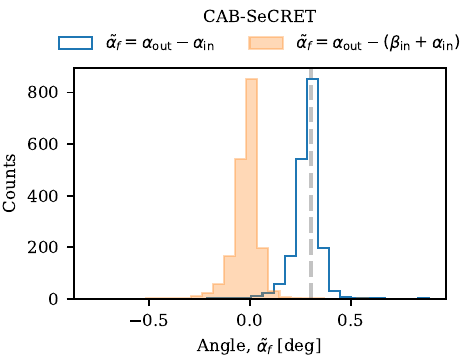}  
    \end{subfigure}
    \begin{subfigure}{.49\textwidth}
        \centering
    \includegraphics[width=\textwidth]{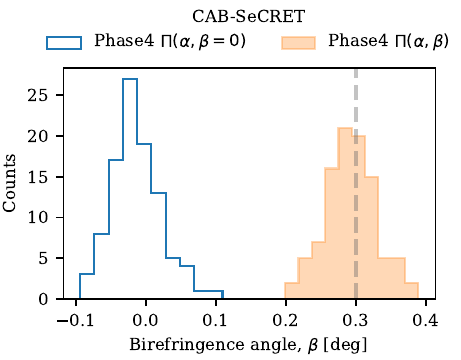}
    \end{subfigure}
\caption{Left: Distribution difference between the estimated $\alpha_{\rm out}$ for all \textit{LiteBIRD} channels and the input $\alpha_f$ (blue line histogram), and the input $\alpha_f$ plus the injected $\beta$ angle (orange filled histogram) when we fit using the $\Pi(\alpha_f, \beta=0)$ priors. Right: Distribution of the estimated $\beta$  when we apply the priors assuming $\beta=0$ (blue line histogram) and with no constraints on $\beta$ (orange filled histogram). The dashed grey line indicates the input $\beta$ value.}   
\label{fig:comparison_angles_prior}
\end{figure}

    \item \textbf{Phase 4:} In this set of simulations, $\beta$ is no longer zero, rendering the method to estimate $\alpha_f$ from ref.~\cite{delaHoz:032D} inaccurate as it estimates the combined angle $\tilde{\alpha}_f = \alpha_f + \beta$. Since $\beta$ represents a common rotation across all frequency channels, $\tilde{\alpha}_f$ also provides a reasonable fit to the data. Consequently, the recovered $\alpha_f$ are biased. This is evident from the left panel of \cref{fig:comparison_angles_prior}, where we compare the distribution of the difference between the recovered $\tilde{\alpha}_f$ and, firstly, the injected $\alpha_f^{\rm in}$ in the simulations, and secondly, $\alpha_f^{\rm in} + \beta$. The former distribution peaks around $0.3^\circ$, corresponding to the input $\beta$ value, while the latter peaks around zero.

    To address this issue, one can use prior information regarding $\alpha_f$ calibration. In particular, we use the $\mathcal{A}^\mathrm{sync} + \mathcal{A}^\mathrm{sync\times dust} + \mathcal{A}^\mathrm{dust}$ (\texttt{d1s1}) case from \cref{pipelineB analysis} in the following. It is important to highlight that we do not explicitly incorporate any $\beta$ constraints from the MK estimator. We only make use of the information on $\alpha_f$ calibration that could otherwise be provided by any other absolute calibration method. Additionally, in this approach, we simultaneously marginalise over the foreground model parameters and $\alpha_f$. Thus, we alleviate any potential biases associated with the mismodelling of the foreground templates in the MK estimator.

    With these priors, we obtain $\alpha_f$ and $\beta$ values that are consistent with the input values, as depicted in \cref{fig:pipeline_c_comparison_angles_c1_std_MCMC}, and right panel of \cref{fig:comparison_angles_prior}. Although the uncertainty in the recovered $\alpha_f$ increases due to the less stringent priors, we find that the uncertainty in the recovered $\beta$ is similar to that of the other phases. This is because $\beta$ is derived from the cleaned CMB, whose noise is minimally affected by changes in the $\alpha_f$ prior. However, we observe a slight reduction in the $\delta_\beta$ bias, despite the astrophysical model being incorrect, as in the other Phases 2 and 3.
    
    \end{itemize}

\paragraph{Comparison of CAB-SeCRET and $\beta$CAB-SeCRET.}
\label{subsubsec:comp_pipeline_c1_c2}
\begin{figure}
    \begin{subfigure}{.49\textwidth}
        \centering
    \includegraphics[width=\textwidth]{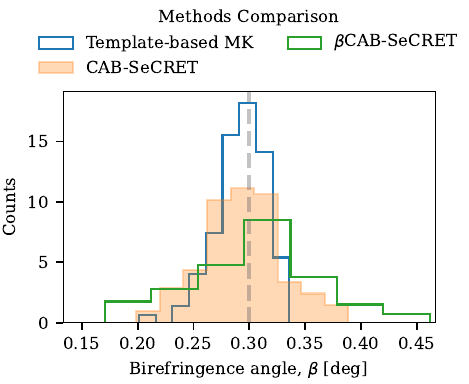}   
    \end{subfigure}
    \begin{subfigure}{.49\textwidth}
        \centering
    \includegraphics[width=\textwidth]{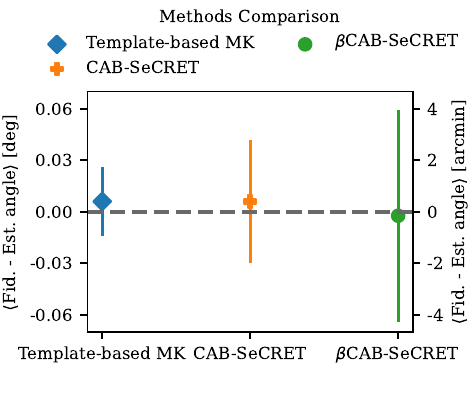}   
    \end{subfigure}
    \caption{Left: Distribution of the estimated $\beta$ from Phase 4 simulations using three methods: \texttt{CAB-SeCRET}; $\beta$\texttt{CAB-SeCRET}; and the template-based MK results ($\mathcal{A}^\mathrm{sync} + \mathcal{A}^\mathrm{sync \times dust} + \mathcal{A}^\mathrm{dust}$ with \texttt{d1s1}) used as prior information in $\beta$\texttt{CAB-SeCRET}. Right: average difference between input and estimated $\beta$. 
    }
    \label{fig:pipeline_c_C1_C2_comp}
\end{figure}

In this section, we compare the results from Phase 4 obtained by applying the \texttt{CAB-SeCRET} and $\beta$\texttt{CAB-SeCRET}. Notice that, in both cases, we use the results from the template-based MK $\mathcal{A}^\mathrm{sync} + \mathcal{A}^\mathrm{sync \times dust} + \mathcal{A}^\mathrm{dust}$ (\texttt{d1s1}) estimator as prior information ($\Pi(\myvector{\alpha_f}, \beta)$). In \texttt{CAB-SeCRET}, we use only the information from $\alpha_f$, while in $\beta$\texttt{CAB-SeCRET} we also incorporate the information from $\beta$. The results are summarised in \cref{tab:pipeline_c1_c2_comp} and illustrated in \cref{fig:pipeline_c_C1_C2_comp}, where we also include the prior estimates for comparison. When comparing the results from \texttt{CAB-SeCRET} with respect to $\beta$\texttt{CAB-SeCRET}, we observe an increase in the uncertainty on $\beta$ and a reduction in the $\delta_{\beta}$ bias.

 \begin{table}[htbp]
    \centering
    \begin{tabular}{|l|cc|}
    \hline
          Method & $\delta_{\beta}/  \sigma_{\beta}$  & $\sigma_{\beta}$ [deg] \\
    \hline
         Template-based MK  & $-0.262$ & $0.0214$ \\
         \texttt{CAB-SeCRET} & $\phantom{-}0.160$ & $0.0359$  \\
         $\beta$\texttt{CAB-SeCRET} &   $-0.037$ & $0.0617$ \\
    \hline
    \end{tabular}
    \caption{Biases and uncertainties from Phase 4 obtained using the template-based MK ($\mathcal{A}^\mathrm{sync} + \mathcal{A}^\mathrm{sync \times dust} + \mathcal{A}^\mathrm{dust}$ with \texttt{d1s1}), \texttt{CAB-SeCRET}, and $\beta$\texttt{CAB-SeCRET} methods. The biases are expressed in terms of the uncertainty on $\beta$. The quantity $\sigma_{\beta}$ 
    represents the standard deviation  
    obtained from simulations.
    }
    \label{tab:pipeline_c1_c2_comp}
\end{table}

The primary distinction between \texttt{CAB-SeCRET} and $\beta$\texttt{CAB-SeCRET} lies in their treatment of $\beta$: in the former, $\beta$ is conditioned on both $\alpha_f$ and the astrophysical component parameters, while in the latter, $\beta$ is marginalised over them. Consequently, the results from $\beta$\texttt{CAB-SeCRET} are more robust and unbiased, as demonstrated in \cref{tab:pipeline_c1_c2_comp}. Furthermore, it is noteworthy that although the uncertainty in $\beta$ obtained with $\beta$\texttt{CAB-SeCRET} is larger, it remains competitive, especially if $\beta \approx 0.3^\circ$, as it would result in a detection of $4.9\sigma$.

Similar conclusions arise from comparing the results of \texttt{CAB-SeCRET} and $\beta$\texttt{CAB-SeCRET} with those from the template-based MK estimator used as prior information. Both \texttt{CAB-SeCRET} and $\beta$\texttt{CAB-SeCRET} simultaneously marginalise over astrophysical model parameters and $\alpha_f$, unlike the template-based MK estimator. This marginalisation results in smaller biases but larger uncertainties. Consequently, the two pipelines can complement each other. For instance, one can utilise the results from the template-based MK estimator to constrain the \ang\ parameters and recover $\beta$ with $\beta$\texttt{CAB-SeCRET} or \texttt{CAB-SeCRET}, and then employ the foreground templates derived from the latter in the template-based MK estimator to mitigate biases arising from incorrect foreground templates.

%% file: PipelineD/pipelineD_analysis.tex
In this section, we discuss the results obtained by the \texttt{J23} pipeline, going phase by phase and discussing the possible shortcomings and extensions to the pipeline in view of its response to the different levels of complexity of each phase. All error bars discussed here correspond to the standard deviation over the set of simulations: $\sigma_x$ ($68\%$ C.L.) for parameter $x$.
\begin{itemize}
    \item \textbf{Phase 1:}

    The foreground cleaning in this Phase should be close to optimal, as the simulations correspond to the data model used in the component separation (\cref{eq:pipeD_DataModel}). 
    Furthermore, the absence of $\alpha_f$ miscalibration should lead to a reduced $\sigma_\beta$ compared to phases with miscalibration.
    
    \Cref{fig:pipeD_MCMC_spec_phase1_truncated} shows the results of the MCMC sampling of the generalised spectral likelihood (\cref{eq:generalised_spectral_likelihood}) for \textit{one} given simulation, corresponding to step 1 in \cref{pipelineD method}. The orange contours correspond to the $68\%$ and $95\%$ contours from the MCMC sampling, while the purple dashed ones represent the Gaussian priors from calibration. Note that, as explained in \cref{pipelineD method}, the priors' central values are chosen at random for each simulation, hence they are not centred at the true values of parameters denoted by the dotted grey lines. We only display $\alpha_f$ of three frequency channels, $\alpha_\mathrm{LFT,68a}$, $\alpha_\mathrm{MFT,100}$, and $\alpha_\mathrm{HFT,235}$, as showing the full triangle plot with $22$ angles would lack clarity. 
    We can see that the foreground parameters are estimated correctly and seem uncorrelated to $\alpha_f$, as expected from ref.~\cite{Jost:2023}. 
    Thanks to the combination of prior information between frequencies and the internal calibration against the foregrounds SED, the statistical error bars associated with $\alpha_f$  estimates are smaller than the initial precision of the priors (which are $\sigma_{\alpha_\mathrm{LFT,68a}} = 0.7^\circ$, $\sigma_{\alpha_\mathrm{MFT,100}} = 0.13^\circ$, and $\sigma_{\alpha_\mathrm{HFT,235}} = 0.21^\circ$) and best-fit values are closer to the true input angle. Although off-centred priors can bias $\alpha_f$, the mean bias on the absolute offset will tend to average out to zero thanks to the randomness of the priors' centres. This can already be partially seen in the figure, with the first angle being biased positively while the two others are negatively biased. 
    Furthermore, the case presented in \cref{fig:pipeD_MCMC_spec_phase1_truncated} only represents one simulation and thus does not encompass the full posterior and, in particular, the variance associated with the random prior centres. 
    
    Therefore, we must look at the results over the 100 simulations. \Cref{fig:pipeD_comp_angle_phase1} shows  
    the results for the estimation of $\beta$ and $\alpha_f$. Central values correspond to the average over simulations of the estimated parameters, and error bars to $\sigma_x$.  
    We can see that all $\alpha_f$ are retrieved without significant bias, with the error bars on $\alpha_f$ ranging from $0.028^\circ$ to $0.085^\circ$. This leads to an estimation of $\beta = -0.002^\circ \pm 0.042^\circ$, hence demonstrating that, in this set of simple simulations, the \texttt{J23} pipeline can correctly retrieve CB and $\alpha_f$. 
    The histograms of best-fit values shown in \cref{fig:pipeD_spec_params_hist_phase1} demonstrate that foreground parameters are also retrieved correctly, with average values of $\beta_\mathrm{d} = 1.5401 \pm 0.0028$, $T_\mathrm{d} = 19.99 \pm 0.10$, and $\beta_\mathrm{s} = -3.0001 \pm 0.0039$.

\begin{figure}
    \centering
    \includegraphics[width=1\linewidth]{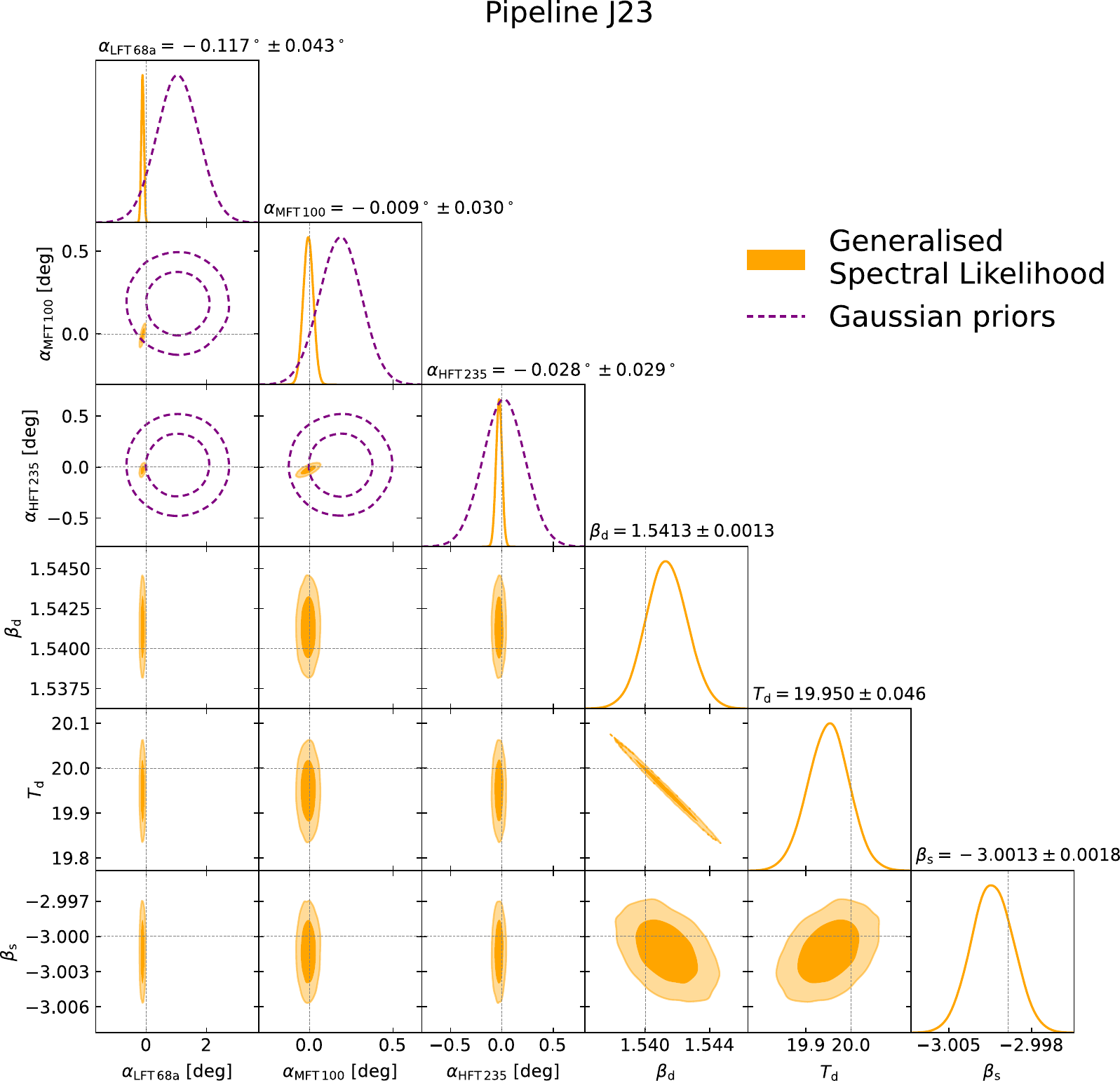}
    \caption{MCMC sampling of the generalised spectral likelihood (\cref{eq:generalised_spectral_likelihood}) for one Phase~1 simulation for pipeline \texttt{J23}. Out of the total 22 $\alpha_f$, only $\alpha_\mathrm{LFT,68a}$, $\alpha_\mathrm{MFT,100}$, and $\alpha_\mathrm{HFT,235}$ are displayed for clarity. Orange contours represent the $68\%$ and $95\%$ contours. Purple dashed contours represent the Gaussian priors applied to the angles, as explained in \cref{paragraph:pipeD_compsep} (Step 1) 
    the central value of the priors are drawn randomly for each simulation. True input values are displayed as the grey dotted lines.}
    \label{fig:pipeD_MCMC_spec_phase1_truncated}
\end{figure}

\begin{figure}
    \centering
    \includegraphics[width=1\linewidth]{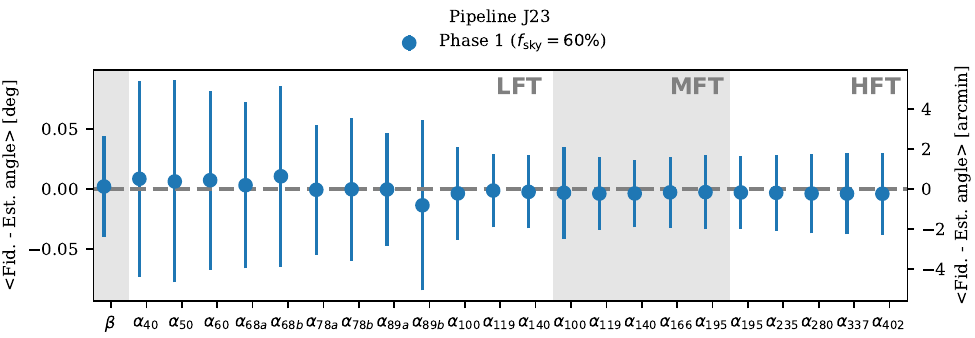}
    \caption{Average difference between the input and estimated angles over Phase 1 simulations obtained with pipeline \texttt{J23}. Here, $\alpha_f$ are estimated in the component-separation step (\cref{paragraph:pipeD_compsep}, Step 1),
    while $\beta$ is obtained using the cosmological likelihood with the angular power spectra of the foreground-cleaned CMB map (\cref{paragraph:pipeD_cosmolikelihood}, Step 5).  
    Error bars represent the standard deviation of the values retrieved over the 100 simulations ($68\%$ C.L.).}
    \label{fig:pipeD_comp_angle_phase1}
\end{figure}

\begin{figure}
    \centering
    \includegraphics[width=1\linewidth]{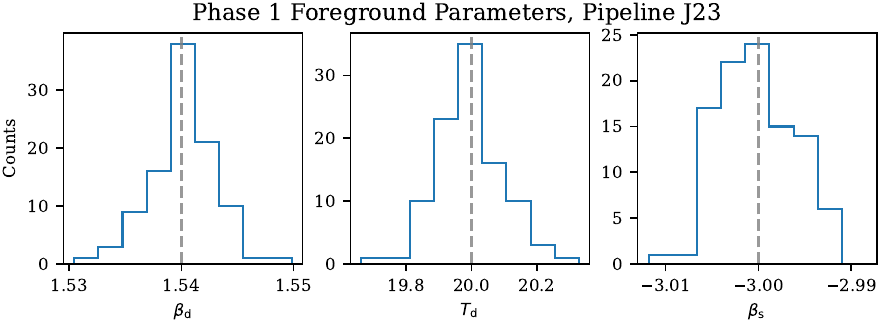}
    \caption{From left to right, distribution of Phase 1 results for $\beta_\mathrm{d}$, $T_\mathrm{d}$, and $\beta_\mathrm{s}$ for pipeline \texttt{J23}. Grey dashed lines represent the true input parameters used in the simulations.}
    \label{fig:pipeD_spec_params_hist_phase1}
\end{figure}

    \item \textbf{Phase 2:}  
    To handle the more complex case of spatially varying spectral parameters, \texttt{CAB-SeCRET} and $\beta$\texttt{CAB-SeCRET} allow for parameters to vary across superpixels during component separation.
    Contrary to these other pipelines, the current implementation of \texttt{J23} does not have yet a way to handle cases with spatially varying spectral parameters and is therefore more susceptible to the biases caused by more complex foregrounds. Ref.~\cite{Jost:2023} explored complex foregrounds (including \texttt{d1s1} and \texttt{d7s3}\footnote{\texttt{d7} uses a non-parametric dust frequency scaling based on dust population modelling \cite{Draine_2013}. \texttt{s3} uses a power law with a curved index for the frequency scaling, where the index map is the same as for \texttt{s1}.}), but the smaller survey footprint of a Simons Observatory SAT-like survey ($f_\mathrm{sky}\sim 10\%$) limits the impact of spatially varying foregrounds. This work is therefore the first to test this method on the bigger sky fraction typical of space missions. 

    \Cref{fig:pipeD_comp_angle_phase2} summarises the Phase 2 results, showing that $\alpha_f$ at low frequencies are biased. In particular, the estimation of $\alpha_{40}$, the lowest frequency band, is biased by almost $5 \sigma$. The bias on $\alpha_f$ reduces as the frequency increases, potentially indicating that this method is sensitive to spatially varying synchrotron emission. However, the estimation of $\alpha_f$ without a significant bias at middle and high frequencies partially compensates for the low-frequency bias, leading to an estimation of $\beta = -0.033^\circ \pm 0.047^\circ$. Hence, even without a way to handle complex foregrounds, the bias is below $1\sigma$ at $\delta_\beta = 0.708\, \sigma_\beta$. 
    
    To confirm that the spatial variability of spectral parameters is indeed the reason for this bias, we perform a second analysis on Phase 2 simulations. In this case, we use a larger mask corresponding to $f_\mathrm{sky} = 40\%$, since the smaller sky fraction should limit the impact of the foregrounds' spatial variation. The results are displayed in orange in \cref{fig:pipeD_comp_angle_phase2} and in \cref{tab:pipeD_summary_of_results}. The bias on low-frequency $\alpha_f$ is indeed greatly reduced, with the bias on $\alpha_{40}$ now being $\lesssim 1 \sigma$, and resulting in a CB estimate of $\beta=-0.010^\circ \pm 0.055^\circ$. The smaller sky fraction leads to a slight increase in $\sigma_\beta$ of about $15 \%$ and a drop in the bias to $\delta_\beta = 0.179\, \sigma_\beta$. 
    
    Reducing the sky fraction to deal with complex foregrounds gives us insight into our analysis, but it should be treated as a diagnosis tool rather than an analysis tool. While having statistically consistent results on $\beta$ between $f_\mathrm{sky}=60\%$ and $f_\mathrm{sky}=40\%$ is promising, the large variation observed in the estimated $\alpha_f$ is a clear indication that complex foregrounds need to be dealt with. Possible extensions to the framework include the multi-patch approach~\cite{Errard:2018, 2022LB_ptep}, which allows for spatial variability of foreground parameters, or foreground marginalisation~\cite{Errard:2018,wolz2023simons}, using the spectra of the foreground maps estimated by the component separation in the cosmological likelihood. These extensions are beyond the scope of this paper and are left for future work. 
    
    For completeness, we also performed the analysis at $f_\mathrm{sky}=40\%$ on Phase 1 simulations, including the results in \cref{tab:pipeD_summary_of_results}. The smaller sky fraction leads again to an increase of around $20\%$ in $\sigma_\beta$, but there is no significant difference in the bias thanks to the absence of spatial variability in these simulations.

\begin{figure}
    \centering
    \includegraphics[width=1\linewidth]{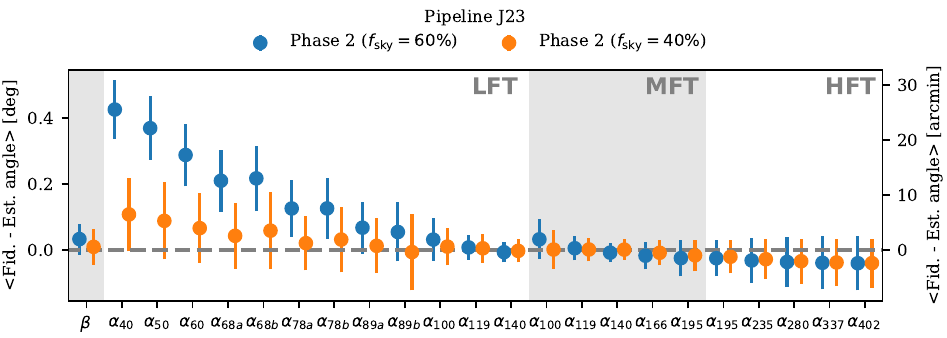}
    \caption{Similar to \cref{fig:pipeD_comp_angle_phase1}, but comparing Phase 2 with the standard $f_\mathrm{sky}=60\%$ mask (in blue) to Phase 2 with a larger mask $f_\mathrm{sky}=40\%$ (in orange) for pipeline \texttt{J23}.}
    \label{fig:pipeD_comp_angle_phase2}
\end{figure}

    \item \textbf{Phase 3:} For this phase 
    we proceed similarly, with a version of the analysis performed with the fiducial mask at $f_\mathrm{sky}=60\%$ and another with $f_\mathrm{sky}=40\%$. Results are displayed in \cref{fig:pipeD_comp_angle_phase3} in blue and orange, respectively, and are summarized in \cref{tab:pipeD_summary_of_results}. Note that, since \cref{fig:pipeD_comp_angle_phase3} shows the mean and standard deviation of the difference between true and estimated angles. 
    Overall, this results in an increase of $3.3\%$ of the error bar on $\beta$ compared to the Phase 2 result for the $f_\mathrm{sky}=60\%$ case, and $1.7\%$ for the $f_\mathrm{sky}=40\%$ cases. The retrieved mean value of $\beta$ is also affected, increasing by $22\%$ ($f_\mathrm{sky}=60\%$) and $77\%$ ($f_\mathrm{sky}=40\%$) going from Phase 2 to 3. Results are still within the $1\sigma$ limit, with relative bias going to $\delta_\beta = 0.849\, \sigma_\beta$  ($f_\mathrm{sky}=60\%$) and $\delta_\beta = 0.320\, \sigma_\beta$  ($f_\mathrm{sky}=40\%$). Hence, in the presence of additional systematic effects, pipeline \texttt{J23} seems to give robust results. However, the discussion in Phase 2 about requiring the extension of the pipeline to handle complex foregrounds holds here. 
    
\begin{figure}
    \centering
    \includegraphics[width=1\linewidth]{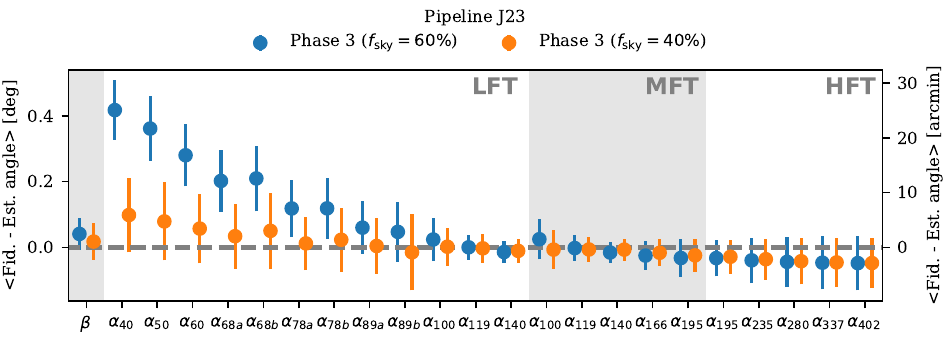}
    \caption{Similar to \cref{fig:pipeD_comp_angle_phase2}, comparing results of Phase 3 with $f_\mathrm{sky}=60\%$ (blue) and $f_\mathrm{sky}=40\%$ (orange) for pipeline \texttt{J23}.}
    \label{fig:pipeD_comp_angle_phase3}
\end{figure}
    \item \textbf{Phase 4:} Finally, for Phase 4,
    we display the results in \cref{fig:pipeD_comp_angle_phase4} and in \cref{tab:pipeD_summary_of_results}. Results are similar to Phase 3, with no significant impact on the estimation of $\alpha_f$. CB is retrieved with $\beta = 0.258^\circ \pm 0.049^\circ$ ($f_\mathrm{sky}=60\%$) and $\beta = 0.282^\circ \pm 0.056^\circ$ ($f_\mathrm{sky}=40\%$), corresponding to $\delta_\beta = 0.852\, \sigma_\beta$ and $\delta_\beta = 0.322\, \sigma_\beta$, respectively, and demonstrating the capacity of pipeline \texttt{J23} to detect $\beta=0.3^\circ$ at the $5\sigma$ level.
\begin{figure}
    \centering
    \includegraphics[width=1\linewidth]{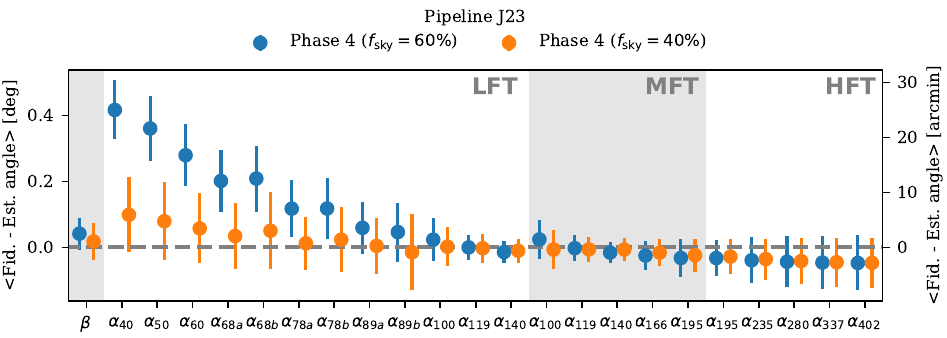}
    \caption{Similar to \cref{fig:pipeD_comp_angle_phase2,fig:pipeD_comp_angle_phase3}, comparing results of Phase 4 with $f_\mathrm{sky}=60\%$ (blue) and $f_\mathrm{sky}=40\%$ (orange) for pipeline \texttt{J23}.}
    \label{fig:pipeD_comp_angle_phase4}
\end{figure}
 \begin{table}[htb]
    \centering
\begin{tabular}{|c|c|cc|}
    \hline
Phase & $f_\mathrm{sky}$ & $\beta$ [deg] & $\delta_\beta / \sigma_\beta$ \\
    \hline
    1 &        0.6 &  $-0.002 \pm 0.042$ &                    $0.049$ \\
    1 &        0.4 &  $-0.003 \pm 0.051$ &                    $0.051$ \\
    \hline
    2 &        0.6 &   $-0.033 \pm 0.047$ &                     $0.708$ \\
    2 &        0.4 & $-0.010 \pm 0.055$ &                     $0.179$ \\
    \hline
    3 &        0.6 &   $-0.041 \pm 0.048$ &                     $0.849$ \\
    3 &        0.4 &  $-0.018 \pm 0.056$ &                      $0.320$ \\
    \hline
    4 &        0.6 &    $\phantom{-}0.258 \pm 0.049$ &                     $0.852$ \\
    4 &        0.4 &    $\phantom{-}0.282 \pm 0.056$ &                     $0.322$ \\
    \hline
\end{tabular}
    \caption{Summary of the results on $\beta$ obtained by pipeline \texttt{J23} across the different phases and sky fractions.
    }
    \label{tab:pipeD_summary_of_results}
\end{table}
\end{itemize}
These results show that \LB, with the use of pipeline \texttt{J23}, is capable of measuring $\beta$ with a precision of about $0.05^\circ$. This is possible thanks to our knowledge of the foreground SEDs and provided the requirements on absolute $\alpha_f$ calibration are met.
However, while those results are promising and seem robust against complex foregrounds and systematic effects, they also show the limits of the simple spatially constant foreground SED used. Indeed, results from Phase 2 onward show clear biases in $\alpha_f$ at low frequencies caused by the foregrounds' complexity. The estimation on $\beta$ is not significantly biased thanks to the compensation by precise $\alpha_f$ calibration at higher frequencies. 
Assessing the stability of systematic, foreground, and cosmological parameters across different sky fractions can serve as a robustness test against spatial variations in future uses of this pipeline.
However, this demonstrates the need for an extension to the method which includes spatially variable foreground parameters to handle more complex foregrounds in the future. 
In particular, since \texttt{J23} is based on \fgbuster, a multi-patch like extension \cite{Errard:2018} constitute a natural next step.
Indeed, it has been used in ref.~\cite{2022LB_ptep} for \LB~forecasts as well as for the estimation of the component separated maps used for the $D$-estimator and the pixel-based estimator in this paper. 
Other extensions capable of handling more complex foregrounds and systematic effects can be added to parametric methods in general although at the price of higher statistical uncertainty.

%% file: alternative_conclusions.tex
\begin{figure}
    \centering
    \includegraphics[width=\textwidth]{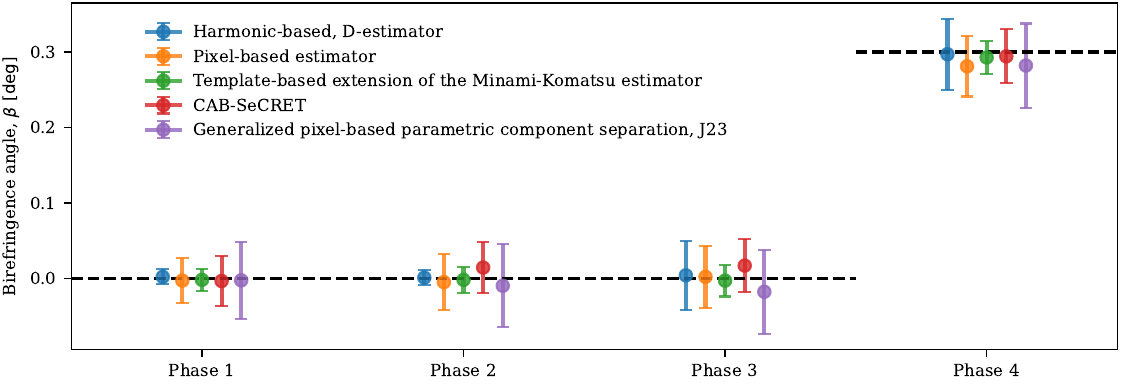}
    \caption{Summary of this paper's results, showing the constraints derived from the five pipelines against the 4 phases of simulations.}
    \label{fig:one}
\end{figure}

We generated four sets of 100 simulations, each containing CMB, foregrounds, and noise, with increasing complexity, as summarised in \cref{tab:phase_description}. These simulations include cases with constant and spatially varying spectral indices of the foregrounds, with and without instrumental $\alpha_f$ rotation, and with and without a non-zero isotropic CB (see \cref{simulations} for details). Using five different pipelines, we recovered the mean $\beta$ summarised in \cref{tab:one} and shown in \cref{fig:one}.

Our results show that the constraints on $\beta$ are consistent across the four sets of simulations and all estimators, without significant bias from the instrumental or astrophysical systematic effects included in the simulations. More specifically, in most cases, the expected value for $\beta$ is recovered within 3 standard deviations of the mean. However, for certain specific pipelines and phases, we observe deviations ranging from 3 to almost 5 standard deviations of the mean, which correspond, in the worst case, to approximately $50\%$ of the statistical uncertainty.
Note that in Phases 3 and 4, the inclusion of non-zero $\alpha_f$ meets the requirements presented in ref.~\citep{Vielva2022:jcna} and introduces both relative miscalibrations between frequency channels and a small global offset across all frequencies. Given the small value of the global offset, pipelines without control over $\alpha_f$ remain unbiased but reflect an increased uncertainty in $\beta$ due to the relative miscalibrations compared to Phases 1 and 2,  which do not include these systematic effects. 

The statistical efficiency with which we recover CB depends on the complexity of the simulations considered and the pipeline used. In general, we find that, for each pipeline, the error bar increases with the complexity of the simulations, with the uncertainty of $\beta$ ranging between $0.01^\circ$ to $0.06^\circ$ depending on the case considered. However, this effect is more pronounced for pipelines that ignore $\alpha_f$. Specifically, when we consider Phase 4, i.e., the case with $\beta=0.3^\circ$ motivated by recent observational hints~\cite{Diego-Palazuelos:2022dsq}, the five pipelines reach a detection with a significance from $5$ to $13$$\sigma$. This is a remarkable result, since these estimates have been obtained considering realistic conditions containing the relevant systematic effects, see also discussion in \cref{discussionone}. To improve our grasp of these astrophysical and instrumental effects, it will be crucial for future research to investigate more advanced sky models (e.g., those by refs.~\cite{HerviasCaimapo2024, Vansyngel_dust_model, MKD_dust_model, Vacher2023, Vacher2024}) and consider other instrumental influences, including those from a non-ideal half-wave plate~\cite{Monelli2023}.


Still focusing on Phase 4, we note that, among all the pipelines, the one providing the tightest constraints is the method based on the Minami-Komatsu approach (see \cref{pipelineB method,pipelineB analysis}). Compared to the $D$-estimator, which is also defined in harmonic space, the statistical power of the MK estimator comes from exploiting the cross-correlations between frequency bands instead of constraining $\beta$ from the cleaned CMB $EB$ spectrum. It also achieves smaller error bars than the \texttt{($\beta$)CAB-SeCRET} and \texttt{J23} pipelines as it does not marginalise over the uncertainties in foreground modeling. Although the MK, \texttt{($\beta$)CAB-SeCRET}, and \texttt{J23} pipelines explicitly depend on our choice of foreground model (the $D$-estimator and stacking analysis indirectly do so through the previous component-separation step), they have so far adapted satisfactorily to the measurement of CB in realistic skies. We expect their performance to improve in the future as we continue to improve our knowledge of polarised foreground emission and implement clustering or foreground marginalisation techniques to adapt to more complex skies.

\begin{table}[h]
    \centering
    \begin{tabular}{|c|cccc|}
    \hline
Pipeline & Phase 1 & Phase 2 & Phase 3 & Phase 4  \\ 
&$\beta\ [\times 10^{-2}\,\deg]$ & $\beta\ [\times 10^{-2}\,\deg]$ & $\beta\ [\times 10^{-2}\,\deg]$&$\beta\ [\times 10^{-2}\,\deg]$\\ \hline
$D$-estimator & $\phantom{-}0.2 \pm 1.0$ & $ \phantom{-}0.1 \pm 1.0 $ & $\phantom{-}0.4 \pm 4.6$ & $\phantom{-}29.7 \pm 4.7 $  \\
 Pixel-based estimator & $-0.3 \pm 3.0$& $-0.5 \pm 3.7$ & $\phantom{-}0.2 \pm 4.1$ & $\phantom{-}28.1 \pm 4.0$  \\
 Template-based MK & $-0.2 \pm 1.5$ & $-0.2 \pm 1.7$ & $-0.3 \pm 2.1$ & $\phantom{-}29.3 \pm 2.2$ \\
 \texttt{CAB-SeCRET} & $-0.4 \pm 3.4$ & $\phantom{-}1.4 \pm 3.4$ & $\phantom{-}1.7 \pm 3.5$ & $\phantom{-}29.4 \pm 3.6$  \\ 
  \texttt{J23} & $-0.2 \pm 5.1$ & $-1.0 \pm 5.5$ & $-1.8 \pm 5.6$ & $\phantom{-}28.2 \pm 5.6$ \\ \hline
    \end{tabular}
    \caption{Summary of the average performances of the five pipelines. Error bars show the standard deviation over simulations.}
    \label{tab:one}
\end{table}

This MK technique, along with \texttt{($\beta$)CAB-SeCRET} and \texttt{J23}, demonstrates the capability to calibrate $\alpha_f$ and simultaneously detect CB under realistic conditions as those considered in the present analysis. The inclusion of information from instrumental calibration, whether from astrophysical sources~\cite{Aumont2020} and laboratory or space-based instruments as proposed in refs.~\cite{Casas2021, Ritacco:2024kug}, could improve the statistical efficiency of these methods and unambiguously break the degeneracy between isotropic CB and absolute miscalibration angles. It would also make the techniques that do not control for this systematic effect more robust.

Having five semi-independent pipelines aimed at estimating $\beta$ that produce consistent results is highly beneficial for ensuring the robustness of the analysis. This is because these pipelines respond differently to systematic effects, reducing the likelihood that a specific artefact affects all of them in the same way. Consequently, even though the pipelines have varying levels of statistical efficiency (see \cref{tab:one}), running all of them is essential for ensuring consistency and cross-validating the results.
This approach significantly enhances the reliability of the analysis and the robustness of the results.

%% file: PipelineB/pipelineB_appendix.tex
`In this work, we have extended the formalism presented in ref.~\cite{Diego-Palazuelos:2023} to include both synchrotron and dust templates. From \cref{eq:pipelineBrotatedEB}, we can build a Gaussian likelihood (introducing our \cref{eq:pipelineBrotatedEB} into equation A2 from ref.~\cite{Diego-Palazuelos:2023}) and simultaneously calculate the maximum-likelihood solution for all $\mathbf{x}_i=(\mathcal{A}^\mathrm{sync}$, $\mathcal{A}^\mathrm{sync \times dust}$, $\mathcal{A}^\mathrm{dust}$, $\beta$, $\alpha_i)$ parameters in a semi-analytical iterative approach. The algorithm assumes that the $\mathbf{x}_i$ parameters in the covariance matrix are known and fixed, starting at $(1, 1, 1, 0, 0)$ values. We then obtain a system of linear equations to analytically calculate the maximum-likelihood solution by differentiating the log-likelihood with respect to each parameter. This first estimate is used to update the covariance matrix and recalculate a new best-fit solution, starting an iterative process that converges after only a few iterations.

For brevity, we introduce the $\sin (x)=\Sin{x}$ and $\cos (x)=\Cos{x}$ notation and the `o', `s', `d', and `s$\times$d' shorthand for `obs', `sync', `dust', and `sync$\times$dust' superscripts in the remainder of this section. Except for \cref{eq:pipeline B CMB contribution to covariance}, all foreground and CMB power spectra are implicitly convolved by the instrumental beam and pixel window functions associated with each frequency band. In the following equations, the `CMB' angular power spectra correspond to the $\Lambda$CDM prediction for the CMB $EE$ and $BB$ spectra in the abse\label{}nce of CB.

The minimisation of the log-likelihood built from \cref{eq:pipelineBrotatedEB} leads to a linear system of the form $\mathbf{A}_{mn}\mathbf{x}_m=\mathbf{b}_m$. The $\mathbf{A}_{mn}$ system matrix is symmetrical, with diagonal elements
\small
\begin{align} 
\begin{split}
    \mathbf{A}_\mathrm{s,s} = & \phantom{-2} \sum\limits_{\substack{b\\i, j\neq i\\p,q\neq p}}   C_b^{E_i^\mathrm{s}B_j^\mathrm{s}} \mathbf{C}_{ijpqb}^{-1} C_b^{E_p^\mathrm{s}B_q^\mathrm{s}}, \\ 
    \mathbf{A}_\mathrm{s\times d,s\times d} =& \phantom{-4}  \sum\limits_{\substack{b\\i, j\neq i\\p,q\neq p}}   \left[ C_b^{E_i^\mathrm{s}B_j^\mathrm{d}} \mathbf{C}_{ijpqb}^{-1} C_b^{E_p^\mathrm{s}B_q^\mathrm{d}} + C_b^{E_i^\mathrm{d}B_j^\mathrm{s}} \mathbf{C}_{ijpqb}^{-1} C_b^{E_p^\mathrm{d}B_q^\mathrm{s}} +2 C_b^{E_i^\mathrm{s}B_j^\mathrm{d}} \mathbf{C}_{ijpqb}^{-1} C_b^{E_p^\mathrm{d}B_q^\mathrm{s}} \right],  \\
    \mathbf{A}_\mathrm{d,d} = & \phantom{-2} \sum\limits_{\substack{b\\i, j\neq i\\p,q\neq p}}    C_b^{E_i^\mathrm{d}B_j^\mathrm{d}} \mathbf{C}_{ijpqb}^{-1} C_b^{E_p^\mathrm{d}B_q^\mathrm{d}},  \\ 
    \mathbf{A}_{\beta,\beta} =& \phantom{-} 4 \sum\limits_{\substack{b\\i, j\neq i\\p,q\neq p}}   \left( C_b^{E_i^\mathrm{CMB}E_j^\mathrm{CMB}} - C_b^{B_i^\mathrm{CMB}B_j^\mathrm{CMB}} \right) \mathbf{C}_{ijpqb}^{-1} \left( C_b^{E_p^\mathrm{CMB}E_q^\mathrm{CMB}} - C_b^{B_p^\mathrm{CMB}B_q^\mathrm{CMB}} \right), 
\end{split}
\end{align}
\normalsize
and off-diagonal terms
\small
\begin{align} 
\begin{split}
    \mathbf{A}_\mathrm{s,s\times d} = & \phantom{-2} \sum\limits_{\substack{b\\i, j\neq i\\p,q\neq p}}  C_b^{E_i^\mathrm{s}B_j^\mathrm{s}} \mathbf{C}_{ijpqb}^{-1} \left( C_b^{E_p^\mathrm{s}B_q^\mathrm{d}} + C_b^{E_p^\mathrm{d}B_q^\mathrm{s}} \right), \\
    \mathbf{A}_\mathrm{s,d} = & \phantom{-2} \sum\limits_{\substack{b\\i, j\neq i\\p,q\neq p}}  C_b^{E_i^\mathrm{s}B_j^\mathrm{s}} \mathbf{C}_{ijpqb}^{-1} C_b^{E_p^\mathrm{d}B_q^\mathrm{d}}, \\
    \mathbf{A}_{\mathrm{s,}\beta} = & \phantom{-} 2 \sum\limits_{\substack{b\\i, j\neq i\\p,q\neq p}} C_b^{E_i^\mathrm{s}B_j^\mathrm{s}} \mathbf{C}_{ijpqb}^{-1} \left( C_b^{E_p^\mathrm{CMB}E_q^\mathrm{CMB}} - C_b^{B_p^\mathrm{CMB}B_q^\mathrm{CMB}} \right),  \\
    \mathbf{A}_{\mathrm{s,}\alpha_n} = & \phantom{-} 2 \sum\limits_{\substack{b\\i\neq n\\p,q\neq p}}  \left[ C_b^{E_i^\mathrm{o}E_n^\mathrm{o}} \mathbf{C}_{inpqb}^{-1} C_b^{E_p^\mathrm{s}B_q^\mathrm{s}} - C_b^{B_n^\mathrm{o}B_i^\mathrm{o}} \mathbf{C}_{nipqb}^{-1} C_b^{E_p^\mathrm{s}B_q^\mathrm{s}} \right],  \\
    \mathbf{A}_\mathrm{s\times d,d} = & \phantom{-2} \sum\limits_{\substack{b\\i, j\neq i\\p,q\neq p}}  C_b^{E_i^\mathrm{d}B_j^\mathrm{d}} \mathbf{C}_{ijpqb}^{-1} \left( C_b^{E_p^\mathrm{s}B_q^\mathrm{d}} + C_b^{E_p^\mathrm{d}B_q^\mathrm{s}} \right), \\ 
    \mathbf{A}_{\mathrm{s\times d,}\beta} = & \phantom{-} 2 \sum\limits_{\substack{b\\i, j\neq i\\p,q\neq p}}   \left( C_b^{E_i^\mathrm{s}B_j^\mathrm{d}} + C_b^{E_i^\mathrm{d}B_j^\mathrm{s}} \right) \mathbf{C}_{ijpqb}^{-1} \left( C_b^{E_p^\mathrm{CMB}E_q^\mathrm{CMB}} - C_b^{B_p^\mathrm{CMB}B_q^\mathrm{CMB}} \right),\\
    \mathbf{A}_{\mathrm{sd,}\alpha_n} = & \phantom{-} 2 \sum\limits_{\substack{b\\i\neq n\\p,q\neq p}}   \left[ C_b^{E_i^\mathrm{o}E_n^\mathrm{o}} \mathbf{C}_{inpqb}^{-1} \left(C_b^{E_p^\mathrm{s}B_q^\mathrm{d}} + C_b^{E_p^\mathrm{d}E_q^\mathrm{s}} \right) - C_b^{B_n^\mathrm{o}B_i^\mathrm{o}} \mathbf{C}_{nipqb}^{-1} \left( C_b^{E_p^\mathrm{s}B_q^\mathrm{d}}+ C_b^{E_p^\mathrm{s}B_q^\mathrm{d}}\right)\right], 
\end{split}
\end{align}

\begin{align} 
\begin{split}
    \mathbf{A}_{\mathrm{d,}\beta} = & \phantom{-} 2 \sum\limits_{\substack{b\\i, j\neq i\\p,q\neq p}}   C_b^{E_i^\mathrm{d}B_j^\mathrm{d}} \mathbf{C}_{ijpqb}^{-1} \left( C_b^{E_p^\mathrm{CMB}E_q^\mathrm{CMB}} - C_b^{B_p^\mathrm{CMB}B_q^\mathrm{CMB}} \right),\\
    \mathbf{A}_{\mathrm{d,}\alpha_n} = & \phantom{-} 2 \sum\limits_{\substack{b\\i\neq n\\p,q\neq p}}   \left[ C_b^{E_i^\mathrm{o}E_n^\mathrm{o}} \mathbf{C}_{inpqb}^{-1} C_b^{E_p^\mathrm{d}B_q^\mathrm{d}} - C_b^{B_n^\mathrm{o}B_i^\mathrm{o}} \mathbf{C}_{nipqb}^{-1} C_b^{E_p^\mathrm{d}B_q^\mathrm{d}} \right], \\
    \mathbf{A}_{\beta,\alpha_n}=& \phantom{-} 4\sum\limits_{\substack{b\\i\neq n\\p,q\neq p}}  \left[ C_b^{E_i^\mathrm{o}E_n^\mathrm{o}} \mathbf{C}_{inpqb}^{-1} C_b^{E_p^\mathrm{CMB}E_q^\mathrm{CMB}} - C_b^{B_n^\mathrm{o}B_i^\mathrm{o}} \mathbf{C}_{nipqb}^{-1} C_b^{B_p^\mathrm{CMB}B_q^\mathrm{CMB}} \right], \\
    \mathbf{A}_{\alpha_m,\alpha_n} =& \phantom{-} 4 \sum\limits_{\substack{b\\i\neq n\\ j\neq m}}  \left[  C_b^{E_i^\mathrm{o}E_n^\mathrm{o}} \mathbf{C}_{injmb}^{-1} C_b^{E_j^\mathrm{o}E_m^\mathrm{o}} + C_b^{B_n^\mathrm{o}B_i^\mathrm{o}} \mathbf{C}_{nimjb}^{-1} C_b^{B_m^\mathrm{o}B_j^\mathrm{o}} \right] \\
    & -4 \sum\limits_{\substack{b\\i\neq n\\ j\neq m}} \left[ C_b^{B_n^\mathrm{o}B_i^\mathrm{o}} \mathbf{C}_{nijmb}^{-1} C_b^{E_j^\mathrm{o}E_m^\mathrm{o}} + C_b^{B_m^\mathrm{o}B_j^\mathrm{o}} \mathbf{C}_{mjinb}^{-1} C_b^{E_i^\mathrm{o}E_n^\mathrm{o}} \right]. 
    \end{split}
\end{align}
\normalsize
The elements of the independent term $\mathbf{b}_m$ are
%
\begin{align}
\begin{split}
    \mathbf{b}_\mathrm{s} = & \phantom{2} \sum\limits_{\substack{b\\i, j\neq i\\p,q\neq p}}   C_b^{E_i^\mathrm{o}B_j^\mathrm{o}} \mathbf{C}_{ijpqb}^{-1} C_b^{E_p^\mathrm{s}B_q^\mathrm{s}}, \\
    \mathbf{b}_\mathrm{s\times d} = & \phantom{2} \sum\limits_{\substack{b\\i, j\neq i\\p,q\neq p}}  C_b^{E_i^\mathrm{o}B_j^\mathrm{o}} \mathbf{C}_{ijpqb}^{-1} \left( C_b^{E_p^\mathrm{d}B_q^\mathrm{s}} + C_b^{E_p^\mathrm{s}B_q^\mathrm{d}} \right),  \\
    \mathbf{b}_\mathrm{d} = & \phantom{2} \sum\limits_{\substack{b\\i, j\neq i\\p,q\neq p}}   C_b^{E_i^\mathrm{o}B_j^\mathrm{o}} \mathbf{C}_{ijpqb}^{-1} C_b^{E_p^\mathrm{d}B_q^\mathrm{d}},  \\
    \mathbf{b}_\beta =& 2 \sum\limits_{\substack{b\\i, j\neq i\\p,q\neq p}}   C_b^{E_i^\mathrm{o}B_j^\mathrm{o}} \mathbf{C}_{ijpqb}^{-1} \left( C_b^{E_p^\mathrm{CMB}E_q^\mathrm{CMB}} -  C_b^{B_p^\mathrm{CMB}B_q^\mathrm{CMB}} \right), \\
    \mathbf{b}_{\alpha_m} =& 2 \sum\limits_{\substack{b\\i\neq m\\p,q\neq p}}  \left[ C_b^{E_i^\mathrm{o}E_m^\mathrm{o}} \mathbf{C}_{impqb}^{-1} C_b^{E_p^\mathrm{o}B_q^\mathrm{o}} - C_b^{B_m^\mathrm{o}B_i^\mathrm{o}} \mathbf{C}_{mipqb}^{-1} C_b^{E_p^\mathrm{o}B_q^\mathrm{o}} \right].
    \end{split}
\end{align}
%
We exclude frequency auto-spectra ($i=j$ and $p=q$) from all summations on $\mathbf{A}_{mn}$ and $\mathbf{b}_m$ to avoid noise bias.

In our likelihood, we use an approximate covariance matrix derived from the observed signal on a realisation-by-realisation basis. For each combination of $i$, $j$, $p$, and $q$ frequency bands, the covariance can be divided into terms that depend only on the observed, foreground, and CMB spectra, and on their cross-correlations:
%
\begin{equation}
    \mathbf{C}_{ijpq\ell} = \cfrac{1}{(2\ell +1)f_\mathrm{sky}} \left[ \mathbf{C}_{ijpq\ell}^{\mathrm{o*o}}  + \mathbf{C}_{ijpq\ell}^{\mathrm{fg*fg}} - \mathbf{C}_{ijpq\ell}^{\mathrm{fg*o}} + \mathbf{C}_{ijpq\ell}^{\mathrm{CMB*CMB}} + \mathbf{C}_{ijpq\ell}^{\mathrm{CMB*o}} \right],
\end{equation}
%
where we have neglected any potential chance correlations between foreground and CMB signals ($\mathbf{C}^{\mathrm{fg*CMB}}=0$) and included a sky-fraction factor ($f_\mathrm{sky}$) to account for partial sky coverage. We make the following approximations when calculating the covariance matrix: we use observed power spectra instead of theoretical models; we neglect $\ell$-to-$\ell'$ correlations, even in the presence of a mask; we assume that the spherical harmonic coefficients of CMB, noise, and foreground signals are Gaussian and isotropic; and we discard the $C_\ell^{EB}$ terms suppressed by $\sin (x)^n$ with $n\geq1$, since they rapidly fluctuate around zero and make the covariance matrix unstable. 

The covariance of the observed signal is calculated as
\begin{align}
   \begin{split}
   \mathbf{C}_{ijpq\ell}^{\mathrm{o*o}} \approx &  \phantom{+} C_\ell^{E_i^\mathrm{o}E_p^\mathrm{o}} C_\ell^{B_j^\mathrm{o}B_q^\mathrm{o}} + C_\ell^{E_i^\mathrm{o}B_q^\mathrm{o}} C_\ell^{B_j^\mathrm{o}E_p^\mathrm{o}}\\
   & +\cfrac{ \Sin{4\alpha_j}\Sin{4\alpha_q}}{[\Cos{4\alpha_i}+\Cos{4\alpha_j}][\Cos{4\alpha_p}+\Cos{4\alpha_q}]} \left( C_\ell^{E_i^\mathrm{o}E_p^\mathrm{o}} C_\ell^{E_j^\mathrm{o}E_q^\mathrm{o}} + C_\ell^{E_i^\mathrm{o}E_q^\mathrm{o}} C_\ell^{E_j^\mathrm{o}E_p^\mathrm{o}} \right)\\ 
   & +\cfrac{ \Sin{4\alpha_i}\Sin{4\alpha_p}}{[\Cos{4\alpha_i}+\Cos{4\alpha_j}][\Cos{4\alpha_p}+\Cos{4\alpha_q}]} \left( C_\ell^{B_i^\mathrm{o}B_p^\mathrm{o}} C_\ell^{B_j^\mathrm{o}B_q^\mathrm{o}} + C_\ell^{B_i^\mathrm{o}B_q^\mathrm{o}} C_\ell^{B_j^\mathrm{o}B_p^\mathrm{o}}\right) .
   \end{split}
\end{align}

The contribution of all CMB-related terms is
%
\begin{equation}\label{eq:pipeline B CMB contribution to covariance}
   \mathbf{C}_{ijpq\ell}^{\mathrm{CMB*CMB}} + \mathbf{C}_{ijpq\ell}^{\mathrm{CMB*o}} = - \cfrac{\SinS{4\beta}b_\ell^i b_\ell^j b_\ell^p b_\ell^q \omega^{4}_{\mathrm{pix},\ell} }{2\Cos{2\alpha_i+2\alpha_j}\Cos{2\alpha_p+2\alpha_q}} \left[ \left(C_\ell^{EE,\mathrm{CMB}}\right)^2 + \left(C_\ell^{BB,\mathrm{CMB}}\right)^2 \right],
\end{equation}
%
where the combination of frequency bands is specified through the different beam and pixel window functions, $b^i_\ell$ and $\omega_{\mathrm{pix},\ell}$, respectively.

In turn, the covariance associated with both foreground templates can be divided into four terms,
%
\begin{align}
   \begin{split}
    \mathbf{C}_{ijpq\ell}^{\mathrm{fg*fg}} = & \phantom{+} \cfrac{4\Cos{2\alpha_i} \Cos{2\alpha_j} \Cos{2\alpha_p} \Cos{2\alpha_q}}{ [\Cos{4\alpha_i}+ \Cos{4\alpha_j}][\Cos{4\alpha_p}+ \Cos{4\alpha_q}]} \mathbf{C}_{ijpq\ell}^{\mathrm{fg_1}} +  \cfrac{4\Cos{2\alpha_i}\Cos{2\alpha_j}\Sin{2\alpha_p}\Sin{2\alpha_q}}{ [\Cos{4\alpha_i}+\Cos{4\alpha_j}][\Cos{4\alpha_p}+\Cos{4\alpha_q}]} \mathbf{C}_{ijpq\ell}^{\mathrm{fg_2}}  \\
   & + \cfrac{4\Sin{2\alpha_i}\Sin{2\alpha_j}\Cos{2\alpha_p}\Cos{2\alpha_q}}{[\Cos{4\alpha_i}+\Cos{4\alpha_j}][\Cos{4\alpha_p}+\Cos{4\alpha_q}]} \mathbf{C}_{ijpq\ell}^{\mathrm{fg_3}} +  \cfrac{4\Sin{2\alpha_i}\Sin{2\alpha_j}\Sin{2\alpha_p}\Sin{2\alpha_q}}{ [\Cos{4\alpha_i}+\Cos{4\alpha_j}][\Cos{4\alpha_p}+\Cos{4\alpha_q}]} \mathbf{C}_{ijpq\ell}^{\mathrm{fg_4}},
   \end{split}
\end{align}
%
which are calculated from the frequency cross-spectra of synchrotron and dust templates, and their cross-correlations, 
\small
\begin{align}
   \begin{split}
   \mathbf{C}_{ijpq\ell}^{\mathrm{fg_1}} \approx & \phantom{+} (\mathcal{A}^{\mathrm{s}})^2 \left(  C_\ell^{E_i^\mathrm{s}E_p^\mathrm{s}} C_\ell^{B_j^\mathrm{s}B_q^\mathrm{s}} + C_\ell^{E_i^\mathrm{s}B_q^\mathrm{s}} C_\ell^{B_j^\mathrm{s}E_p^\mathrm{s}} \right) + (\mathcal{A}^\mathrm{d})^2 \left(  C_\ell^{E_i^\mathrm{d}E_p^\mathrm{d}} C_\ell^{B_j^\mathrm{d}B_q^\mathrm{d}} + C_\ell^{E_i^\mathrm{d}B_q^\mathrm{d}} C_\ell^{B_j^\mathrm{d}E_p^\mathrm{d}} \right)\\ 
    & + (\mathcal{A}^\mathrm{s\times d})^2 \left(  C_\ell^{E_i^\mathrm{s}E_p^\mathrm{s}} C_\ell^{B_j^\mathrm{d}B_q^\mathrm{d}} + C_\ell^{E_i^\mathrm{s}B_q^\mathrm{d}} C_\ell^{B_j^\mathrm{d}E_p^\mathrm{s}} + C_\ell^{E_i^\mathrm{d}E_p^\mathrm{d}} C_\ell^{B_j^\mathrm{s}B_q^\mathrm{s}} + C_\ell^{E_i^\mathrm{d}B_q^\mathrm{s}} C_\ell^{B_j^\mathrm{s}E_p^\mathrm{d}} \right)\\ 
    & + \mathcal{A}^\mathrm{s}\mathcal{A}^\mathrm{d}\left(  C_\ell^{E_i^\mathrm{s}E_p^\mathrm{d}} C_\ell^{B_j^\mathrm{s}B_q^\mathrm{d}} + C_\ell^{E_i^\mathrm{s}B_q^\mathrm{d}} C_\ell^{B_j^\mathrm{s}E_p^\mathrm{d}} + C_\ell^{E_i^\mathrm{d}E_p^\mathrm{s}} C_\ell^{B_j^\mathrm{d}B_q^\mathrm{s}} + C_\ell^{E_i^\mathrm{d}B_q^\mathrm{s}} C_\ell^{B_j^\mathrm{d}E_p^\mathrm{s}} \right)\\
    & + (\mathcal{A}^\mathrm{s\times d})^2 \left(  C_\ell^{E_i^\mathrm{s}E_p^\mathrm{d}} C_\ell^{B_j^\mathrm{d}B_q^\mathrm{s}} + C_\ell^{E_i^\mathrm{s}B_q^\mathrm{s}} C_\ell^{B_j^\mathrm{d}E_p^\mathrm{d}} + C_\ell^{E_i^\mathrm{d}E_p^\mathrm{s}} C_\ell^{B_j^\mathrm{s}B_q^\mathrm{d}} + C_\ell^{E_i^\mathrm{d}B_q^\mathrm{d}} C_\ell^{B_j^\mathrm{s}E_p^\mathrm{s}} \right)\\
    & + \mathcal{A}^\mathrm{s}\mathcal{A}^\mathrm{s\times d} \left(  C_\ell^{E_i^\mathrm{s}E_p^\mathrm{s}} C_\ell^{B_j^\mathrm{s}B_q^\mathrm{d}} + C_\ell^{E_i^\mathrm{s}B_q^\mathrm{d}} C_\ell^{B_j^\mathrm{s}E_p^\mathrm{s}} + C_\ell^{E_i^\mathrm{s}E_p^\mathrm{s}} C_\ell^{B_j^\mathrm{d}B_q^\mathrm{s}} + C_\ell^{E_i^\mathrm{s}B_q^\mathrm{s}} C_\ell^{B_j^\mathrm{d}E_p^\mathrm{s}} \right)\\
    & + \mathcal{A}^\mathrm{s}\mathcal{A}^\mathrm{s\times d} \left(  C_\ell^{E_i^\mathrm{s}E_p^\mathrm{d}} C_\ell^{B_j^\mathrm{s}B_q^\mathrm{s}} + C_\ell^{E_i^\mathrm{s}B_q^\mathrm{s}} C_\ell^{B_j^\mathrm{s}E_p^\mathrm{d}} + C_\ell^{E_i^\mathrm{d}E_p^\mathrm{s}} C_\ell^{B_j^\mathrm{s}B_q^\mathrm{s}} + C_\ell^{E_i^\mathrm{d}B_q^\mathrm{s}} C_\ell^{B_j^\mathrm{s}E_p^\mathrm{s}} \right)\\
    & + \mathcal{A}^\mathrm{d}\mathcal{A}^\mathrm{s\times d} \left(  C_\ell^{E_i^\mathrm{d}E_p^\mathrm{s}} C_\ell^{B_j^\mathrm{d}B_q^\mathrm{d}} + C_\ell^{E_i^\mathrm{d}B_q^\mathrm{d}} C_\ell^{B_j^\mathrm{d}E_p^\mathrm{s}} + C_\ell^{E_i^\mathrm{s}E_p^\mathrm{d}} C_\ell^{B_j^\mathrm{d}B_q^\mathrm{d}} + C_\ell^{E_i^\mathrm{s}B_q^\mathrm{d}} C_\ell^{B_j^\mathrm{d}E_p^\mathrm{d}} \right)\\
    & + \mathcal{A}^\mathrm{d}\mathcal{A}^\mathrm{s\times d} \left(  C_\ell^{E_i^\mathrm{d}E_p^\mathrm{d}} C_\ell^{B_j^\mathrm{d}B_q^\mathrm{s}} + C_\ell^{E_i^\mathrm{d}B_q^\mathrm{s}} C_\ell^{B_j^\mathrm{d}E_p^\mathrm{d}} + C_\ell^{E_i^\mathrm{d}E_p^\mathrm{d}} C_\ell^{B_j^\mathrm{s}B_q^\mathrm{d}} + C_\ell^{E_i^\mathrm{d}B_q^\mathrm{d}} C_\ell^{B_j^\mathrm{s}E_p^\mathrm{d}} \right),
    \end{split}
\end{align}
\normalsize
\small
\begin{align}
\begin{split}
   \mathbf{C}_{ijpq\ell}^{\mathrm{fg_2}} \approx & \phantom{+}  (\mathcal{A}^\mathrm{s})^2 C_\ell^{E_i^\mathrm{s}E_q^\mathrm{s}} C_\ell^{B_j^\mathrm{s}B_p^\mathrm{s}} + (\mathcal{A}^\mathrm{d})^2 C_\ell^{E_i^\mathrm{d}E_q^\mathrm{d}} C_\ell^{B_j^\mathrm{d}B_p^\mathrm{d}} + \mathcal{A}^\mathrm{s}\mathcal{A}^\mathrm{d} \left(
   C_\ell^{E_i^\mathrm{s}E_q^\mathrm{d}} C_\ell^{B_j^\mathrm{s}B_p^\mathrm{d}} +  C_\ell^{E_i^\mathrm{d}E_q^\mathrm{s}} C_\ell^{B_j^\mathrm{d}B_p^\mathrm{s}}  \right)  \\  
    & + (\mathcal{A}^\mathrm{s\times d})^2 \left( C_\ell^{E_i^\mathrm{s}E_q^\mathrm{d}} C_\ell^{B_j^\mathrm{d}B_p^\mathrm{s}} +  C_\ell^{E_i^\mathrm{d}E_q^\mathrm{s}} C_\ell^{B_j^\mathrm{s}B_p^\mathrm{d}} +  C_\ell^{E_i^\mathrm{s}E_q^\mathrm{s}} C_\ell^{B_j^\mathrm{d}B_p^\mathrm{d}} +  C_\ell^{E_i^\mathrm{d}E_q^\mathrm{d}} C_\ell^{B_j^\mathrm{s}B_p^\mathrm{s}} \right)\\
    & + \mathcal{A}^\mathrm{s}\mathcal{A}^\mathrm{s\times d} \left( C_\ell^{E_i^\mathrm{s}E_q^\mathrm{d}} C_\ell^{B_j^\mathrm{s}B_p^\mathrm{s}} +  C_\ell^{E_i^\mathrm{s}E_q^\mathrm{s}} C_\ell^{B_j^\mathrm{d}B_p^\mathrm{s}} +  C_\ell^{E_i^\mathrm{s}E_q^\mathrm{s}} C_\ell^{B_j^\mathrm{s}B_p^\mathrm{d}} +  C_\ell^{E_i^\mathrm{d}E_q^\mathrm{s}} C_\ell^{B_j^\mathrm{s}B_p^\mathrm{s}} \right)\\
    & + \mathcal{A}^\mathrm{d}\mathcal{A}^\mathrm{s\times d} \left( C_\ell^{E_i^\mathrm{d}E_q^\mathrm{d}} C_\ell^{B_j^\mathrm{d}B_p^\mathrm{s}} +  C_\ell^{E_i^\mathrm{s}E_q^\mathrm{d}} C_\ell^{B_j^\mathrm{d}B_p^\mathrm{d}} +  C_\ell^{E_i^\mathrm{d}E_q^\mathrm{s}} C_\ell^{B_j^\mathrm{d}B_p^\mathrm{d}} +  C_\ell^{E_i^\mathrm{d}E_q^\mathrm{d}} C_\ell^{B_j^\mathrm{s}B_p^\mathrm{d}} \right), 
    \end{split}
\end{align}
\normalsize
\small
\begin{align}
   \begin{split}
   \mathbf{C}_{ijpq\ell}^{\mathrm{fg_3}} \approx & \phantom{+} (\mathcal{A}^\mathrm{s})^2 C_\ell^{B_i^\mathrm{s}B_q^\mathrm{s}} C_\ell^{E_j^\mathrm{s}E_p^\mathrm{s}} + (\mathcal{A}^\mathrm{d})^2 C_\ell^{B_i^\mathrm{d}B_q^\mathrm{d}} C_\ell^{E_j^\mathrm{d}E_p^\mathrm{d}} + \mathcal{A}^\mathrm{s}\mathcal{A}^\mathrm{d} \left( C_\ell^{B_i^\mathrm{d}B_q^\mathrm{s}}C_\ell^{E_j^\mathrm{d}E_p^\mathrm{s}} + C_\ell^{E_j^\mathrm{s}E_p^\mathrm{d}} C_\ell^{B_i^\mathrm{s}B_q^\mathrm{d}} \right) \\ 
   & + (\mathcal{A}^\mathrm{s\times d})^2 \left( C_\ell^{B_i^\mathrm{s}B_q^\mathrm{d}}C_\ell^{E_j^\mathrm{d}E_p^\mathrm{s}} + C_\ell^{E_j^\mathrm{s}E_p^\mathrm{d}} C_\ell^{B_i^\mathrm{d}B_q^\mathrm{s}} +  C_\ell^{B_i^\mathrm{d}B_q^\mathrm{d}}C_\ell^{E_j^\mathrm{s}E_p^\mathrm{s}} + C_\ell^{E_j^\mathrm{d}E_p^\mathrm{d}} C_\ell^{B_i^\mathrm{s}B_q^\mathrm{s}} \right) \\
   & + \mathcal{A}^\mathrm{s}\mathcal{A}^\mathrm{s\times d} \left( C_\ell^{B_i^\mathrm{s}B_q^\mathrm{s}}C_\ell^{E_j^\mathrm{d}E_p^\mathrm{s}} + C_\ell^{E_j^\mathrm{s}E_p^\mathrm{s}} C_\ell^{B_i^\mathrm{s}B_q^\mathrm{d}} +  C_\ell^{B_i^\mathrm{d}B_q^\mathrm{s}}C_\ell^{E_j^\mathrm{s}E_p^\mathrm{s}} + C_\ell^{E_j^\mathrm{s}E_p^\mathrm{d}} C_\ell^{B_i^\mathrm{s}B_q^\mathrm{s}} \right) \\
    & + \mathcal{A}^\mathrm{d}\mathcal{A}^\mathrm{s\times d} \left( C_\ell^{B_i^\mathrm{s}B_q^\mathrm{d}}C_\ell^{E_j^\mathrm{d}E_p^\mathrm{d}} + C_\ell^{E_j^\mathrm{d}E_p^\mathrm{s}} C_\ell^{B_i^\mathrm{d}B_q^\mathrm{d}} +  C_\ell^{B_i^\mathrm{d}B_q^\mathrm{d}}C_\ell^{E_j^\mathrm{s}E_p^\mathrm{d}} + C_\ell^{E_j^\mathrm{d}E_p^\mathrm{d}} C_\ell^{B_i^\mathrm{d}B_q^\mathrm{s}} \right) ,
\end{split}
\end{align}
\normalsize
and
\small
\begin{align}
\begin{split}
   \mathbf{C}_{ijpq\ell}^{\mathrm{fg_4}} \approx & \phantom{+}   (\mathcal{A}^\mathrm{s})^2 C_\ell^{B_i^\mathrm{s}B_p^\mathrm{s}} C_\ell^{E_j^\mathrm{s}E_q^\mathrm{s}} + (\mathcal{A}^\mathrm{d})^2 C_\ell^{B_i^\mathrm{d}B_p^\mathrm{d}} C_\ell^{E_j^\mathrm{d}E_q^\mathrm{d}} + \mathcal{A}^\mathrm{s}\mathcal{A}^\mathrm{d}\left(  C_\ell^{B_i^\mathrm{s}B_p^\mathrm{d}} C_\ell^{E_j^\mathrm{s}E_q^\mathrm{d}} + C_\ell^{B_i^\mathrm{d}B_p^\mathrm{s}} C_\ell^{E_j^\mathrm{d}E_q^\mathrm{s}} \right) \\
    & + (\mathcal{A}^\mathrm{s\times d})^2\left(  C_\ell^{B_i^\mathrm{s}B_p^\mathrm{s}} C_\ell^{E_j^\mathrm{d}E_q^\mathrm{d}} +  C_\ell^{B_i^\mathrm{d}B_p^\mathrm{d}} C_\ell^{E_j^\mathrm{s}E_q^\mathrm{s}} +  C_\ell^{B_i^\mathrm{s}B_p^\mathrm{d}} C_\ell^{E_j^\mathrm{d}E_q^\mathrm{s}} +  C_\ell^{B_i^\mathrm{d}B_p^\mathrm{s}} C_\ell^{E_j^\mathrm{s}E_q^\mathrm{d}} \right)\\
    & + \mathcal{A}^\mathrm{s}\mathcal{A}^\mathrm{s\times d}\left(  C_\ell^{B_i^\mathrm{s}B_p^\mathrm{s}} C_\ell^{E_j^\mathrm{s}E_q^\mathrm{d}} +  C_\ell^{B_i^\mathrm{s}B_p^\mathrm{s}} C_\ell^{E_j^\mathrm{d}E_q^\mathrm{s}} +  C_\ell^{B_i^\mathrm{s}B_p^\mathrm{d}} C_\ell^{E_j^\mathrm{s}E_q^\mathrm{s}} +  C_\ell^{B_i^\mathrm{d}B_p^\mathrm{s}} C_\ell^{E_j^\mathrm{s}E_q^\mathrm{s}} 
      \right)\\
    & + \mathcal{A}^\mathrm{d}\mathcal{A}^\mathrm{s\times d}\left(  C_\ell^{B_i^\mathrm{d}B_p^\mathrm{s}} C_\ell^{E_j^\mathrm{d}E_q^\mathrm{d}} +  C_\ell^{B_i^\mathrm{s}B_p^\mathrm{d}} C_\ell^{E_j^\mathrm{d}E_q^\mathrm{d}} +  C_\ell^{B_i^\mathrm{d}B_p^\mathrm{d}} C_\ell^{E_j^\mathrm{d}E_q^\mathrm{s}} +  C_\ell^{B_i^\mathrm{d}B_p^\mathrm{d}} C_\ell^{E_j^\mathrm{s}E_q^\mathrm{d}} 
      \right).
\end{split}
\end{align}
\normalsize

Finally, the cross-correlation between the observations and both foreground templates is given as
\small
\begin{align}
\begin{split}
   \mathbf{C}_{ijpq\ell}^{\mathrm{fg*o}} \approx & \phantom{+} \cfrac{2\Cos{2\alpha_i}\Cos{2\alpha_j}}{\Cos{4\alpha_i}+\Cos{4\alpha_j}} \left[ {\cal A}^\mathrm{s}\left( C_\ell^{E_i^\mathrm{s}E_p^\mathrm{o}} C_\ell^{B_j^\mathrm{s}B_q^\mathrm{o}} + C_\ell^{E_i^\mathrm{s}B_q^\mathrm{o}} C_\ell^{B_j^\mathrm{s}E_p^\mathrm{o}}\right) + {\cal A}^\mathrm{d}\left( C_\ell^{E_i^\mathrm{d}E_p^\mathrm{o}} C_\ell^{B_j^\mathrm{d}B_q^\mathrm{o}} + C_\ell^{E_i^\mathrm{d}B_q^\mathrm{o}} C_\ell^{B_j^\mathrm{d}E_p^\mathrm{o}}\right)\right]\\
    & + \cfrac{2\Cos{2\alpha_i}\Cos{2\alpha_j}}{\Cos{4\alpha_i}+\Cos{4\alpha_j}} {\cal A}^\mathrm{s\times d}\left( C_\ell^{E_i^\mathrm{s}E_p^\mathrm{o}} C_\ell^{B_j^\mathrm{d}B_q^\mathrm{o}} + C_\ell^{E_i^\mathrm{s}B_q^\mathrm{o}} C_\ell^{B_j^\mathrm{d}E_p^\mathrm{o}} +  C_\ell^{E_i^\mathrm{d}E_p^\mathrm{o}} C_\ell^{B_j^\mathrm{s}B_q^\mathrm{o}} + C_\ell^{E_i^\mathrm{d}B_q^\mathrm{o}} C_\ell^{B_j^\mathrm{s}E_p^\mathrm{o}}\right)\\
   & + \cfrac{2\Cos{2\alpha_p}\Cos{2\alpha_q}}{\Cos{4\alpha_p}+\Cos{4\alpha_q}} \left[ {\cal A}^\mathrm{s}\left( C_\ell^{E_i^\mathrm{o}E_p^\mathrm{s}} C_\ell^{B_j^\mathrm{o}B_q^\mathrm{s}} + C_\ell^{E_i^\mathrm{o}B_q^\mathrm{s}}C_\ell^{B_j^\mathrm{o}E_p^\mathrm{s}} \right) + {\cal A}^\mathrm{d}\left( C_\ell^{E_i^\mathrm{o}E_p^\mathrm{d}} C_\ell^{B_j^\mathrm{o}B_q^\mathrm{d}} + C_\ell^{E_i^\mathrm{o}B_q^\mathrm{d}}C_\ell^{B_j^\mathrm{o}E_p^\mathrm{d}} \right)\right]\\  
   & + \cfrac{2\Cos{2\alpha_p}\Cos{2\alpha_q}}{\Cos{4\alpha_p}+\Cos{4\alpha_q}} {\cal A}^\mathrm{s\times d}\left( C_\ell^{E_i^\mathrm{o}E_p^\mathrm{s}} C_\ell^{B_j^\mathrm{o}B_q^\mathrm{d}} + C_\ell^{E_i^\mathrm{o}B_q^\mathrm{d}}C_\ell^{B_j^\mathrm{o}E_p^\mathrm{s}} + C_\ell^{E_i^\mathrm{o}E_p^\mathrm{d}} C_\ell^{B_j^\mathrm{o}B_q^\mathrm{s}} + C_\ell^{E_i^\mathrm{o}B_q^\mathrm{s}}C_\ell^{B_j^\mathrm{o}E_p^\mathrm{d}} \right)\\    
   & + \cfrac{2\Sin{2\alpha_i}\Sin{2\alpha_j}}{\Cos{4\alpha_i}+\Cos{4\alpha_j}} \left[ {\cal A}^\mathrm{s}C_\ell^{B_i^\mathrm{s}B_q^\mathrm{o}} C_\ell^{E_j^\mathrm{s}E_p^\mathrm{o}} + {\cal A}^\mathrm{d}C_\ell^{B_i^\mathrm{d}B_q^\mathrm{o}} C_\ell^{E_j^\mathrm{d}E_p^\mathrm{o}} + {\cal A}^\mathrm{s\times d}\left( C_\ell^{B_i^\mathrm{s}B_q^\mathrm{o}} C_\ell^{E_j^\mathrm{d}E_p^\mathrm{o}} + C_\ell^{B_i^\mathrm{d}B_q^\mathrm{o}}C_\ell^{E_j^\mathrm{s}E_p^\mathrm{o}}\right)\right] \\ 
   & + \cfrac{2\Sin{2\alpha_p}\Sin{2\alpha_q}}{\Cos{4\alpha_p}+\Cos{4\alpha_q}} \left[  {\cal A}^\mathrm{s}C_\ell^{E_i^\mathrm{o}E_q^\mathrm{s}}C_\ell^{B_j^\mathrm{o}B_p^\mathrm{s}} + {\cal A}^\mathrm{d}C_\ell^{E_i^\mathrm{o}E_q^\mathrm{d}}C_\ell^{B_j^\mathrm{o}B_p^\mathrm{d}} + {\cal A}^\mathrm{s\times d}\left( C_\ell^{E_i^\mathrm{o}E_q^\mathrm{d}} C_\ell^{B_j^\mathrm{o}B_p^\mathrm{s}} + C_\ell^{E_i^\mathrm{o}E_q^\mathrm{s}}C_\ell^{B_j^\mathrm{o}B_p^\mathrm{d}}\right) \right]
\end{split}
\end{align}
\normalsize